\documentclass[a4paper,11pt]{article}
\pdfoutput=1

\usepackage{jheppub}
\usepackage{graphicx}
\usepackage{amsmath}
\usepackage{bm}
\usepackage{cancel}
\usepackage{booktabs}
\usepackage[small]{subfigure}
\usepackage{url}

\usepackage[normalem]{ulem} 
\usepackage[usenames,dvipsnames]{xcolor}

\newcommand{\eq}[1]{eq.~\eqref{eq:#1}}
\newcommand{\eqs}[2]{eqs.~\eqref{eq:#1} and \eqref{eq:#2}}
\renewcommand{\sec}[1]{section~\ref{sec:#1}}

\newcommand{\subsec}[1]{section~\ref{subsec:#1}}
\newcommand{\subsubsec}[1]{section~\ref{subsubsec:#1}}
\newcommand{\subsecs}[2]{sections~\ref{subsec:#1} and \ref{subsec:#2}}
\newcommand{\app}[1]{Appendix~\ref{app:#1}}
\newcommand{\fig}[1]{figure~\ref{fig:#1}}

\newcommand{\mycite}[1]{ref.~\cite{#1}}
\newcommand{\mycites}[1]{refs.~\cite{#1}}

\newcommand{\ord}[1]{\mathcal{O}(#1)}
\newcommand{\Ord}[1]{\mathcal{O}\biggl(#1\biggr)}

\newcommand{\Mae}[3]{\bigl\langle#1 \bigl\lvert#2\bigr\rvert#3\bigr\rangle}

\newcommand{\plus}[1]{\left[#1\right]_{+}}

\newcommand{\ABr}[3]{\left\langle #1 \left\lvert #2 \right\rvert #3 \right\rangle}

\newcommand{\mylog}[1]{\ln\left( #1 \right)}

\newcommand{\df}{\mathrm{d}}

\newcommand{\MSb}{\overline{\rm MS}}

\newcommand{\muf}{\mu_F}
\newcommand{\mur}{\mu_R}
\newcommand{\muH}{\mu_H}
\newcommand{\muL}{\mu_{\Lambda}}
\newcommand{\mum}{\mu_{m}}
\newcommand{\as}{\alpha_s}

\newcommand{\mh}{m_H}
\newcommand{\mb}{m_b}

\renewcommand{\Im}{\mathrm{Im}}

\def\bea#1\eea{\begin{align}#1\end{align}}
\def\beq{\begin{equation}}
\def\eeq{\end{equation}}
\let\originalleft\left
\let\originalright\right
\renewcommand{\left}{\mathopen{}\mathclose\bgroup\originalleft}
\renewcommand{\right}{\aftergroup\egroup\originalright}
\def\({\left(}
\def\){\right)}
\def\[{\left[}
\def\]{\right]}

\newcommand{\GeV}{\,\mathrm{GeV}}
\newcommand{\TeV}{\,\mathrm{TeV}}

\newcommand{\nn}{\nonumber}

\newcommand{\lqcd}{\Lambda_\mathrm{QCD}}

\newcommand{\DIS}{\mathrm{DIS}}

\newcommand{\Cm}{D}

\newcommand{\Mm}{\mathcal{M}}

\newcommand{\nfour}{{{[4]}}}
\newcommand{\nfive}{{{[5]}}}

\newcommand{\FO}{\mathrm{FO}}
\newcommand{\resum}{\mathrm{resum}}
\newcommand{\nons}{\mathrm{nons}}
\newcommand{\sing}{\mathrm{sing}}

\newcommand{\f}{\tilde{f}}

\allowdisplaybreaks[2]

\title{\boldmath Resummation and Matching of $b$-quark Mass Effects in $b\bar bH$ Production}

\author[a]{Marco Bonvini,}
\author[b]{Andrew S.~Papanastasiou,}
\author[b]{and Frank J.~Tackmann}
\affiliation[a]{Rudolf Peierls Center for Theoretical Physics, 1 Keble Road, University of Oxford, OX1 3NP Oxford, UK}
\affiliation[b]{Theory Group, Deutsches Elektronen-Synchrotron (DESY), Notkestra\ss e 85, D-22607 Hamburg, Germany}

\emailAdd{marco.bonvini@physics.ox.ac.uk}
\emailAdd{andrew.papanastasiou@desy.de}
\emailAdd{frank.tackmann@desy.de}

\abstract{
We use a systematic effective field theory setup
to derive the $b\bar b H$ production cross section. Our result combines the merits of both
fixed 4-flavor and 5-flavor schemes. It contains the full 4-flavor result, including the exact
dependence on the $b$-quark mass, and improves it with a resummation of collinear logarithms of $\mb/\mh$.
In the massless limit, it corresponds to a reorganized 5-flavor result.
While we focus on $b\bar bH$ production,
our method applies to generic heavy-quark initiated processes at hadron colliders.
Our setup resembles the variable flavor number schemes known from heavy-flavor production
in deep-inelastic scattering, but also differs in some key aspects.
Most importantly, the effective $b$-quark PDF appears as part of the perturbative expansion of the final result
where it effectively counts as an $\ord{\as}$ object.
The transition between the fixed-order (4-flavor) and resummation (5-flavor) regimes is governed by
the low matching scale at which the $b$-quark is integrated out. Varying this scale
provides a systematic way to assess the perturbative uncertainties associated with the
resummation and matching procedure and reduces by going to higher orders.
We discuss the practical implementation and present numerical results for the $b\bar bH$ production
cross section at NLO+NLL. We also provide a comparison to the corresponding predictions
in the fixed 4-flavor and 5-flavor results and the Santander matching prescription.
Compared to the latter, we find a slightly reduced uncertainty and a larger central value,
with its central value lying at the lower edge of our uncertainty band.
}


\preprint{
\begin{flushright}
OUTP-15-16P\\
DESY 15-137\\
August 28, 2015
\end{flushright}
}

\begin{document}

\maketitle

\section{Introduction}
\label{sec:intro}

The formulation of reliable predictions for heavy-quark initiated processes has been the subject of 
much study over many years, in particular for the determination
of parton distribution functions (PDFs) in the context of deep-inelastic scattering (DIS)~\cite{Aivazis:1993kh, Aivazis:1993pi, Thorne:1997ga, Kretzer:1998ju, Collins:1998rz, Kramer:2000hn,Tung:2001mv, Thorne:2006qt, Bierenbaum:2008yu, Bierenbaum:2009zt, Forte:2010ta, Alekhin:2010sv, Guzzi:2011ew, Kawamura:2012cr, Alekhin:2012vu, Ablinger:2014vwa, Ablinger:2014nga}.
At the LHC, important examples of heavy-quark initiated processes are Higgs or vector-boson
production in association with heavy quarks.
In this paper, we are interested in Higgs production in association with $b$ quarks, i.e., the
inclusive $b\bar b H$-induced cross section.

In a typical hard-scattering process with protons in the initial state, there are at least two
parametrically separate scales. First, the hard scale $\muH \sim Q$, where $Q$ denotes the physical quantity
that determines the momentum transfer in the hard interaction, e.g., $Q = \sqrt{-q^2}$ in DIS or $Q = m_H$ for the case of
Higgs production that we will be interested in. Second, the low scale $\muL\sim \lqcd$, which separates
the perturbative and nonperturbative regimes and is typically taken to be of order the proton mass,
$\muL \sim 1\GeV$. In the limit $\muL \ll Q$ we can apply the standard QCD factorization
theorem~\cite{Bodwin:1984hc, Collins:1985ue, Collins:1988ig} to
compute the hadronic cross section in terms of the partonic cross section convolved with PDFs.

\begin{figure}[t]
\hfill%
\subfigure[$m_b\sim Q$]{\includegraphics[scale=0.75]{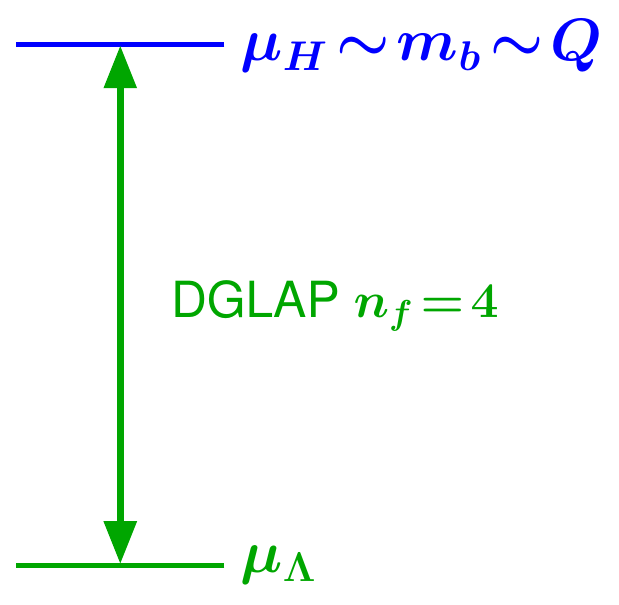}%
\label{fig:FO}}%
\hfill%
\subfigure[$m_b \ll Q$]{\includegraphics[scale=0.75]{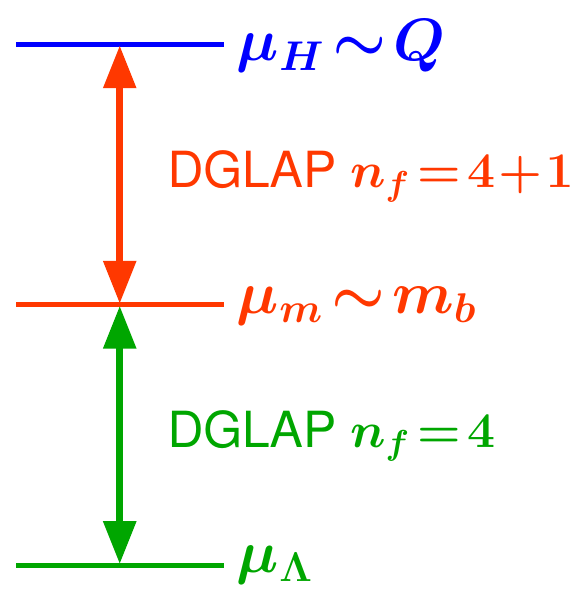}%
\label{fig:resum}}%
\hspace*{\fill}%
\caption{The two parametric scale hierarchies for inclusive cross sections for heavy-quark initiated processes.}
\label{fig:scale-hierarchy}
\end{figure}

For heavy-quark initiated processes, the mass $m$ of the heavy quark introduces another physical scale.
Depending on its value, we can distinguish two parametrically different cases, shown in \fig{scale-hierarchy}:%
\footnote{In principle, there is a third parametric limit $m \gg Q$, which we are not interested in. In this case, when the heavy quark appears as an external state, $m$ itself is the physical quantity that sets the hard interaction scale, so $Q\equiv m$. The relevant setup is then determined by what other parametrically smaller physical scales are present in the process. Otherwise, when the heavy quark only appears in internal loops, it can simply be integrated out.}
\begin{itemize}
\item[(a)] $m \sim Q$: There is a single parametric scale $\muH \sim m\sim Q$ in addition to $\muL$.
\item[(b)] $m \ll Q$: There are two parametric scales $\muH \sim Q$ and $\mum \sim m$ in addition to $\muL$.
\end{itemize}
When working in the limit $m\sim Q$, the heavy quark never appears in
the initial state of the hard partonic process. Instead, it is
produced as part of the hard interaction at $\muH$ by an incoming
gluon splitting into a pair of heavy quarks. The partonic calculation
contains the exact dependence on $m$, including the correct
$m$-dependent phase space. The gluon splitting into a heavy-quark pair
contains a collinear singularity, which is regulated by $m$, and as a
result produces logarithms $\ln(m/Q)$. For $m\sim Q$, these collinear
logarithms are counted as small and are included at fixed order in the
$\as$ expansion.

When working in the limit $m\ll Q$, the heavy quark explicitly appears
in the initial state of the hard partonic process, and the collinear
logarithms are resummed to all orders in $\as$ into an effective heavy-quark
PDF. The quark mass $m$ only appears in the boundary condition of the
PDF's DGLAP evolution, which starts at the scale $\mum\sim m$. The
hard process itself is computed in the $m\to 0$ limit. That is,
finite-mass effects of $\ord{m/Q}$, including the exact phase space of
the gluon splitting into a massive quark pair, are power corrections
and are neglected.

Predictions obtained in strictly one of the above two limits are usually referred to as obtained in a fixed-flavor number scheme.
Which of these limits is more appropriate in practice depends on the process and the numerical size of the corrections.
For $b$-initiated processes at hadron colliders, the relative importance of $\ln(\mb/Q)$ corrections has been discussed
for example in \cite{Maltoni:2003pn, Maltoni:2005wd, Maltoni:2012pa}.

To obtain the best possible theoretical predictions, it is often desirable to have a complete description that incorporates the 
results from both limits. In this way, the final result is valid in each limit as well as in the transition region in between, 
and hence one can be agnostic about which parametric regime is the more appropriate one.

For $b\bar b H$, predictions exist in the 4-flavor scheme (4FS)~\cite{Dittmaier:2003ej,Dawson:2003kb}, which works in the limit
$\mb\sim m_H$, and in the 5-flavor scheme (5FS)~\cite{Dicus:1998hs,Balazs:1998sb,Harlander:2003ai,Buehler:2012cu}, which works
in the limit $\mb \ll m_H$.  Currently, both predictions are combined
using the pragmatic ``Santander Matching'' prescription~\cite{Harlander:2011aa},
which is a weighted average of the 4FS and 5FS predictions, where the relative weighting depends on
the numerical size of $\ln(\mb/m_H)$.

There are various methods available in the literature, referred to as variable-flavor number schemes (VFNS), which
aim to combine the virtues of both limits in a more systematic fashion.
That is, they include the full $m$ dependence in the limit $m\sim Q$ and the
resummation of collinear logarithms $\ln(m/Q)$ in the limit $m \ll Q$. There are a number of such
schemes available, namely the ACOT scheme \cite{Aivazis:1993kh, Aivazis:1993pi}
(and its simplified variants S-ACOT \cite{Kramer:2000hn}, S-ACOT-$\chi$ \cite{Tung:2001mv, Guzzi:2011ew}, and the more recent
m-ACOT~\cite{Han:2014nja} for hadron-hadron collisions),
the TR scheme \cite{Thorne:1997ga, Thorne:2006qt}, and the FONLL scheme \cite{CacciariFONLL, Forte:2010ta}.
The differences between the schemes essentially amount to how the two limits are combined.

Effective field theories (EFTs) are the standard tool to describe processes
with parametrically separated scales, allowing to systematically resum
the logarithms of ratios of these scales.  In this paper, we discuss the
EFT formulation of heavy-quark initiated processes for the case of inclusive
cross sections. All the basic ingredients are actually well known in this case.
Nevertheless, we find it worthwhile to discuss the EFT formulation in detail,
as it provides a conceptually clear field-theoretic derivation, including
the transition between the two parametric regimes and a way to assess the associated
theoretical uncertainties. This setup can also be extended to more differential cross
sections, which we leave for future work.
A similar setup has also been used to incorporate quark-mass effects for final-state jets
in Refs.~\cite{Gritschacher:2013pha,Pietrulewicz:2014qza,Hoang:2015iva}.

Our final result for DIS resembles the aforementioned schemes in several ways,
but also differs in some key aspects. Most importantly, the $b$-quark
PDF is not treated as an external $\ord{1}$ quantity. Rather, it contributes as part of the perturbative
series of the final result, where it effectively counts as an $\ord{\alpha_s}$ object.
(In this work, we follow the assumption made in all available PDF sets that there is
no intrinsic bottom in the proton, such that the effective bottom-quark PDF is generated purely perturbatively.)

The application of our method to hadron-hadron collisions is completely straightforward.
Our final result for $b\bar b H$ encompasses the merits of both 4F and 5F schemes. It
contains the full 4FS result at NLO, including the exact $\mb$ dependence and phase space. In addition,
it improves the 4FS result with the all-order resummation of collinear logarithms up to NLL order.
In the $\mb\to 0$ limit, our result corresponds to a reorganized 5FS result, where the perturbative series
is expanded to NLO with the $b$-quark PDF counted as $\ord{\alpha_s}$.

In the next section, we discuss the general setup in detail, focussing on DIS to be specific.
In \sec{comparison} we briefly discuss the similarities and differences with respect
to other heavy-flavor schemes in the literature. Then in \sec{bbh}, we apply this framework to
$b\bar bH$ production. We discuss in detail the perturbative uncertainties and present our final
numerical results at LO$+$LL and NLO$+$NLL. We also compare to the predictions in the
4F and 5F schemes using a consistent set of inputs. We conclude in \sec{conclusions}.

\section{EFT formulation of heavy-quark initiated processes}
\label{sec:EFT}

In this section, we discuss the EFT formulation in detail.
For simplicity and to be specific we frame the discussion in the context of heavy-quark production in DIS,
where we have to deal with only one strongly-interacting initial
state. In this case we associate $Q = \sqrt{-q^2}$.  The extension to hadron-hadron collisions is straightforward
and will be discussed in \sec{bbh}.
For definiteness we consider the heavy quark to be the $b$ quark,%
\footnote{Our setup can be equally applied to processes involving the
  top quark.  For the charm quark, the low value of its
  mass might not justify the treatment $m_c\gg\muL$, which would mean
  that $\muL/m_c$ corrections are important. In this case, a better
  treatment would be to take $m_c\sim \muL$ and not integrate out the charm
  quark, but instead consider a nonperturbative charm PDF.}
and treat the four lighter quarks as massless. We take
$Q < m_t$, so we can essentially ignore the top quark (i.e., we either
integrate it out at the scale $\muH\sim Q$ or it has already been
integrated out at a higher scale).

In \subsec{FO}, we review the case $\muL \ll \mb\sim Q$, corresponding
to \fig{FO}, where a single matching step at the hard scale
$\muH\sim Q\sim \mb$ is required. We will refer to this as the
fixed-order region or limit.  This also serves to introduce our
notation and language.  In \subsec{resum}, we discuss the case
$\muL \ll \mb \ll Q$, corresponding to \fig{resum}, where two separate
matching steps, at $\muH\sim Q$ and $\mum\sim \mb$, are performed. We
will refer to this as the resummation region or limit. In
\subsec{powercounting}, we discuss the appropriate perturbative
counting in our result, and in \subsec{FOtransition}, we combine the
results in both limits to yield our final predictions valid in both
limits and anywhere in between.

Throughout this paper, we use roman indices $i, j, k$ to denote the light flavors, i.e., the four light quarks 
and the gluon. We also use the convention that any repeated indices are implicitly summed over
(also a repeated index $b$ implies a sum over $b$ and $\bar b$).
For clarity, we will focus on the dependence on the relevant physical and renormalization scales, but suppress all other
kinematic dependences. In particular, we will not write out the dependence on the momentum fractions and the Mellin-type 
convolutions in them. We will denote the number $n_f$ of light active flavors as superscripts for quantities where the 
distinction is relevant, e.g., $\alpha_s^\nfour$ vs.\ $\as^\nfive$.

\subsection{$\mb \sim Q$: Fixed order}
\label{subsec:FO}

In this case, shown in \fig{FO}, the $b$-quark mass is treated parametrically as of the same size as $Q$.
At the scale $\muH \sim Q \sim \mb$, all degrees of freedom with virtualities $\sim Q^2 \sim \mb^2$, including the heavy $b$ quark,
are integrated out. We match full QCD onto a theory of collinear gluons and collinear light quarks with
typical virtuality $\lqcd^2$.\footnote{%
In SCET, this is the purely $n$-collinear quark and gluon sector, which is equivalent to a boosted version of QCD, where
$n^\mu = (1, \vec{n})$ and $\vec{n}$ is the direction of the incoming proton.
In lightcone coordinates, the momentum of the collinear modes scales as $p_c \sim (Q, \lqcd^2/Q, \lqcd)$.
In principle, there could also be soft modes with momentum scaling $p_s \sim (\lqcd, \lqcd, \lqcd)$, and also Glauber modes.
Since their contributions cancel in the inclusive cross section~\cite{Collins:1988ig}, they are not needed here.}
This matching step is precisely equivalent to the standard operator product expansion (OPE) in DIS~\cite{Wilson:1969zs, Christ:1972ms, Gross:1973ju, Gross:1974cs, Georgi:1951sr, Witten:1975bh, Georgi:1976ve, Altarelli:1977zs, Bardeen:1978yd, Collins:1981uw}, which we briefly review now.

We define the DIS operator $O_\DIS(Q, \mb)$, whose 
proton matrix element determines the DIS cross section (or equivalently the hadronic tensor or DIS structure functions),
\begin{align} \label{eq:ODIS}
\df\sigma(Q, \mb) &= \Mae{p}{O_\DIS(Q, \mb)}{p}\,.
\end{align}
At the scale $\muH$, it is matched onto a sum of nonlocal PDF operators
\begin{equation} \label{eq:4Fmatching}
O_{\rm DIS}(Q,\mb)
= \Bigl[ \Cm_i(Q,\mb,\muH) \otimes O_i^\nfour(\muH) \Bigr]
\biggl[1 + \Ord{\frac{\lqcd^2}{Q^2}}\biggr]
\,,\end{equation}
where a sum over light quarks and gluons $i=u,d,s,c,g$ is understood, and ``$\otimes$'' denotes the Mellin-type convolutions in the momentum fractions. The Wilson coefficients $\Cm_i(Q,\mb,\muH)$ are also called coefficient functions. The $O_i^\nfour(\mu)$ are the standard $\MSb$-renormalized quark and gluon PDF operators~\cite{Collins:1981uw}%
\footnote{For corresponding operator definitions in SCET and a discussion of their equivalence see e.g.~\mycites{Bauer:2002nz, Stewart:2010qs, Gaunt:2014xga}.}, whose proton matrix elements define the nonperturbative PDFs,
\begin{equation} \label{eq:4FPDFs}
f^\nfour_i(\mu) = \Mae{p}{O_i^\nfour(\mu)}{p}
\,.\end{equation}
Since the $b$ quark is being integrated out and not present in the theory below $\muH$, there is also no $O_b$ operator and no $\Cm_b$ coefficient on the right-hand side of \eq{4Fmatching}. As indicated, the right-hand side of \eq{4Fmatching} is the leading term in an expansion in $\lqcd^2/Q^2$, where the low scale $\lqcd^2$ is set by the external proton state we are eventually interested in. For ease of notation, we will not indicate these power corrections in the rest of this section.

In the above, $\alpha_s(\mu) \equiv \alpha_s^\nfour(\mu)$ and $O_i^\nfour(\mu)$ are renormalized with $n_f = 4$ active quark flavors. That is, we use $\MSb$ with dimensional regularization with respect to the four light quark flavors, while $b$-quark loops are renormalized in the decoupling scheme, such that the $b$-quark decouples from the theory below $\muH$ (see \app{ren}).

Since $O_\DIS$ determines the full-theory cross section, it does not have an explicit dependence on $\muH$, i.e., it does not
receive additional operator renormalization. It only has an implicit dependence on $\muH$ through the renormalization of
$\alpha_s(\muH)$, which cancels order by order in perturbation theory. On the other hand, the coefficients $\Cm_i$
are explicitly $\mu$ dependent, and their $\mu$ dependence cancels against the explicit $\mu$ dependence of the operators $O_i(\mu)$.

The full dependence on the physical scales $Q$ and $\mb$, which are treated as hard scales, resides in the Wilson coefficients 
$\Cm_i(Q, \mb, \muH)$. The coefficients $\Cm_i$ at some scale $\mu$ contain logarithms $\ln(\mu/Q)\sim\ln(\mu/\mb)$. 
Therefore, they are computed by a perturbative matching calculation (see below) at the hard scale $\mu_H\sim Q\sim \mb$,
where they contain no large logarithms.

The PDFs $f_i(\mu)$ at some scale $\mu$ contain logarithms $\ln(\mu/\lqcd)$. Hence, the input PDFs that are determined from the
experimental data are defined at a low scale $\muL \gtrsim \lqcd$, which should still be large enough for perturbation
theory to be valid. All contributions from lower scales, including the nonperturbative regime, are absorbed into the input 
PDFs $f_i(\muL)$. The renormalization of the PDF operators leads to their renormalization group equation (RGE)
\begin{align}
\mu \frac{\df}{\df\mu}\, O_i^\nfour(\mu) &= \gamma_{ij}^\nfour(\mu) \otimes O_j^\nfour(\mu)
\,,\end{align}
where $\gamma_{ij}$ are the PDF anomalous dimensions, which are given in terms of the standard Altarelli-Parisi splitting
functions~\cite{Gross:1973ju, Georgi:1951sr, Altarelli:1977zs}. The solution of this RGE yields the standard DGLAP evolution~\cite{Gribov:1972ri, Altarelli:1977zs, Dokshitzer:1977sg} relating the operators (and PDFs) at a scale $\mu_0$ to the operators (and PDFs) at the scale $\mu$,
\begin{equation} \label{eq:Oevolution}
O_i^\nfour(\mu) = U_{ij}^\nfour(\mu,\mu_0) \otimes O_j^\nfour(\mu_0)
\,.\end{equation}
As denoted, the anomalous dimensions and evolution factors involve $n_f = 4$ light quark flavors.
By definition, the coefficients and operators in \eq{4Fmatching} must be evaluated at the same scale, which means the operators 
on the right-hand side give PDFs at $\mu_H$ containing large logarithms $\ln(\muH/\muL)$. These logarithms are resummed by 
using \eq{Oevolution} to evolve the PDFs from the low scale $\muL$ up to $\muH$,
\begin{equation} \label{eq:PDFevolution}
f_i^\nfour(\mu_H) = U_{ij}^\nfour(\mu_H,\muL) \otimes f_j^\nfour(\muL)
\,.\end{equation}

Equivalently, we can perform the resummation for the Wilson coefficients. The coefficients and operators obey inverse RGEs, 
since their scale dependences must cancel each other. After performing the matching at the scale $\mu_H$, the coefficients 
are evolved from $\muH$ down to $\muL$,
\begin{equation} \label{eq:Cevolution}
\Cm_j(Q, \mb, \muL) = \Cm_i(Q, \mb, \muH) \otimes U_{ij}^\nfour(\muH,\muL)
\,.\end{equation}
The evolution factor is precisely the same as in \eq{PDFevolution}. Evolving the coefficients down corresponds to successively 
integrating out virtualities between $\muH^2$ and $\muL^2$. After evolving down to $\muL$, we can take the proton matrix 
element to obtain the final DIS cross section
\begin{equation} \label{eq:EFT1_factorization}
\df\sigma^\FO(Q, \mb)
= \Cm_i(Q,\mb,\muH) \otimes U_{ij}^\nfour(\muH,\muL) \otimes f^\nfour_{j}(\muL)
\,.\end{equation}
The full cross section on the left-hand side contains large logarithms $\ln(Q/\lqcd)$ and $\ln(\mb/\lqcd)$, which on the right-hand side are 
factorized into logarithms $\ln(Q/\muH)$ and $\ln(\mb/\muH)$, which are considered small and reside in the coefficients, large logarithms 
$\ln(\muH/\muL)$, which are resummed into the evolution factor, and logarithms $\ln(\muL/\lqcd)$, which are absorbed into the PDFs. 
This result for the cross section is precisely the 4FS result. Since the $\mb$ dependence is included at fixed order in \eq{EFT1_factorization}, we will refer to it as the fixed-order (``FO'') result.

\subsubsection{Matching at $\muH \sim Q \sim \mb$}
\label{subsubsec:4FmuHmatching}

\begin{figure}[t]
\centering
$\Im$ \parbox{12ex}{\includegraphics[trim=5.0cm 19.6cm 10.9cm 4.4cm,clip,width=2cm]{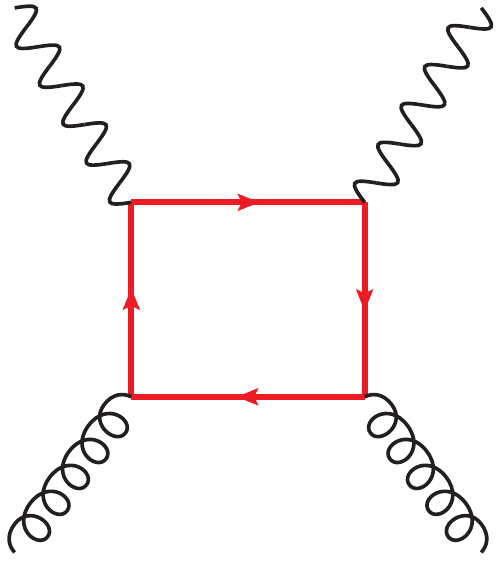}}
$= \Cm_g^{(1)}(Q,\mb,\muH)$
\parbox{7ex}{\includegraphics[trim=5.0cm 22.6cm 13.3cm 4.4cm,clip,width=1.3cm]{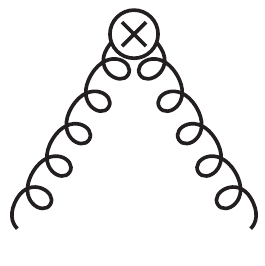}}
\caption{Schematic leading-order matching for heavy-quark production in DIS for $\mb \sim Q$.}
\label{fig:LO4FSDIS}
\end{figure}

\begin{figure}[t]
\centering
$\Im$ \parbox{12ex}{\includegraphics[trim=5.0cm 19.6cm 10.9cm 4.4cm,clip,width=2cm]{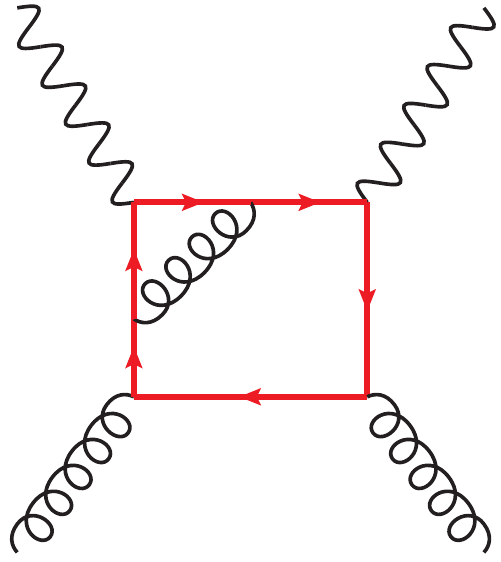}} $+ \dotsb$
$= \Cm_g^{(1)}(Q,\mb,\muH) \biggl($
\parbox{7ex}{\includegraphics[trim=5.0cm 22.6cm 13.3cm 4.4cm,clip,width=1.3cm]{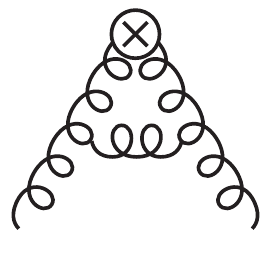}} 
$+ \dotsb \biggr) + \Cm_g^{(2)}(Q,\mb,\muH)$
\parbox{7ex}{\includegraphics[trim=5.0cm 22.6cm 13.3cm 4.4cm,clip,width=1.3cm]{figures/pdf_g_0.pdf}}
\\
$\Im$ \parbox{12ex}{\includegraphics[trim=5.0cm 19.6cm 10.9cm 4.0cm,clip,width=2cm]{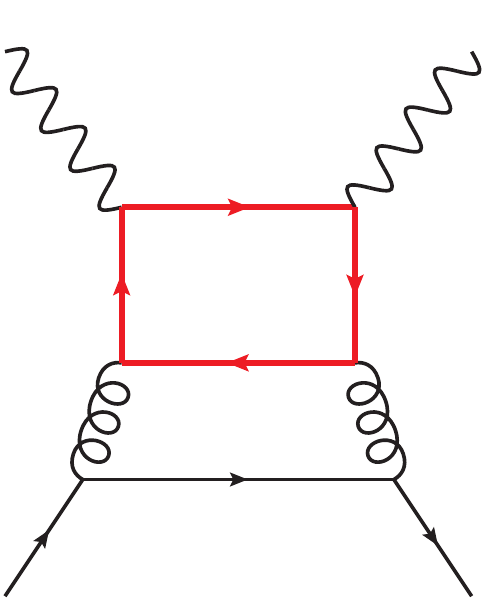}} $+ \dotsb$
$= \Cm_g^{(1)}(Q,\mb,\muH) \biggl($
\parbox{7ex}{\includegraphics[trim=5.0cm 22.6cm 13.3cm 4.4cm,clip,width=1.3cm]{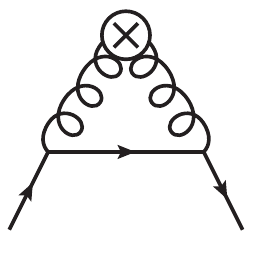}}
$+ \dotsb \biggr) + \Cm_q^{(2)}(Q,\mb,\muH) $
\parbox{7ex}{\includegraphics[trim=5.0cm 22.6cm 13.3cm 4.4cm,clip,width=1.3cm]{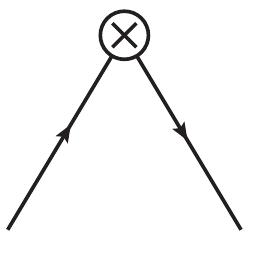}}
\caption{Schematic NLO matching for heavy-quark production in DIS for $\mb \sim Q$.}
\label{fig:NLO4FSDIS}
\end{figure}

The Wilson coefficients  $\Cm_i(Q,\mb, \muH)$ are determined in perturbation theory by matching the matrix elements of both 
sides of \eq{4Fmatching} between the same partonic external states $j$,
\begin{equation} \label{eq:Cm4computation}
\Mae{j}{O_{\rm DIS}(Q,\mb)}{j} = \Cm_i(Q,\mb,\muH) \otimes \Mae{j}{O_i^\nfour(\muH)}{j}
\,.\end{equation}
The left-hand side corresponds to the full-theory matrix element.
The partonic matrix element of the $\MSb$-renormalized PDF operators on the right-hand side are the partonic PDFs,
\begin{equation} \label{eq:4FSpartonicPDF}
f_{i/j}^\nfour(\muH) = \Mae{j}{O_i^\nfour(\muH)}{j} \equiv \Gamma_{ij}^\nfour
\,.\end{equation}
They are equivalent to the collinear $\MSb$ subtractions often denoted as $\Gamma_{ij}^\nfour$.

The matching in \eq{Cm4computation} is performed order by order in $\alpha_s(\mu_H) \equiv \alpha_s^\nfour(\mu_H)$, for which we expand each of the pieces as
\begin{align} \label{eq:pert_exp}
\Mae{j}{O_{\rm DIS}(Q,\mb)}{j}
&= \sum_k \Mae{j}{O_{\rm DIS}(Q,\mb)}{j}^{(k)}\, \Bigl[\frac{\alpha_s(\mu_H)}{4\pi}\Bigr]^k
\,, \nn \\
f^\nfour_{i/j}(\muH)
&= \sum_k f^{\nfour(k)}_{i/j}\, \Bigl[\frac{\alpha_s(\mu_H)}{4\pi}\Bigr]^k
\,, \nn \\
\Cm_i(Q, \mb, \muH)
&= \sum_k \Cm_i^{(k)}(Q, \mb, \muH)\, \Bigl[\frac{\alpha_s(\mu_H)}{4\pi}\Bigr]^k
\,.\end{align}
The leading-order matching is shown schematically in \fig{LO4FSDIS}.
At the lowest order, only the gluon external state contributes. Light quarks in the external state first contribute at NLO.
The $b$-quark does not appear as external state in the matching calculation, since it cannot appear anymore on the right-hand
side. Writing out the dependence on the momentum fraction $z$ explicitly, the partonic PDFs at LO are simply given by
\begin{equation}
f^{\nfour(0)}_{i/j}(z) = \delta_{ij}\, \delta(1 - z)
\,,\end{equation}
so the LO gluon coefficient $\Cm_g^{(1)}$ is directly given by the LO diagram on the left of \fig{LO4FSDIS},
\begin{align} \label{eq:DiLO}
\Cm_g^{(1)}(Q, \mb, \muH) &= \Mae{g}{O_{\rm DIS}(Q,\mb)}{g}^{(1)}
\,,\nn \\
\Cm_q^{(1)}(Q, \mb, \muH) &= 0
\,.\end{align}

The schematic matching at NLO is shown in \fig{NLO4FSDIS}. There are virtual and real emission corrections to the gluon channel as well as the contribution with light (anti)quarks in the external state. Using pure dimensional regularization to regulate both the UV and IR, the $\MSb$-renormalized partonic PDFs at NLO are
\begin{align} \label{eq:fijNLO}
f_{g/g}^{\nfour(1)}(z) &= -\frac{1}{\epsilon}\,
\Bigl[2C_A\, \theta(z) P_{gg}(z) + \beta_0(4)\,\delta(1-z) \Bigr]
\,,\nn \\
f_{g/q}^{\nfour(1)}(z) &= -\frac{1}{\epsilon}\, 2C_F\,\theta(z)\, P_{gq}(z)
\,,\end{align}
where $\beta_0(n_f) = (11C_A-4T_Fn_f)/3$ (with $n_f=4$ here) and the one-loop (LO) gluon splitting functions are
\begin{align} \label{eq:Pgi}
P_{gg}(z)
&= 2 \, \frac{(1 - z + z^2)^2}{z} \biggl[\frac{\theta(1-z)}{1-z}\biggr]_+
\,,\nn\\
P_{gq}(z) &= \theta(1-z)\, \frac{1+(1-z)^2}{z}
\,.\end{align}
Together with the LO coefficients from \eq{DiLO}, the NLO matching coefficients are
\begin{align}
\Cm_g^{(2)}(Q, \mb, \muH)
&= \Mae{g}{O_{\rm DIS}(Q,\mb)}{g}^{(2)} - \Cm_g^{(1)}(Q, \mb, \muH) \otimes f_{g/g}^{\nfour(1)}
\,,\nn \\
\Cm_q^{(2)}(Q, \mb, \muH)
&= \Mae{q}{O_{\rm DIS}(Q,\mb)}{q}^{(2)} - \Cm_g^{(1)}(Q, \mb, \muH) \otimes f_{g/q}^{\nfour(1)}
\,.\end{align}
The $1/\epsilon$ terms in the $f_{i/j}^{(1)}$ in \eq{fijNLO} are collinear IR divergences.
They precisely cancel between the two terms on the right-hand side, such that NLO Wilson coefficients are free from
IR divergences. While the same IR regulator must be used in the full and effective theories, the Wilson coefficients are independent of the specific IR regulator. On the other hand, the coefficients do explicitly depend on the UV renormalization scheme of the PDF
operators, which is where the standard $\MSb$ scheme is used. In other words, the fact that \eq{fijNLO} are a pure pole contribution is
just an artifact from using pure dimensional regularization for both UV and IR divergences.%
\footnote{Technically, all loop corrections to the bare PDF matrix elements are scaleless and vanish, which means the UV and IR divergences are precisely equal with opposite sign and cancel each other. Adding the UV counterterms then leaves the IR divergences. In this case, the $\mu$ dependence in $f_{i/j}(\mu)$ is purely through $\alpha_s(\mu)$ and the coefficients $f_{i/j}^{(k)}$ have no explicit $\mu$ dependence as written in \eq{pert_exp}.} Using any other IR regulator, e.g., putting the external states off shell, the $f_{g/j}^{(1)}$ would look different, but the final results for the $\Cm_i^{(2)}$ would be exactly the same once the IR regulator is taken to zero.

\subsection{$\mb \ll Q$: Resummation}
\label{subsec:resum}

In this case, shown in \fig{resum}, there is a parametric hierarchy between the $b$-quark mass $\mb$ and $Q$. The calculation now proceeds via a two-step matching. First, at the scale $\muH\sim Q$, all degrees of freedom with virtualities $\sim Q^2$
are integrated out and full QCD is matched onto a theory of collinear gluons, collinear light quarks, and in addition collinear massive $b$-quarks, all with typical virtualities $p_c^2 \sim m_b^2$.%
\footnote{In SCET this would be a theory containing massive collinear fermions \cite{Leibovich:2003jd}. The collinear modes have momentum scaling $p_c \sim (Q, m_b^2/Q, m_b)$. The corresponding soft modes with momentum scaling $p_s\sim (m_b, m_b, m_b)$ are again not needed since they cancel. This implies that the production of secondary $b$-quarks can only arise from the splitting of collinear gluons.}
Next, we evolve from $\muH$ down to the intermediate scale $\mum\sim \mb$. At $\mum$, all degrees of freedom with virtualities $\sim \mb^2$ are integrated out, including the massive $b$-quark, and the theory is matched onto the same theory as in the previous \subsec{FO} of collinear gluons and collinear light quarks with typical virtuality $\lqcd^2$.

The matching at the scale $\muH$ proceeds as before, except that
above $\mum$ the bottom quark is still a dynamical degree of freedom. Analogous to \eq{4Fmatching}, the DIS operator is matched onto a sum of nonlocal PDF operators, which now includes the bottom-PDF operator $O_b$,
\begin{equation} \label{eq:5Fmatching}
O_{\rm DIS}(Q,\mb) = \Bigl[C_i(Q,\muH) \otimes O_i^\nfive(\muH) +  C_b(Q,\muH) \otimes O_b^\nfive(\muH)\Bigr]
\biggl[1 + \Ord{\frac{\mb^2}{Q^2}} \biggr]
\,.\end{equation}
Here, we use the notation $C_{i,b}$ for the Wilson coefficients to distinguish them from the $\Cm_i$ coefficients in the previous subsection. The PDF operators, $O_i^\nfive$ and $O_b^\nfive$, have the same structure as those in \eqs{4Fmatching}{4FPDFs}. The essential difference is that they are now renormalized with $n_f = 5$ active flavors. That is, $b$-quark loops are now renormalized using $\MSb$ with dimensional regularization.

Eq.~\eqref{eq:5Fmatching} corresponds again to the standard OPE in DIS. It is important to note, however, that the expansion performed in \eq{5Fmatching} is by construction an expansion in $p^2/Q^2$, where $p^2$ is the typical virtuality of the external states in the theory below $\muH$. Compared to the fixed-order case in \eq{4Fmatching}, where we had $p^2\sim\lqcd^2$, we now have $p^2\sim m_b^2$, which not only includes $b$-quarks but also collinear gluons of that virtuality. Thus, as indicated in \eq{5Fmatching}, it is \emph{always} an expansion in $m_b^2/Q^2$. In particular, matrix elements with external $b$ quarks (e.g.\ in the matching calculation below) are expanded in the $m_b\to 0$ limit. The coefficients $C_{i,b}(Q,\muH)$ contain the full dependence on the physical scale $Q$ but are independent of the low scale $\mb$. On the other hand, the operators do contain an implicit $m_b$ dependence since they involve massive $b$-quark fields.

In principle, one is free to reabsorb some of the neglected $\ord{m_b^2/Q^2}$ corrections in \eq{5Fmatching} into the coefficients. This corresponds to including some subleading power (subleading twist) corrections in the leading-power (leading-twist) result and letting the leading-power resummation act on them. However, we stress that to correctly include the subleading power corrections in the resummation requires extending the factorization in \eq{5Fmatching} to the subleading order. As mentioned before, the power corrections in $m_b^2/Q^2$ can be important in practice and should be added back such that in the fixed-order limit $\mum \to \muH$ we recover the fixed-order result of the previous subsection. This is discussed in detail in \subsec{FOtransition}. In the rest of this section, we do not indicate the power corrections for ease of notation.

After the matching at $\muH$ in \eq{5Fmatching}, we want to evolve the theory from $\muH$ down to $\mum$. The renormalization of the PDF operators again gives rise to their RGE, the solution of which is given by DGLAP evolution, relating the operators at different scales,
\begin{align} \label{eq:Oevolution5f}
O_i^\nfive(\mu) &= U_{ij}^\nfive(\mu,\mu_0) \otimes O_j^\nfive(\mu_0) +  U_{ib}^\nfive(\mu,\mu_0) \otimes O_b^\nfive(\mu_0) \,, \nn \\
O_b^\nfive(\mu) &= U_{bj}^\nfive(\mu,\mu_0) \otimes O_j^\nfive(\mu_0) +  U_{bb}^\nfive(\mu,\mu_0) \otimes O_b^\nfive(\mu_0)
\,.\end{align}
The difference to \eq{Oevolution} is that now $n_f = 5$ and $O_b^\nfive$ contributes to the evolution, which we have written out explicitly. Taking the proton matrix elements on both sides yields the corresponding evolution of the PDFs from $\mu_0$ to $\mu$ in the theory above $\mum$. Equivalently, we can use \eq{Oevolution5f} to evolve the Wilson coefficients from $\muH$ down to $\mum$,
\begin{align} \label{eq:Cevolution5F}
C_j(Q, \mum) &= C_i(Q, \muH) \otimes U_{ij}^\nfive(\muH,\mum) + C_b(Q, \muH) \otimes U_{bj}^\nfive(\muH,\mum)
\,, \nn \\
C_b(Q, \mum) &= C_i(Q, \muH) \otimes U_{ib}^\nfive(\muH,\mum) + C_b(Q, \muH) \otimes U_{bb}^\nfive(\muH,\mum)
\,.\end{align}

Next, at the scale $\mum$, the operators $O_i^\nfive(\mum)$ and $O_b^\nfive(\mum)$ are matched onto the set of operators $O_i^\nfour(\mum)$, which are precisely the ones appearing in \eq{4Fmatching} and do not include a $b$-quark operator,
\begin{align}
\label{eq:lpdf-operator-matching}
O_j^\nfive(\mum) &= \Mm_{jk}(\mb, \mum) \otimes O_k^\nfour(\mum)
\,, \\
\label{eq:bpdf-operator-matching}
O_b^\nfive(\mum) &= \Mm_{bk}(\mb, \mum) \otimes O_k^\nfour(\mum)
\,.\end{align}
By integrating out the $b$ quark, the $m_b$ dependence implicit in the $O_{j,b}^\nfive(\mum)$ is now fully contained in the matching coefficients $\Mm_{jk}(\mb, \mum)$ and $\Mm_{bk}(\mb, \mum)$. In particular, the $O_b$ operator does not exist in the theory below $\mum$ and its effects are moved into the $\Mm_{bj}$ coefficient. In addition, secondary $b$-quark loops are integrated out, which corresponds to switching the UV renormalization scheme for $b$ quarks at the scale $\mum$ from $\MSb$ to the decoupling scheme, and the $\Mm_{ij}$ and $\Mm_{bj}$ contain the associated matching (threshold) corrections.

The remaining steps now proceed as in the previous subsection. The operators (or PDFs) at $\mum$ still contain logarithms $\ln(\mum/\lqcd)$, which are resummed by using \eqs{Oevolution}{PDFevolution} to evolve them from $\muL$ up to $\mum$. Equivalently, we can think of evolving the products $C_x(Q, \mum) \Mm_{xk}(m_b, \mum)$ from $\mum$ further down to $\muL$ (with $x=j,b$). The final expression for the DIS cross section is then given by
\begin{align} \label{eq:DIS_factorization5F}
&\df\sigma^\resum(Q, m_b)
\nn \\ & \qquad
= \,\Bigl\{\Bigl[C_i(Q,\muH) \otimes U_{ij}^\nfive(\muH,\mum) + C_b(Q,\muH) \otimes U_{b j}^\nfive(\muH,\mum) \Bigr] \otimes \Mm_{jk}(\mb, \mum)
\nn \\ & \qquad\quad
+ \Bigl[C_i(Q,\muH) \otimes U_{i b}^\nfive(\muH,\mum) + C_b(Q,\muH) \otimes U_{b b}^\nfive(\muH,\mum) \Bigr] \otimes \Mm_{bk}(\mb, \mum) \Bigr\}
\nn \\ & \qquad\quad
\otimes U_{kl}^\nfour(\mum, \muL)\otimes f_{l}^\nfour(\muL)
\,.\end{align}
The full cross section on the left-hand side contains large logarithms $\ln(Q/\mb)$ and $\ln(\mb/\lqcd)$.
On the right-hand side these are factorized into logarithms $\ln(Q/\muH)$ and $\ln(\mb/\mum)$, which are considered small and reside in the coefficients $C_{i,b}(Q, \muH)$ and $\Mm(\mb, \mum)$, large logarithms $\ln(\muH/\mum)$ and $\ln(\mum/\muL)$, which are resummed into the evolution factors $U^\nfive(\muH,\mum)$ and $U^{\nfour}(\mum,\muL)$, and finally logarithms $\ln(\muL/\lqcd)$, which are absorbed into the PDFs at $\muL$. We will refer to \eq{DIS_factorization5F} as the resummed (``resum'') result, since it has all logarithms $\ln(Q/\mb)$ resummed.

In the traditional 5F scheme, the resummed result in \eq{DIS_factorization5F} is written as
\begin{align} \label{eq:EFT2_factorization}
\df\sigma^{\rm 5F}(Q, \mb)
&= C_b(Q,\muH) \otimes f_b^\nfive(\muH, \mb) + C_i(Q,\muH) \otimes f_i^\nfive(\muH,\mb)
\,,\end{align}
where the combinations
\begin{align} \label{eq:PDFmu_m}
f_b^\nfive(\mb, \muH)
&= \Bigl[ U_{b j}^\nfive(\muH,\mum) \otimes \Mm_{jk}(\mb, \mum)
+ U_{b b}^\nfive(\muH,\mum) \otimes \Mm_{b k}(\mb, \mum) \Bigr] \otimes f_k^\nfour(\mum)
\,, \nn \\
f_i^\nfive(\mb, \muH)
&= \Bigl[ U_{ij}^\nfive(\muH,\mum) \otimes \Mm_{jk}(\mb, \mum)
+ U_{i b}^\nfive(\muH,\mum) \otimes \Mm_{b k}(\mb, \mum) \Bigr] \otimes f_k^\nfour(\mum)
\,,\end{align}
are interpreted as the evolved 5F PDFs including a PDF for the bottom quark $f_b^\nfive$.
To all orders in $\alpha_s$, \eqs{EFT2_factorization}{PDFmu_m} are simply a different way to write \eq{DIS_factorization5F}. In practice, however, the evolution and matching corrections are always carried out to a certain finite order, where the different interpretations lead to different perturbative countings yielding different results. This is discussed in detail in \subsec{powercounting}. Basically, in \eq{EFT2_factorization} the 5F PDFs are traditionallly regarded as external $O(1)$ inputs, and the perturbative order counting in $\alpha_s$ is only applied to the coefficients $C_i$ and $C_b$. In contrast, in \eq{DIS_factorization5F}, we only regard the $f_l^\nfour(\muL)$ as external $O(1)$ quantities, while the perturbative order counting is applied to all terms in curly brackets. As we will see, one advantage of doing so is that this renders the order counting consistent between the resummed and fixed-order results, which facilitates their combination, as discussed in detail in \subsec{FOtransition}.

We also note that the publicly available 5F PDF sets are constructed as in \eq{PDFmu_m}, with the notable difference that the matching scale $\mum$ is commonly identified with and fixed to the heavy-quark mass, $\mum\equiv m_b$. However, it is clear from our discussion that $\mum$ is a (in principle arbitrary) perturbative matching scale and it is important to keep it conceptually distinct from the parametric $m_b$ dependence. In our results, we will utilize the $\mum$ dependence to estimate the intrinsic resummation uncertainties.

\subsubsection{Matching at $\muH \sim Q$}
\label{sec:WilsonResmuH}

The Wilson coefficients $C_i(Q,\muH)$ and $C_b(Q,\muH)$ are computed in perturbation theory by taking partonic matrix elements of both sides of \eq{5Fmatching},
\begin{align}
\label{eq:me-matching1}  
\ABr{b}{O_{\rm DIS}(Q,\mb)}{b}
  &= C_b(Q,\muH) \otimes \Mae{b}{O_b^\nfive(\muH)}{b} + C_i(Q,\muH) \otimes \Mae{b}{O_i^\nfive(\muH)}{b}
\,, \\
\label{eq:me-matching2}  
\ABr{j}{O_{\rm DIS}(Q,\mb)}{j}
  &= C_b(Q,\muH) \otimes \Mae{j}{O_b^\nfive(\muH)}{j} + C_i(Q,\muH) \otimes \Mae{j}{O_i^\nfive(\muH)}{j}
\,.\end{align}
The calculation proceeds analogous to \subsubsec{4FmuHmatching}. The essential difference is that now $b$ quarks are present in the theory below $\muH$ and so we also have to consider external $b$-quark states to determine the $C_b$ matching coefficient. As discussed earlier, the full-theory matrix elements on the left-hand side are expanded to leading order in $\mb^2/Q^2$. The partonic matrix elements on the right-hand side now lead to partonic PDFs similar to those of \eq{4FSpartonicPDF}, now including also $b$-quarks,
\begin{align}  \label{eq:5FSpartonicPDF}
f_{b/b}^\nfive(\mb, \muH) &= \Mae{b}{O_b^\nfive(\muH)}{b}
\,,\quad&
f_{i/b}^\nfive(\mb, \muH) &= \Mae{b}{O_i^\nfive(\muH)}{b}
\,, \nonumber \\
f_{i/j}^\nfive(\mb, \muH) &= \Mae{j}{O_i^\nfive(\muH)}{j}
\,,\quad&
f_{b/j}^\nfive(\mb, \muH) &= \Mae{j}{O_b^\nfive(\muH)}{j}
\,.\end{align}

\begin{figure}[t]
\centering
$\Im$ \parbox{12ex}{\includegraphics[trim=5.0cm 21.6cm 11.5cm 4.4cm,clip,width=2cm]{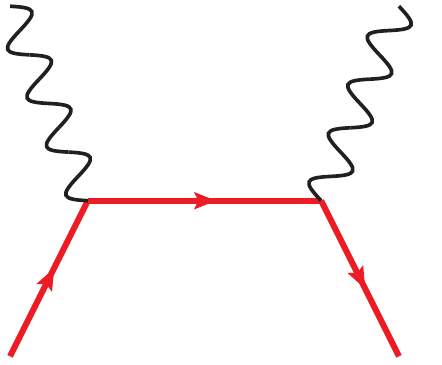}}
$= C_b^{(0)}(Q,\muH)$
\parbox{7ex}{\includegraphics[trim=5.0cm 22.6cm 13.3cm 4.4cm,clip,width=1.3cm]{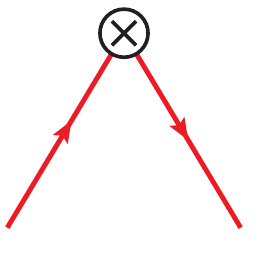}}
\caption{Schematic leading-order matching at $\muH$ for heavy-quark production in DIS for $\mb \ll Q$.}
\label{fig:LO5FSDIS}
\end{figure}

\begin{figure}[t]
\centering
$\Im$ \parbox{12ex}{\includegraphics[trim=5.0cm 21.6cm 11.5cm 4.4cm,clip,width=2cm]{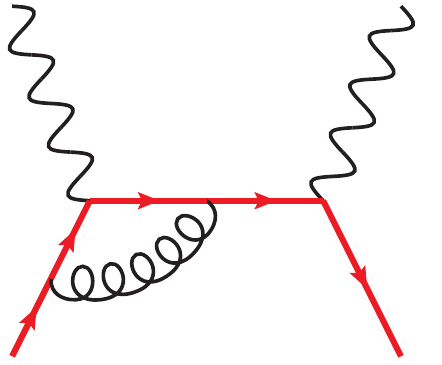}} $+\dotsb$
$= C_b^{(0)}(Q,\muH) \biggl($
\parbox{7ex}{\includegraphics[trim=5.0cm 22.6cm 13.3cm 4.4cm,clip,width=1.3cm]{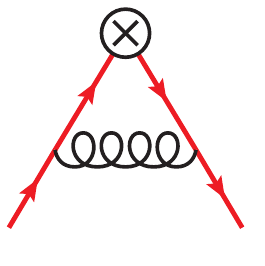}}
$+ \dotsb \biggr) + C_b^{(1)}(Q,\muH)$
\parbox{7ex}{\includegraphics[trim=5.0cm 22.6cm 13.3cm 4.4cm,clip,width=1.3cm]{figures/pdf_b_0.pdf}}
\\
$\Im$ \parbox{12ex}{\includegraphics[trim=5.0cm 19.6cm 10.9cm 4.4cm,clip,width=2cm]{figures/dis_born_fo.pdf}}
$= C_g^{(1)}(Q,\muH) $
\parbox{7ex}{\includegraphics[trim=5.0cm 22.6cm 13.3cm 4.4cm,clip,width=1.3cm]{figures/pdf_g_0.pdf}}
$+\,  C_b^{(0)}(Q,\muH) $
\parbox{7ex}{\includegraphics[trim=5.0cm 22.6cm 13.3cm 4.4cm,clip,width=1.3cm]{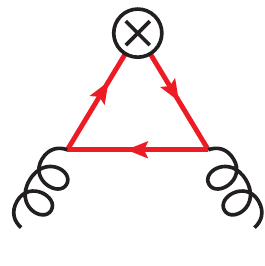}}
\caption{Schematic NLO matching at $\muH$ for heavy-quark production in DIS for $\mb \ll Q$.}
\label{fig:NLO5FSDIS}
\end{figure}

To perform the matching, we expand both sides of
\eqs{me-matching1}{me-matching2} in powers of
$\as(\muH) \equiv \as^\nfive(\muH)$, where all the pieces are expanded
analogously to \eq{pert_exp}. The leading-order matching is illustrated in \fig{LO5FSDIS}.
At LO, the partonic bottom PDF is
\begin{equation}
f^{\nfive(0)}_{b/b}(z, \mb, \muH) = \delta(1 - z)
\,,\end{equation}
so the LO bottom-quark coefficient $C_b^{(0)}$ is directly given by the LO diagram on the left of \fig{LO5FSDIS},
\begin{align} \label{eq:Di5FLO}
C_b^{(0)}(Q, \muH) &= \lim_{\mb\to0}\Mae{b}{O_{\rm DIS}(Q,\mb)}{b}^{(0)}
\,.\end{align}
The limit $\mb\to 0$ explicitly highlights that the full-theory matrix element is expanded in $\mb^2/Q^2$.

The NLO matching is illustrated schematically in \fig{NLO5FSDIS}. At this order, there are 1-loop and real-emission corrections to the bottom-quark LO contribution as well as a contribution from a gluon channel. The partonic PDFs at NLO for finite $\mb$ are (see e.g.~\mycite{Kniehl:2005mk})
\begin{align} \label{eq:fijNLOmb}
f_{b/b}^{\nfive(1)}(z, \mb, \muH) &= 2C_F \theta(z) \plus{\frac{1+z^2}{1-z} \biggl(\ln\frac{\muH^2}{\mb^2(1-z)^2}-1\biggr) }
\,, \nn \\
f_{b/g}^{\nfive(1)}(z, \mb, \muH) &= 2T_F\theta(z)\, P_{qg}(z) \ln\frac{\muH^2}{\mb^2},
\end{align}
with
\begin{align} \label{eq:Pqg}
P_{qg}(z) &= \theta(1-z)\, [(1-z)^2 + z^2]
\,.\end{align}
Together with the LO $b$-quark coefficient in \eq{Di5FLO}, the NLO matching coefficients are
\begin{align} \label{eq:5F-muH-matching}
C_b^{(1)}(Q, \muH)
&= \lim_{\mb\to0}\Bigl[\Mae{b}{O_{\rm DIS}(Q,\mb)}{b}^{(1)} - C_b^{(0)}(Q, \muH) \otimes f_{b/b}^{\nfive(1)}(\mb, \muH) \Bigr]
\,,\nn \\
C_g^{(1)}(Q, \muH)
&= \lim_{\mb\to0} \Bigl[\Mae{g}{O_{\rm DIS}(Q,\mb)}{g}^{(1)} - C_b^{(0)}(Q, \muH) \otimes f_{b/g}^{\nfive(1)}(\mb, \muH) \Bigr]
\,.\end{align}
The logarithms of $\mb$ inside the matrix elements of the DIS operator precisely match those in the partonic PDFs in \eq{fijNLOmb}, such that the $\mb\to0$ limit is finite. The reason is that for the matching at $\muH$, the finite bottom mass is nothing but an IR regulator for the collinear divergences associated with bottom quarks, which cancels in the matching.

Since the matching coefficients are independent of the IR regulator, we can also take the $\mb\to 0$ limit at the beginning, as long as we use another IR regulator, such as dimensional regularization. In this case, the computation of the coefficients $C_{i,b}$ becomes much simpler since there is one less scale involved. The partonic PDFs are then the usual ones in pure dimensional regularization, completely analogous to \eq{fijNLO},
\begin{align} \label{eq:fijNLOb}
f_{b/b}^{\nfive(1)}(z) &= -\frac{1}{\epsilon}\, 2C_F\, \theta(z) P_{qq}(z)
\,,\nn \\
f_{b/g}^{\nfive(1)}(z) &= -\frac{1}{\epsilon}\, 2T_F \,\theta(z)\, P_{qg}(z)
\,,\end{align}
with
\begin{equation}
P_{qq}(z) = \biggl[\theta(1-z)\frac{1+z^2}{1-z}\biggr]_+
\,.\end{equation}
and $P_{qg}(z)$ as in \eq{Pqg}. The NLO coefficients are then given by
\begin{align} \label{eq:5F-muH-matching-massless}
C_b^{(1)}(Q, \muH) &= \Mae{b}{O_{\rm DIS}(Q,0)}{b}^{(1)} - C_b^{(0)}(Q, \muH) \otimes f_{b/b}^{\nfive(1)}
\,,\nn \\
C_g^{(1)}(Q, \muH) &= \Mae{g}{O_{\rm DIS}(Q,0)}{g}^{(1)} - C_b^{(0)}(Q, \muH) \otimes f_{b/g}^{\nfive(1)}
\,,\end{align}
and are precisely the same as in \eq{5F-muH-matching}.


\subsubsection{Matching at $\mum \sim \mb$}

To compute the matching coefficients $\Mm_{ij}$ at the low scale $\mum$, we calculate matrix elements of both sides of \eqs{lpdf-operator-matching}{bpdf-operator-matching} with the same external partonic states. Using the definitions in \eq{4FSpartonicPDF} and \eq{5FSpartonicPDF}, we have
\begin{align} \label{eq:mum-matching}
f_{i/k}^\nfive(\mb, \mum) &= \Mm_{i j}(\mb, \mum) \otimes f_{j/k}^\nfour(\mum)
\,, \nn \\ 
f_{b/k}^\nfive(\mb, \mum) &= \Mm_{b j}(\mb, \mum) \otimes f_{j/k}^\nfour(\mum)
\,.\end{align}
Now, the $b$ quark cannot appear anymore as an external state, since it is integrated out on the right-hand side. (Hence, there are no equivalent matching equations for $f_{b/b}^\nfive$ or $f_{i/b}^\nfive$.) At the same time, $\mb$ is now the hard scale which appears in the matching coefficients (and cannot be set to zero). The matching coefficients $\Mm_{ij}$ are known fully to $\ord{\as^2}$ \cite{Buza:1996wv}. (They are also known partially to $\ord{\as^3}$, see e.g.\ \mycites{Blumlein:2012vq,Ablinger:2014uka} and references therein.)

\begin{figure}[t]
\centering
\parbox{9ex}{\includegraphics[trim=5.0cm 22.6cm 13.3cm 4.4cm,clip,width=1.5cm]{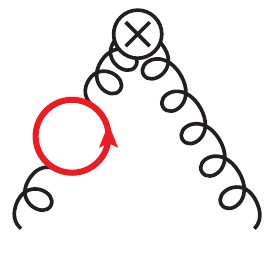}} $+\dotsb$
$= \Mm_{gg}^{(1)}(\mb,\mum)$ \parbox{9ex}{\includegraphics[trim=5.0cm 22.6cm 13.3cm 4.4cm,clip,width=1.5cm]{figures/pdf_g_0.pdf}}
\\
\parbox{9ex}{\includegraphics[trim=5.0cm 22.6cm 13.3cm 4.4cm,clip,width=1.5cm]{figures/pdf_b_3.pdf}}
$= \Mm_{bg}^{(1)}(\mb,\mum)$ \parbox{9ex}{\includegraphics[trim=5.0cm 22.6cm 13.3cm 4.4cm,clip,width=1.5cm]{figures/pdf_g_0.pdf}}
\caption{NLO matching at $\mum$.}
\label{fig:pdf-mum-matching}
\end{figure}

Expanding \eq{mum-matching}, the LO matching is simply
\begin{equation} \label{eq:5F-mum-matching-LO}
\Mm_{gg}^{(0)}(z) = \Mm_{qq}^{(0)}(z) = \delta(1-z)
\,, \qquad
\Mm_{gq}^{(0)} = \Mm_{qg}^{(0)} = \Mm_{bq}^{(0)} = \Mm_{bg}^{(0)} = 0
\,.\end{equation}
At NLO, there are nontrivial matching conditions for $\Mm_{gg}^{(1)}$ and $\Mm_{bg}^{(1)}$, which are illustrated in \fig{pdf-mum-matching},
\begin{align} \label{eq:5F-mum-matching}
\Mm_{gg}^{(1)}(z, \mb, \mum)
&= f_{g/g}^{\nfive (1)}(z, \mb, \mum) - f_{g/g}^{\nfour (1)}(z) = -\frac{4T_F}3 \ln\frac{\mum^2}{\mb^2}\, \delta(1-z)
\,, \nn \\
\Mm_{bg}^{(1)}(z, \mb, \mum)
&= f_{b/g}^{\nfive (1)}(z, \mb, \mum) = 2T_F\theta(z)\, P_{qg}(z) \ln\frac{\mum^2}{\mb^2}
\,.\end{align}
Note that the precise number of flavors used in $\alpha_s$ here is an $\ord{\alpha_s^2}$ effect, which will then also generate nontrivial matching conditions for $\Mm_{gg}^{(2)}$ and $\Mm_{bg}^{(2)}$.

\subsection{Perturbative expansion and order counting}
\label{subsec:powercounting}

We now discuss the perturbative counting for the cross section for the two scale hierarchies in \fig{scale-hierarchy}.
Since the gluon and light quark PDFs at the scale $\muL$ are nonperturbative objects
fitted from data, we make the standard assumption and count them as external $\ord{1}$ quantities,
\begin{equation} \label{eq:fmuLcounting}
f_q^\nfour(\muL) \sim f_{\bar q}^\nfour(\muL) \sim f_g^\nfour(\muL) \sim \ord1\,,
\qquad q=d,u,s,c
\,.\end{equation}
To determine the cross section to a certain perturbative accuracy, a perturbative counting should be applied to
all remaining terms in the cross section that are computed in perturbation theory. In the fixed-order case $\mb \sim Q$, this implies that the standard perturbative counting in terms of powers of $\as$ appearing in the hard matching coefficients applies. On the other hand, for $\mb \ll Q$, the perturbative counting should be applied also to the matching coefficients at $\mum$ and the evolution factors between $\muH$ and $\mum$. We will argue that for phenomenologically relevant hard scales this implies that an appropriate perturbative counting takes the effective bottom PDF to be an $\ord{\as}$ object.

\subsubsection{Fixed order}

First, recall that the DGLAP evolution factors $U_{ij}(\mu_1,\mu_2)$ resum single logarithms of the ratio
$\mu_1/\mu_2$ to all orders in $\as$. For this purpose, one performs a logarithmic counting where one expands in powers of $\as$ while counting $\as\ln(\mu_1/\mu_2) \equiv \as L \sim 1$. That is, one formally counts $L\sim 1/\alpha_s$. We can then write
\begin{align} \label{eq:Ucounting}
U(\mu_1,\mu_2) &= U^{\rm LL}(\as L)
+ \as\, U^{\rm NLL} (\as L)
+ \as^2\, U^{\rm NNLL} (\as L) + \dotsb
\nn \\
&\sim \quad\ord{1}\quad\, + \qquad\ord{\as}\quad\,\,\, + \qquad\ord{\as^2}\qquad + \dotsb
\,,\end{align}
where $U^{\text{N$^k$LL}}$ are functions of $\as L$ to all orders in $\as$. Combining this with \eq{fmuLcounting}, we can also count the evolved 4F PDFs as $\ord{1}$ quantities
\begin{equation} \label{eq:fmucounting}
f^\nfour_i(\mu) = U^\nfour_{ij}(\mu, \muL) \otimes f_j^\nfour(\muL) \sim \ord{1}
\,.\end{equation}
The PDFs evolved at N$^k$LL are then usually called N$^k$LO PDFs. Note that while the evolution mixes the PDFs, it does not induce a parametric difference between the light-parton PDFs. Also, in the limit $\mu_1 \to \mu_2$ we have $U^\nfour_{ij} \to \delta_{ij}$. Hence, we can generically treat $f^\nfour_i(\mu)$ as external $\ord{1}$ quantities for any $\mu$ regardless of how large the logarithms $\ln(\mu_1/\mu_2)$ actually are, and this is the standard praxis.

For the fixed-order (4F) cross section in \eq{EFT1_factorization}, the perturbative counting in $\as$ is then directly applied to the $\Cm_i$ coefficients, so it has the perturbative expansion [with $a_s \equiv \alpha_s^\nfour(\muH)/(4\pi)$ as in \eq{pert_exp}]
\begin{align} \label{eq:rescount-fixed-ord}
\text{LO (FO, 4F)}&&
\df\sigma^\FO(Q, \mb)
  &= \quad a_s \,   \Cm_g^{(1)}(Q,\mb,\muH) \otimes f_g^\nfour(\muH)
\nn \\
\text{NLO (FO, 4F)}&&
& \quad + a_s^2\, \Cm_i^{(2)}(Q,\mb,\muH) \otimes f_i^\nfour(\muH)
\nn \\
\text{NNLO (FO, 4F)}&&
& \quad + a_s^3\, \Cm_i^{(3)}(Q,\mb,\muH) \otimes f_i^\nfour(\muH)
\nn \\
&& & \quad + \dotsb
\,.\end{align}
Taking into account \eq{Ucounting}, obtaining the cross section at an accuracy of order $\as^{k}$ (N$^k$LO) then requires the N$^k$LO matching coefficient with the N$^k$LL evolution (i.e. N$^k$LO PDFs). Here the order is counted relative to the lowest nonvanishing order, which for heavy-quark production in DIS is $\ord{\as}$.

When combining the expansion in \eq{Ucounting} with the $\as$ expansion of $\Cm_i(Q,\mb,\muH)$, one could in principle reexpand the product of the two series. In practice, this is usually not done, since the $f_i^\nfour(\muH)$ are treated as external $\ord{1}$ inputs as mentioned above.

\subsubsection{Resummation}

We now discuss the perturbative counting for the resummed cross section in \eq{DIS_factorization5F}.
First, we can use the same arguments as in \eq{fmucounting} to treat the 4F PDFs at $\mum$ as $\ord{1}$ inputs.
We then have to consider the perturbative counting for the terms in curly brackets in \eq{DIS_factorization5F}.
The situation is more subtle now due to the presence of the additional scale $\mum\sim \mb$. Depending on the hierarchy between $\muH$ and $\mum$, there are two different options of how to count the evolution factors $U_{ij}^\nfive(\muH,\mum)$.
\begin{itemize}
\item For very large hierarchies, we can use a strict logarithmic counting, in which case $\as L \sim1$ and \eq{Ucounting} generically applies to all the evolution kernels so
\begin{equation} \label{eq:5FUcounting}
U^\nfive_{ij} \sim U^\nfive_{bj} \sim U^\nfive_{ib} \sim U^\nfive_{bb} \sim 1
\,.\end{equation}
\item For intermediate hierarchies $\mum\lesssim\muH$, the resummation can still be important, so we still use \eq{Ucounting} to organize the logarithmic order of the resummation. However, we should also take into account that in the limit $\mum\to \muH$ the off-diagonal mixing evolution kernels vanish $U^\nfive_{bg}(\mum\to \muH, \muH)\to 0$ and similarly for $U^\nfive_{gb}$. This is because their fixed-order expansion starts at order $\as L$ rather than $1$, so they are suppressed by an overall factor of $\as L$ relative to the diagonal $U_{bb}^\nfive$ and $U_{gg}^\nfive$. Therefore we count
\begin{align} \label{eq:resumUcounting}
U^\nfive_{ij}(\muH, \mum) &\sim U^\nfive_{bb}(\muH, \mum) \sim 1
\,, \nn \\
U^\nfive_{bg}(\muH, \mum) &\sim U^\nfive_{gb}(\muH, \mum) \sim \as
\,.\end{align}
\end{itemize}

The counting in \eq{5FUcounting} corresponds to the traditional 5F scheme. With this counting and using \eqs{5F-mum-matching-LO}{5F-mum-matching}, the evolved 5F PDFs in \eq{PDFmu_m} have the perturbative expansion
\begin{align} \label{eq:5FPDF_exp}
f_g^\nfive(\mb, \muH)
&= \Bigl\{ U_{gg}^\nfive(\muH,\mum)
+ \frac{\as(\mum)}{4\pi} \Bigl[ U_{g b}^\nfive(\muH,\mum) \otimes \Mm_{bg}^{(1)}(\mb, \mum) + \dotsb \Bigr] \Bigr\} \otimes f_g^\nfour(\mum)
\nn \\
&\sim \qquad\, \ord{1} \qquad + \qquad \qquad \ord{\alpha_s}
\,, \nn \\
f_b^\nfive(\mb, \muH)
&= \Bigl\{ U_{bg}^\nfive(\muH,\mum)
+ \frac{\as(\mum)}{4\pi} \Bigl[ U_{b b}^\nfive(\muH,\mum) \otimes \Mm_{bg}^{(1)}(\mb, \mum) + \dotsb \Bigr] \Bigr\} \otimes f_g^\nfour(\mum)
\nn \\
&\sim \qquad\, \ord{1} \qquad + \qquad \qquad \ord{\alpha_s}
\,.\end{align}
Hence, they are treated as external $\ord{1}$ quantities. The resummed result is then written as in \eq{EFT2_factorization} and has the perturbative expansion [with $a_H = \as^\nfive(\muH)/(4\pi)$]
\begin{align} \label{eq:5Fcounting}
\text{LO (5F)}&&
\df\sigma(Q, \mb)
  &= C_b^{(0)}(Q,\muH) \otimes f_b^\nfive(\mb, \muH)
\nn \\
\text{NLO (5F)}&&
& \quad + a_H \Bigl[ C_b^{(1)}(Q,\muH) \otimes f_b^\nfive(\mb, \muH) + C_i^{(1)}(Q,\muH) \otimes f_i^\nfive(\mb, \muH) \Bigr]
\nn \\
\text{NNLO (5F)}&&
& \quad + a_H^2\, \Bigl[ C_b^{(2)}(Q, \muH) \otimes f_b^\nfive(\mb, \muH) + C_i^{(2)}(Q, \muH) \otimes f_i^\nfive(\mb, \muH) \Bigr]
\nn \\
&& & \quad + \dotsb
\,.\end{align}
The N$^k$LO cross section then requires using the N$^k$LO 5F PDFs, which are given by the expansion of \eq{5FPDF_exp} to $\ord{\as^k}$ together with the N$^k$LL evolution factors.

A rough numerical estimate shows that for $\mum\sim\mb\sim 5\GeV$ the first case \eq{5FUcounting} applies for hard scales $\muH\gtrsim1\TeV$. Thus, the second case in \eq{resumUcounting} is more appropriate for our purposes. This is also confirmed by the fact that for $\muH\sim\ord{100\GeV}$ and standard PDF sets one finds that numerically $f_b^\nfive(\muH) \ll f_g^\nfive(\muH)$.
Adopting this counting, the resummed result in \eq{DIS_factorization5F} has the perturbative expansion
\begin{align} \label{eq:resum_exp}
&& &\df\sigma^\resum(Q, m_b)
\nn \\
\text{LL (resum)}&& &\qquad
= \Bigl\{a_H\, C_g^{(1)}\, U_{gg}^\nfive
+ C_b^{(0)} \Bigl[ U_{b g}^\nfive + a_m\, U_{b b}^\nfive\, \Mm_{bg}^{(1)} \Bigr] \Bigr\} f_g^\nfour
\nn \\
\text{NLL (resum)}&& & \qquad\quad
+ a_H \Bigl\{a_H\, C_g^{(2)}\, U_{gg}^\nfive
+ C_b^{(1)} \Bigl[ U_{b g}^\nfive + a_m\, U_{b b}^\nfive\, \Mm_{bg}^{(1)} \Bigr] \Bigr\} f_g^\nfour
\nn \\ && & \qquad\quad
+ a_m \Bigl\{a_H\, C_g^{(1)} U_{gg}^\nfive \Mm_{gg}^{(1)}
+ C_b^{(0)} \Bigl[ U_{b g}^\nfive\, \Mm_{gg}^{(1)} + a_m\, U_{b b}^\nfive\, \Mm_{bg}^{(2)} \Bigr] \Bigr\} f_g^\nfour
\nn \\ && & \qquad\quad
+ \Bigl\{a_H^2\, C_q^{(2)}\, U_{qq}^\nfive + a_H\, C_g^{(1)}\, U_{gq}^\nfive
+ C_b^{(0)} \Bigl[ U_{b q}^\nfive + a_m^2\, U_{b b}^\nfive\, \Mm_{bq}^{(2)} \Bigr] \Bigr\} f_q^\nfour
\nn \\ && & \qquad\quad
+ \dotsb
\,.\end{align}
Here, $a_H \equiv \alpha_s^\nfive(\muH)/(4\pi)$ and $a_m \equiv \alpha_s^\nfive(\mum)/(4\pi)$, and for notational simplicity we have suppressed the convolution symbols and all arguments (which are as in \eq{DIS_factorization5F}). In the contributions proportional to $f_q^\nfour$ we have also counted $U_{gq} \sim \as$ and $U_{bq}\sim \as^2$.
Note that, in the region where this counting applies, $a_H$ and $a_m$ can be regarded as being parametrically (and practically) of the same size.

From \eq{resum_exp} we see that the counting in \eq{resumUcounting} leads us to include the matching terms $C_g^{(1)}$ and $\Mm_{bg}^{(1)}$, which provide the boundary conditions for the RGE, already at the lowest order, i.e.\ one order lower compared to the 5F.
Furthermore, any cross terms in \eq{resum_exp} from the matching at $\muH$ and $\mum$ are expanded against each other. In other words, compared to \eq{5Fcounting}, we do not have overall 5F PDFs, but rather the contributions $\sim U^\nfive_{ij}(\muH, \mum)\otimes \Mm_{jk}(\mum)$ making up the 5F PDFs in \eq{5FPDF_exp} are expanded together with the hard matching coefficients. As we will see in the next subsection, these features enable us to have an easy and smooth transition to the fixed-order result. Note that this is quite similar to how the primed resummation orders N$^k$LL$'$ are implemented in the resummation for differential spectra, see e.g.\ \mycites{Ligeti:2008ac, Abbate:2010xh, Berger:2010xi, Stewart:2013faa}, where this facilitates a clean and smooth transition to the fixed-order result.

We can of course collect the terms proportional to $C_b$ and $C_i$ in \eq{resum_exp} into effective PDFs, which we denoted as $\f_b$ and $\f_i$ to distinguish them from the standard 5F PDFs in \eq{5FPDF_exp}. With the counting in \eq{resumUcounting} we then have
\begin{align} \label{eq:5FPDF_exp_resum}
\f_i(\mb, \muH)
&= U_{ii}^\nfive(\muH,\mum) \otimes f_i^\nfour(\mum)
+ \quad \dotsb \quad
\nn\\
&\sim \qquad\qquad \ord{1} \qquad\qquad\, + \,\,\ord{\alpha_s}
\,, \\
\f_b(\mb, \muH)
&=
\Bigl[U_{bg}^\nfive(\muH,\mum)
+ \frac{\as(\mum)}{4\pi} U_{b b}^\nfive(\muH,\mum) \!\otimes\! \Mm_{bg}^{(1)}(\mb, \mum) \Bigr] \!\otimes f_g^\nfour(\mum)
+ \dotsb 
\nn \\ 
&\sim \qquad \ord{\as} \quad\,\, + \qquad \qquad\qquad \ord{\alpha_s} \qquad\qquad\qquad\qquad\qquad\qquad + \ord{\as^2}
\nn\,.\end{align}
Thus, in the region of scales we consider, the effective $b$-quark PDF should be treated as an $\ord{\as}$ object, while the gluon PDF still starts at $\ord{1}$. Note though that the $\ord{\as}$ terms in $\f_g$ are different from those in $f_g^\nfive$. For example, the $U^\nfive_{gb}\otimes \Mm_{bg}^{(1)}$ term in \eq{5FPDF_exp} counts as $\ord{\as}$ in $f_g^\nfive$ while it only appears at $\ord{\as^2}$ in $\f_g$.
Since the light-to-light $\Mm_{gg}$ and the light-to-heavy $\Mm_{bg}$
matching functions are needed at different relative orders in $\f_b(\mb,\muH)$,
this definition of the effective bottom PDF differs with respect to the
usual $f_b^\nfive(\muH)$ in \eq{5FPDF_exp}. Hence, in our numerical implementation we cannot use the $b$-quark PDF from the standard 5F PDF sets. Instead, we need to construct $\f_b(\mb,\muH)$ ourselves. We do so by creating PDF grids that have the matching coefficients at the required order but the same order in the evolution factors. The technical details are discussed in \app{PDFs}.

Denoting with $\f_{i,b}^{\{k\}}$ the truncation of the effective PDF $\f_{i,b}$ to $\ord{\as^k}$, we can write the NLL result 
in \eq{resum_exp} using \eq{5FPDF_exp_resum} in a compact form as
\begin{align}
\df\sigma^{\rm NLL}(Q, m_b)
&= \quad\! a_H C_g^{(1)}(Q, \muH) \otimes \f_g^{\{1\}}(\mb, \muH)
+ \phantom{a_H} C_b^{(0)}(Q, \muH) \otimes  \f_b^{\{2\}}(\mb, \muH)
\nn\\ & \quad
+ a_H^2 C_i^{(2)}(Q, \muH) \otimes \f_i^{\{0\}}(\mb, \muH)
+ a_H C_b^{(1)}(Q, \muH) \otimes \f_b^{\{1\}}(\mb, \muH)
\label{eq:resum_exp_compact}
\,.\end{align}
Here, we still consistently drop any higher-order $\ord{\as^3}$ cross terms in the product of coefficients and effective PDFs by keeping the effective PDFs to different orders in the different terms. As already mentioned, this is important to ensure a smooth transition to the fixed-order result in the limit $\mum\sim\muH$.

On the other hand, for the $b\bar bH$ hadron collider process we are eventually interested in in \sec{bbh}, we will have two PDFs and the practical implementation of the strict expansion gets quite involved. Therefore, as long as we are only interested in the phenomenologically relevant region $\mum \sim \mb \ll \muH \sim m_H$, we can also keep the higher-order cross terms to simplify the practical implementation. We then have
\begin{align} \label{eq:rescount}
&& &\df\sigma^\resum(Q, m_b)
\nn \\
\text{LL (resum)}&& &\qquad
= \quad\! a_H C_g^{(1)}(Q,\muH) \otimes \f_g(\mb, \muH) + \phantom{a_H} C_b^{(0)}(Q,\muH) \otimes \f_b(\mb, \muH)
\nn\\
\text{NLL (resum)}&& & \qquad\quad
+ a_H^2 C_i^{(2)}(Q,\muH) \otimes \f_i(\mb, \muH)
    + a_H  C_b^{(1)}(Q,\muH) \otimes \f_b(\mb, \muH)
 \nn \\ && &\qquad \quad
+ \dotsb
\,.\end{align}
Once we allow keeping higher-order terms, we can also further simplify the practical implementation by replacing the effective PDFs $\f_{i,b}$ above by standard 5F PDFs $f_{i,b}^\nfive$. These must then be of sufficiently high order such that they include all necessary matching corrections as required by our perturbative counting. However, we note that whenever one keeps higher-order terms for practical convenience, one should check that this does not have a large numerical influence on the results in the kinematic region of interest. We will come back to this in \sec{bbh}.

We stress, that even when keeping higher-order cross terms, the perturbative counting is still performed for both $C_{i,b}$ and $\f_{i,b}$ with $\f_b$ counted as $\ord{\as}$. So even though the leading term in $C_b^{(0)}$ is $\ord{\as^0}$, the resummed result starts at $\ord{\as}$. Comparing to \eq{rescount-fixed-ord}, the resummed result has a perturbative counting consistent with the fixed-order result. It precisely corresponds to a resummed version of the fixed-order result in the $\mb\to0$ limit. This organization and implementation of the resummation is one of the main ways in which our approach differs with other approaches. This will be discussed further in \sec{comparison}.

\subsection{Combination of resummation and fixed order}
\label{subsec:FOtransition}

In \subsecs{FO}{resum} we have derived results for the heavy-quark production cross section in DIS that are relevant for two different parametric scale hierarchies. The fixed-order result $\df\sigma^\FO$ in \eq{EFT1_factorization} is relevant for $\mb\sim Q$, as it keeps the exact $\mb$ dependence to a given fixed order in $\as$, but does not include the all-order resummation of logarithms $\ln(\mb/Q)$. The resummed result in \eq{DIS_factorization5F} is relevant for $\mb \ll Q$, as it resums the logarithms $\ln(\mb/Q)$ to all orders in $\as$, but neglects any $\mb/Q$ power corrections that vanish for $\mb\to 0$. These two results represent two ways of computing the same cross section. In this section, we combine these two results and obtain our final result accurate for any value of $\mb/Q$.

We follow the usual approach for combining a higher-order resummation with its corresponding fixed-order result. We write the full result for the cross section as
\begin{equation} \label{eq:sigma_full}
\df\sigma = \df\sigma^\resum + \df\sigma^\nons
\,.\end{equation}
Here, the nonsingular cross section $\df\sigma^\nons$ contains all contributions that are suppressed by $\ord{\mb/Q}$ relative to $\df\sigma^\resum$ and vanishes in the limit $\mb\to 0$. With this condition, $\df\sigma$ automatically contains the correct resummation in the $\mb\to 0$ limit.

Furthermore, we require that the fixed-order expansion of \eq{sigma_full} reproduces the correct FO result, including the full $\mb$ dependence. Therefore,
\begin{equation}
\df\sigma^\nons = \df\sigma^\FO - \df\sigma^\sing
\,,\qquad
\df\sigma^\sing = \df\sigma^\resum \bigl\vert_\FO
\,,\end{equation}
where the singular contributions $\df\sigma^\sing$ are obtained from the fixed-order expansion of the resummed result to the desired order in $\as$. For $\df\sigma^\nons$ to indeed be nonsingular and vanish for $\mb\to 0$, $\df\sigma^\sing$ must contain all singular contributions in $\df\sigma^\FO$, i.e.\ all terms that do not vanish as $\mb\to 0$. This in turn requires that the resummation to a given order fully incorporates all these fixed-order singular terms. In this sense, the resummed result should be consistent with the fixed-order result. This condition is precisely satisfied by our resummed result with the perturbative counting used in \eqs{resum_exp}{rescount}, for which the (N)LL result contains the full (N)LO singular terms, as we will see below.

To explicitly identify the nonsingular terms, we need a meaningful and consistent comparison between $\df\sigma^\FO$ and $\df\sigma^\sing$, which means we have to write both in terms of the same external 4F PDFs and expand both in terms of the same $\alpha_s$.
For this purpose, it is most convenient to use $f_i^\nfour(\muH)$ as in \eq{rescount-fixed-ord} but perform the expansion in terms of $\as^\nfive(\muH)$ as in the resummed result \eqs{resum_exp}{rescount}. First, for $\df\sigma^\FO$, we can simply change the $b$-quark renormalization scheme for $\as$ used to computed the $\Cm_i$ matching coefficients in \eq{rescount-fixed-ord} from the decoupling scheme to the $\MSb$ scheme. This leads to
modified Wilson coefficients $\Cm_i(Q,\mb,\muH) \to \Cm_i^{\MSb}(Q,\mb,\muH)$ which are now
expanded in terms of the same $\as^\nfive(\muH)$ as is used in $C_i(Q,\muH), C_b(Q,\muH)$.
From the point of view of the fixed-order calculation, this is actually the more appropriate expansion for $\mb < \muH$.
Next, the singular cross section can be easily obtained by evaluating the resummed result in \eq{resum_exp} or \eq{resum_exp_compact} at $\mum = \muH$,
\begin{align} \label{eq:resum-xs-mum}
\df\sigma^\sing &= \df\sigma^\resum \bigl\vert_\FO = \df\sigma^\resum \bigl\vert_{\mum=\muH}
\nn \\
&= \Bigl[ C_j(Q,\muH) \otimes \Mm_{ji}(\mb, \muH) + C_b(Q,\muH)  \otimes \Mm_{bi}(\mb, \muH) \Bigr]
\otimes f_i^\nfour(\muH)
\,.\end{align}
Finally, the fixed-order nonsingular cross section is given by
\begin{align} \label{eq:nonsing}
\df\sigma^\nons &= \df\sigma^\FO - \df\sigma^\sing
\nonumber\\
&=\Bigl[\Cm^{\MSb}_i(Q,\mb,\muH) - C_j(Q,\muH) \otimes \Mm_{ji}(\mb, \muH)
 - C_b(Q,\muH)  \otimes \Mm_{bi}(\mb, \muH) \Bigr]
\nn \\ & \quad
\otimes f_i^\nfour(\muH)
\,.\end{align}
At each order in $\as$, all singular terms in $\Cm^{\MSb}_i$ are exactly cancelled by the corresponding singular terms from the resummed result, such that $\df\sigma^\nons$ is free of collinear logarithms and vanishes as $\mb\to 0$.

We stress that the statement $\df\sigma^\resum\vert_\FO = \df\sigma^\resum \vert_{\mum=\muH}$ utilized above is quite nontrivial and crucially relies on the fact that with our perturbative counting in the resummed result all the matching corrections $\Mm_{ij}$ are always included to sufficiently high order (basically to the same order in $\as$ to which we have to expand the evolution kernels) such that the $\mum$ dependence precisely cancels in $\df\sigma^\sing$ to the given order in $\as$ to which we expand. Once we know that this is the case, we can pick any $\mum$ we like to perform the fixed-order expansion of $\df\sigma^\resum$. The choice $\mum = \muH$ is then the most convenient, since all the evolution kernels become trivial. For example, at LL we have
\begin{equation}
\Bigl[ U_{bg}^\nfive(\muH, \mum) + a_m U_{bb}^\nfive(\muH, \mum) \Mm_{bg}^{(1)}(\mb, \mum) \Bigr]_{\mum=\muH}
= a_H \Mm_{bg}^{(1)}(\mb, \muH)
\,,\end{equation}
and therefore
\begin{align} \label{eq:nons-LO}
\df\sigma^{\sing\,\rm LO} &= \df\sigma^{\rm LL} \bigl\vert_{\rm LO} = \df\sigma^{\rm LL} \bigl\vert_{\mum=\muH}
\nn \\
&= a_H \Bigl[ C_g^{(1)}(Q,\muH) + C_b^{(0)}(Q,\muH)  \otimes \Mm^{(1)}_{bg}(\mb, \muH) \Bigr]
\otimes f_g^\nfour(\muH)
\,, \nn \\
\df\sigma^{\nons\,\rm LO} &=
a_H \Bigl[\Cm^{\MSb \, (1)}_g(Q,\mb,\muH) - C_g^{(1)}(Q,\muH) - C_b^{(0)}(Q,\muH) \otimes \Mm_{bg}^{(1)}(\mb, \muH)
 \Bigr]
\nn \\ & \quad
\otimes f_g^\nfour(\muH)
\,.\end{align}
Comparing to the matching conditions in \eqs{5F-muH-matching}{5F-mum-matching}, we can see explicitly that the last two terms in square brackets in $\df\sigma^\nons$ precisely reproduce the singular $\mb\to 0$ contributions of $\Cm^{\MSb \,(1)}_g(Q,\mb,\muH)$.
Similarly, at NLL we have
\begin{align} \label{eq:sing-NLO}
\df\sigma^{\sing\,\rm NLO} &= \df\sigma^{\rm NLL} \bigl\vert_{\mum=\muH}
\,.\end{align}
Note that in \eqs{nons-LO}{sing-NLO} we have implicitly assumed that the resummed result is taken as in \eqs{resum_exp}{resum_exp_compact}, with all cross terms consistently expanded. Otherwise, e.g.\ when using \eq{rescount}, any higher-order cross terms then need to be dropped at the level of $\df\sigma^\sing$ to avoid introducing spurious uncancelled singular terms in $\df\sigma^\nons$.

So far, the nonsingular corrections are expressed in terms of 4F PDFs at the
hard scale $\muH$, while the resummed cross section is necessarily written in terms of 4F PDFs at $\mum$ or effective $\f_i(\mb,\muH)$ as in \eq{resum_exp_compact} or \eq{rescount}. To simplify the practical implementation it is desirable to only deal with a single set of PDFs. For this purpose, we can choose to write the nonsingular contributions in terms of only light-parton effective PDFs $\f_i(\mb,\muH)$ as
\begin{equation} \label{eq:DeltaCnons}
\df\sigma^\nons = \Delta C_i^\nons(Q,\mb,\muH) \otimes \f_i(\mb, \muH)
\,,\end{equation}
where the new coefficients $\Delta C_i^\nons(Q,\mb, \muH)$ are fixed by equating this to \eq{nonsing} at each order in $\as$. This has a unique solution, since the nonsingular contributions are by definition a FO contribution, so at $\ord{\as^n}$ the terms in \eq{nonsing} always have the form $[\Cm_i^{(n)} - C_i^{(n)} - \dotsb]\otimes f_i$. Therefore, we can naturally associate them with the gluon and light-quark PDFs $\f_i$. We can then absorb the nonsingular corrections into the light-parton coefficient functions by taking
\begin{equation}
C_i(Q,\muH) \to \bar{C}_i(Q,\mb, \muH) = C_i(Q,\muH) + \Delta C_i^\nons(Q,\mb, \muH)
\,,\end{equation}
while keeping the $C_b$ coefficient unchanged. Equivalently, we can replace $C_i \to \bar{C}_i$ everywhere and impose the condition
\begin{align} \label{eq:defn-cbar}
\bar{C}_i(Q,\mb,\muH) \otimes \Mm_{ij}(\mb, \muH) = \Cm^{\MSb}_j(Q,\mb,\muH) -  C_b(Q,\muH)  \otimes \Mm_{bj}(\mb, \muH)
\,,\end{align}
such that \eq{nonsing} vanishes.

The above shows that we can choose to absorb the nonsingular contributions into the resummed result
by modifiying the matching coefficients at $\muH$. The condition in \eq{defn-cbar} implies that the
light-parton coefficients $\bar C_i(Q,\mb, \muH)$ can be obtained from the matching at $\muH$ in section~\ref{sec:WilsonResmuH} without taking the $\mb\to 0$ limit in the light-parton full-theory matrix elements, while for all bottom contributions and coefficients the $\mb\to 0$ limit is still taken.

We can now write the final result for the cross section as
\begin{align} \label{eq:res-final}
\df\sigma &= \df\sigma^\resum + \df\sigma^\nons
\nonumber\\
&= \bar C_i(Q,\mb,\muH)\otimes \f_i(\mb, \muH) + C_b(Q,\muH)\otimes \f_b(\mb, \muH)
\,,\end{align}
which now uses the effective PDFs $\f_{i,b}$ throughout whilst capturing the full nonsingular corrections. The same perturbative counting as in \eqs{resum_exp_compact}{rescount} still applies, which now gives
\begin{align} \label{eq:res-final-exp}
&& &\df\sigma(Q, m_b)
\nn \\
\text{LO$+$LL}&& &\qquad
= \phantom{+} a_H \bar C_g^{(1)}(Q,\mb,\muH) \otimes \f_g(\mb, \muH) +  \phantom{a_H} C_b^{(0)}(Q,\muH) \otimes \f_b(\mb, \muH)
\nn\\
\text{NLO$+$NLL}&& & \qquad\quad
+ a_H^2 \bar C_i^{(2)}(Q,\mb,\muH) \otimes \f_i(\mb, \muH)
    + a_H  C_b^{(1)}(Q,\muH) \otimes \f_b(\mb, \muH)
 \nn \\ && &\qquad \quad
+ \dotsb
\,,\end{align}
where the choice of expanding the cross terms or not is kept implicit and will determine to what order the PDFs are kept in each term.
We emphasise that in this form the result is very convenient to implement, since it essentially only requires the fixed-order result
(after changing the $b$-quark renormalization scheme for $\as$) and the massless resummed result.
In \sec{bbh} we will apply this strategy to the $b\bar b H$ cross section and provide further details
on the construction of the coefficient functions.

By choosing to absorb $\df\sigma^\nons$ into the matching coefficients in the resummed result, we effectively let the leading-power resummation 
also act on the nonsingular corrections. This introduces power-suppressed higher-order logarithmic terms, which however are beyond the order we 
are working at. In particular, this does not include the correct resummation of power-suppressed logarithmic terms. (This would require the extension of 
$\df\sigma^\resum$ to subleading order in $\mb/Q$, which is well beyond the scope of this work, and also very likely irrelevant at the current precision.) 
Fundamentally, we only have control over the nonsingular corrections at the level of their fixed-order expansion.  The above procedure to include 
the nonsingular contributions is not unique, and while physically motivated, is ultimately driven by practical convenience. 
We would like to underline that alternative choices are in principle possible, provided they do not change the resummation in 
the $\mb\to 0$ limit and reproduce the correct fixed-order expansion, in which case they will effectively differ by power-suppressed 
higher-order logarithmic terms.\footnote
{In general, one could write the nonsingular contribution in terms of both light-parton and bottom PDFs,
$\f_{i,b}(\mb,\muH)$, and in this case there would not be a unique solution to \eq{DeltaCnons}. 
This gives rise to several (equivalent) possibilities of writing the final result, and this generates some of the 
differences between the various VFNSs.}
We will come back to this point in section~\ref{sec:comparison}.

From the discussion so far, it is clear that transition between $\df\sigma^\resum$ and $\df\sigma^\FO$ is
controlled by the scale $\mum$. To provide a smooth transition between the resummation and fixed-order regions,
this scale is promoted to a $\mb$-dependent profile scale $\mum \to \mum(\mb,\muH)$.
It has the properties that in the resummation region for $\mb \ll Q$ it has the canonical resummation scaling $\mum\sim \mb$, while
in the fixed-order region $\mb \sim Q$ it approaches $\mum \to \muH$, such that the resummation is turned off there and
the fixed-order result is recovered, with a smooth transition in between.
The fact that it is possible to control
this transition between limits with a single scale, makes our predictions in the transition region robust
and, moreover, variation of this scale and of its functional form provides a solid handle on the associated
theoretical uncertainties.
The precise definition and variations of the profile function are discussed in detail in \subsec{scaledep} for 
the case of $b\bar{b}H$ production.

\section{Comparison to existing approaches} \label{sec:comparison}

In the previous section we used a systematic field-theory analysis to derive a result for the heavy-quark production cross section
in DIS accurate for all possible scale hierarchies from $m\ll Q$ to $m\sim Q$.
Various approaches to the same problem are available in the literature, which go under the name of variable flavor number schemes (VFNSs). In this section, we briefly compare the existing schemes to our result.
In what follows, we do not attempt to give an in-depth review of the different schemes but rather focus on similarities and differences
with respect to our EFT result. For reviews of the different schemes in the literature we refer to \mycites{Thorne:2008xf,Olness:2008px} and 
sect.~22 of \mycite{Binoth:2010nha}.

VFNSs can in principle differ in various aspects. The first is the general construction,
namely how resummation of collinear logarithms is achieved and how nonsingular power corrections are included.
Secondly, they can differ in how the perturbative counting is performed, that is, which of the various perturbative
ingredients are included at a given order.
Third, they can differ in how the heavy-quark threshold is implemented, which in our language corresponds to the exact choice of the low matching scale $\mum$.
The first aspect is the one that primarily distinguishes the different schemes, while the remaining two aspects
are more related to choices made within each scheme. Here, we compare to the choices often used in the literature. We stress though that these choices correspond to how a particular scheme has been used or implemented in practice, but (in most cases) they do not necessarily represent restrictions of a particular scheme itself.

\subsection{Construction}
We start by discussing the differences in the basic construction of the cross sections.
For $\mb \gtrsim Q$ (``below threshold'') all schemes use the same fixed-order 4F result in \eq{EFT1_factorization}.
For $\mb \lesssim Q$ (``above threshold'') the various schemes construct their cross sections as follows:

\begin{itemize}
\item Zero-Mass (ZM).
  In this approach, the massless resummed result in \eq{EFT2_factorization} is used for $\mb \lesssim Q$,
  while nonsingular power corrections are neglected at any order in $\as$.
  Hence, this scheme is only expected to be accurate for $\mb \ll Q$. Since power-suppressed contributions are not included,
  it is not accurate close to the heavy-quark threshold and does not reproduce the full fixed-order result.
  For this reason, we do not discuss it further.

\item ACOT~\cite{Aivazis:1993kh,Aivazis:1993pi,Collins:1998rz}.
  The ACOT scheme is based on the idea that the power corrections can be fully included in DIS
  at the level of the matching at the hard scale $\muH$, \eq{5Fmatching}, by generalizing it such that
  power corrections in $\mb/Q$ are included in the definition of the Wilson coefficients.
  The heavy quark is considered as an active flavor and the quark mass dependence is retained at each matching
  step, yielding
  \beq \label{eq:ACOTmatching}
  \df\sigma = \tilde C_i(Q,\mb,\muH)\otimes f_i^\nfive(\mb, \muH) + \tilde C_b(Q,\mb,\muH)\otimes f_b^\nfive(\mb,\muH)
  \,. \eeq
  The $\tilde C_{i,b}(Q,\mb,\muH)$ incorporate the nonsingular contributions and reduce to
  the original $C_{i,b}(Q,\muH)$ in the $\mb\to0$ limit.
  In contrast to \eq{res-final}, the heavy-quark contributions in \eq{ACOTmatching} are computed with a massive on-shell heavy quark in the initial state.
  To account for the massive kinematics including the presence of a massive (unresolved) heavy quark in the final state, the heavy-quark Bjorken-$x$ can be rescaled, leading to a variant of this scheme called ACOT-$\chi$~\cite{Tung:2001mv,Amundson:2000vg}.
  While the validity of ACOT in DIS can be based on including heavy-quark masses in the hard-scattering factorization~\cite{Collins:1998rz}, its extension to the case of two incoming hadrons is problematic due to the massive kinematics, see \app{collider_kin}.

\item S-ACOT~\cite{Kramer:2000hn}.
  The fact that $f_b^\nfive$ is not independent of $f_i^\nfive$ (see \eq{PDFmu_m}) allows one to move power corrections between $\tilde C_b$ and $\tilde C_i$ without spoiling the formal accuracy of \eq{ACOTmatching} \cite{Kramer:2000hn,Collins:1998rz}.
  This was used to construct a simplified variant of ACOT, in which the heavy-quark Wilson coefficients
  are computed in the massless limit, $\tilde C_b(Q,\mb, \muH) \to C_b(Q,\muH)$, while the full mass dependence is retained in the
  (modified) light-parton coefficients.
  This is evidently equivalent to how we include the nonsingular corrections in \eqs{defn-cbar}{res-final} for practical purposes.
  To account for the massive heavy-quark kinematics, a $\chi$-rescaling is also applied, leading to S-ACOT-$\chi$~\cite{Tung:2001mv,Guzzi:2011ew}.%
  \footnote{The $\chi$ rescaling is not uniformly used in the literature. In some cases,
    the rescaling only takes into account the resolved $b$ quark,
    whose kinematics is massive in ACOT but massless in S-ACOT (this variant only applies to S-ACOT),
    while in other cases, the $\chi$ rescaling takes also into account the unresolved $b$ quark,
    whose massive kinematics is not taken into account even in ACOT (this variant applies both to ACOT and S-ACOT).}
  In \mycite{Han:2014nja}, a modification of ACOT, dubbed m-ACOT, is used for the case of two incoming hadrons, where
  the massless limit is applied only to channels with two incoming heavy quarks, while the mass dependence is kept in
  heavy-light and light-light channels.

\item TR~\cite{Thorne:1997ga,Thorne:2006qt}.
  The TR scheme is defined by requiring that the fixed-order result, after being expressed in terms of 5F PDFs,
  corresponds to the resummed result up to power-suppressed contributions.
  This requirement fixes the singular contributions. However, there is still freedom for the treatment of nonsingular terms,
  and this is fixed by making a choice such that the coefficient functions obey a sensible threshold limit.
  The result is hence different from both ACOT and S-ACOT.
  Due to the choice of perturbative counting,
  a discontinuity exists at threshold which is removed by adding a $Q$-independent contribution to the result above threshold.
  Though this contribution is formally higher order, it can be sizeable, even far from threshold.
  The presence of the constant terms complicates the generalization of this scheme to higher-orders and to hadron-hadron collisions.

\item FONLL~\cite{CacciariFONLL,Forte:2010ta}.
  This scheme is constructed by adding the massless resummed result to the full fixed-order
  result and consistently subtracting the double counting order-by-order in $\as$.
  The fixed-order contribution is rewritten in terms of 5F PDFs with the resulting ambiguity fixed
  through the choice that only light channels contribute, as we have also done in \subsec{FOtransition}.
  The double-counting terms are equivalent to the singular terms in our notation, and remove from the fixed-order result its
  massless limit, i.e.\ all its terms that do not vanish in the $\mb\to 0$ limit.
  The FONLL procedure is thus equivalent to adding the $\df\sigma^\nons$ to the resummed result.
  This also makes the FONLL construction formally equivalent to S-ACOT.
  Finally, a damping factor, which performs the same function as the $\chi$ rescaling in S-ACOT, is used to suppress
  higher-order spurious contributions and guarantee continuity at threshold.

\end{itemize}

From the point of view of the all-order resummation, all these schemes
are equivalent, as they all include the same resummation.
As discussed in \subsec{FOtransition}, the minimal and formally correct result above threshold is given by \eq{sigma_full}
as $\df\sigma = \df\sigma^\resum + \df\sigma^\nons$. This result
is formally correct in the sense that it correctly resums the collinear massive logarithms and
correctly includes the full mass dependence and kinematics at fixed order.
It is minimal in the sense that the nonsingular corrections $\df\sigma^\nons$
are unambiguous and unique when written in terms of 4F PDFs as in \eq{nonsing},
and are strictly included at fixed order, while the resummation strictly only includes leading-power terms.

As discussed in \subsec{FOtransition}, there is an ambiguity when one tries to partially or fully absorb the
nonsingular contribution into the resummed result, which amounts to expressing them in terms
of 5F PDFs. The primary perturbative ingredients of ACOT, S-ACOT, TR, and FONLL are the same and they only differ
in the way by which they fix this ambiguity. This ambiguity corresponds to power-suppressed higher-order logarithmic terms.
Hence, these schemes can be regarded as formally equivalent up to such terms, which are beyond the considered formal accuracy.

\subsection{Combination of the ingredients}

We now move to the second source of scheme differences, namely how the perturbative counting is performed.
By construction, the coefficient functions of (S-)ACOT, TR, and FONLL differ from each other
and to those in our EFT result by formally higher-order contributions.
Therefore, the largest differences between the approaches arise from the perturbative counting.
In all practical implementations we are aware of, the perturbative counting used by each scheme is as follows:

\begin{itemize}
\item 
ACOT-like schemes, used in the CTEQ family of PDF fits, construct perturbative expansions in the usual way
by counting explicit powers of $\as$ in the coefficient functions. As a result, for DIS at LO ($\as^0$) the result below 
threshold is zero, while above threshold it is nonzero due to the heavy-quark initiated contributions ($C_b^{(0)}$).
At NLO, the gluon-initiated contribution ($C_g^{(1)}$) starts to contribute, as do the $\as$ corrections of the heavy-quark contributions ($C_b^{(1)}$).
\item
The TR scheme, used in MSTW and HERAPDF fits, is somewhat different as it combines the orders such that the lowest nonvanishing order
below and above threshold appear at the same time. This means that at LO the result below threshold is the $\ord{\as}$
gluon-initiated contribution, while above threshold it is the $\ord{\as^0}$ heavy-quark initiated contribution.
The additional $Q$-independent term added above threshold is formally of higher order and does not affect the counting.
\item
FONLL, used in NNPDF fits, also adopts the standard perturbative counting. The NLO and NNLO results are called FONLL-A and FONLL-C respectively.
There is an intermediate result, FONLL-B, where the fixed-order terms are computed to order $\as^2$ (NNLO) but the massless contribution
is only included at order $\as$ (NLO). 
\end{itemize}
\begin{figure}[t]
  \centering
  \includegraphics[width=0.85\textwidth,page=2]{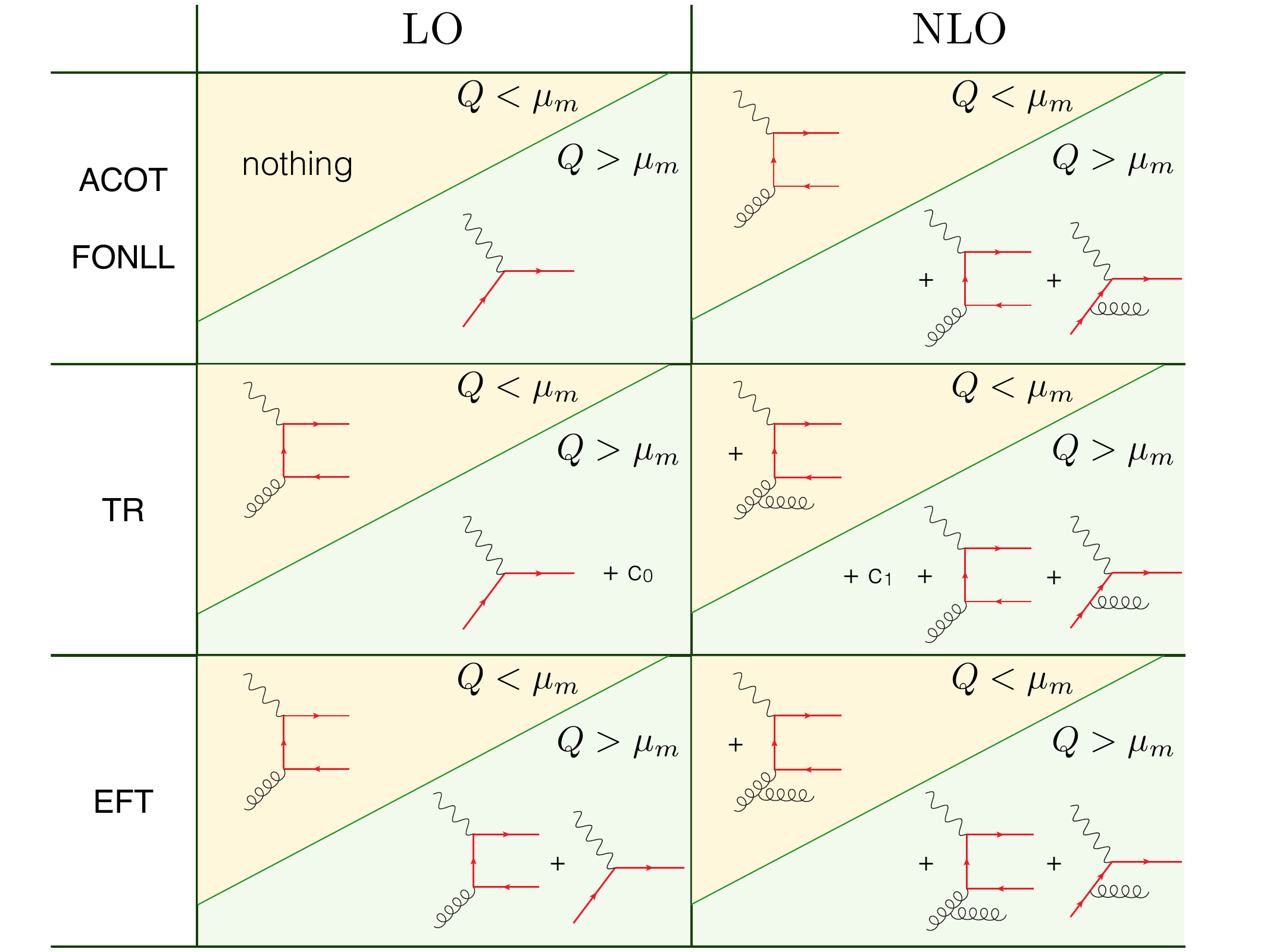}
  \caption{Comparison of the construction of LO and NLO results in different counting schemes.
    Only representative diagrams at a given order and for a given channel are shown.
    Notice that, while the diagrams appearing in the $Q<\mum$ boxes contain collinear logarithms due to the heavy quark,
    the latter are subtracted in the diagrams appearing in the $Q>\mum$ boxes. $c_{0,1}$ are the $Q$-independent terms
    present in the TR-scheme that ensure continuity at threshold. We also point the reader to a very similar table in \mycite{Olness:2008px}.}
  \label{fig:comparison}
\end{figure}
None of the schemes discussed above adopts a perturbative counting which is directly comparable to our approach of
performing the counting on the full perturbative part of the cross section including both evolution and matching.
In particular, it implies that the effective heavy-quark PDF should be counted as an $\ord\as$ object. Since the perturbative counting used by the different VFNSs summarized here does not distinguish between heavy-quark and light-parton initiated contributions, this difference in order counting is a principal difference in our approach.
In \fig{comparison} we summarize the perturbative counting adopted in the (S-)ACOT, TR and FONLL schemes as well as the counting 
we propose.

As argued in \subsec{powercounting}, our order counting is well justified theoretically and appropriate for a wide range of scales (including scales appropriate for DIS experiments in the case of both bottom and charm quarks).
It also has several advantages. As we will see in \sec{bbh}, one of these is that the perturbative convergence tends to be improved with reduced uncertainties from the hard-scale matching.
Another advantage, as highlighted in \subsec{FOtransition}, is that it facilitates a smooth transition to the fixed-order result at the heavy-quark threshold (provided the counting is strictly applied and higher-order cross terms are neglected), without the need of any rescaling or damping factors.

\subsection{Matching scale dependence}

Finally, the last important difference between the existing schemes
and our approach is the position and treatment of the heavy-quark
threshold. In all applications we are aware of, the threshold is
always set equal to the heavy quark mass, i.e.\ effectively the
resummation scale $\mum$ is fixed to $\mum =\mb$. This is both the
scale at which the low-scale matching is performed and also the scale
at which one switches from the fixed-order result to the resummed
result.

Recently~\cite{Kusina:2013slm,Olness:2008px}, it has been suggested to consider an additional
switching scale $\mu_S> \mum$, at which the computation switches from a fixed-order result to a resummed one,
but nevertheless keeping the matching at a different (lower) scale. The effect of this choice is to delay
the use of the resummed result, perhaps to a region where mass effects are negligible,
though the transition between the resummation and fixed-order regions is not guaranteed to be smooth.

In this work, we exploit the dependence on the matching scale $\mum$
to explicitly control the transition to the fixed-order result as
$\mb\to Q$, and furthermore to estimate the intrinsic perturbative
uncertainty in the resummation and matching procedure. This is in fact
the standard practice in resummed calculations involving different
resummation scales. This uncertainty should be taken into account as
part of the total perturbative uncertainty in the result, which is
typically not the case in existing approaches.

\section{Higgs production in association with $b$ quarks}
\label{sec:bbh}

In this section, we extend the framework presented in \sec{EFT} to hadron-hadron collisions and
apply it to the $b\bar b H$ process, i.e.\ Higgs-boson production in association with $b$ quarks.
Specifically, this process can be defined as Higgs production via the bottom Yukawa coupling $Y_b$, with
all other Yukawa couplings set to zero.
(As discussed in \subsec{setup}, we do not include the $b$-quark loop contributions that are usually
included in the gluon-fusion process. There we also comment on the inclusion of $Y_t Y_b$ interference terms
that are usually regarded as part of the $b\bar b H$ process.)

The $b\bar bH$ process makes up only a tiny fraction, $\sim 1\%$, of the total Higgs production
cross section in the Standard Model (SM). It is nevertheless an interesting process within the SM, since
the total $b\bar{b}H$ cross section is comparable to the total $pp\to t\bar{t}H$ cross section for LHC energies,
and because it provides direct access to the bottom Yukawa coupling.
Furthermore, this process may be sensitive to new physics effects, since
in many BSM scenarios, such as two-Higgs-doublet models with large $\tan\beta$, the Higgs coupling
to bottom quarks can be enhanced.

In the SM, the cross section in the massless 5FS is known at NLO~\cite{Dicus:1998hs,Balazs:1998sb}
and NNLO~\cite{Harlander:2003ai, Buehler:2012cu}, and in the 4FS at NLO~\cite{Dittmaier:2003ej, Dawson:2003kb}.
NLO predictions matched to parton showers for $b\bar{b}H$ production have been studied
in both 4F and 5F schemes \cite{Wiesemann:2014ioa}.
The 4FS and 5FS calculations can lead to very different results,
with cross sections differing by as much as an order of magnitude. For appropriate choices
of the factorization scale, the difference can be reduced significantly, leading to more compatible
results within the perturbative uncertainties, see the discussions in Refs.~\cite{Maltoni:2003pn,Campbell:2004pu,Dittmaier:2003ej,Dawson:2003kb,Harlander:2003ai,Maltoni:2012pa}.
As discussed already, the 5FS and 4FS possess different merits, and predictions that combine the advantages of both are
highly desirable. The current combined values by the LHC Higgs Cross Section Working Group~\cite{Dittmaier:2012vm, lhchxswg:bbh} are obtained using the Santander matching prescription~\cite{Harlander:2011aa}, which
amounts to a weighted average of the cross sections obtained in the two schemes.
In contrast, our predictions here are derived from a consistent field-theory setup,
and can thus be regarded as a definite improvement over the currently used prescription.

We start in \subsec{EFTbbh} by extending the EFT result of \sec{EFT} to the case of two incoming
protons.
In \subsec{setup}, we give details about the practical setup of our results. In \subsec{scaledep},
we discuss our procedure to obtain robust estimates of the perturbative uncertainties from separate variations of the $\muH$ and $\mum$ matching scales. In particular, we discuss the profile scales and variations for the matching scale $\mum$.
In \subsec{results}, we present our result for the $b\bar{b}H$ cross section as a function of the $b$-quark 
mass. This serves as a validation of our matching procedure, confirming that our approach satisfies all the required
properties. There, we also discuss the size of nonsingular power corrections suppressed by $\mb/Q$.
Finally, in \subsec{santander}, we present our final results at the physical $b$-quark mass for several Higgs masses and compare
to the existing results obtained in the 4FS, 5FS, and the Santander prescription.

\subsection{Extension of the EFT approach to hadron-hadron colliders}\label{subsec:EFTbbh}

The simplicity of the EFT framework presented in \sec{EFT} for DIS makes it possible to straightforwardly 
extend the setup to the case of two incoming protons. This is certainly not the case for any of 
the schemes discussed in \sec{comparison}, whose consistent generalization to hadron-hadron collisions can
be highly nontrivial. 
We first point out that the evolution of the quark operators
and the matching at $\mum$ are identical and therefore the evolved PDFs of \eq{PDFmu_m} are the same.
Of course, the matching at the hard scale is different. For $\mb \sim Q$ we have
\beq
O_{bbH}(\mh,\mb) = \Cm_{ij}(\mh,\mb,\muH) O^\nfour_i(\muH) O^\nfour_j(\muH)
\eeq
while for $\mb \ll Q$ we find\footnote{In this section we restore the difference between $b$ and $\bar b$,
  and we omit the convolution symbol $\otimes$ for ease of notation.}
\begin{align}
O_{bbH}(\mh,\mb)
&=\bar C_{ij}(\mh,\mb,\muH) O^\nfive_i(\muH) O^\nfive_j(\muH) \nonumber\\
&+C_{bk}(\mh,\muH) \Big[O^\nfive_b(\muH) O^\nfive_k(\muH) + O^\nfive_{\bar b}(\muH) O^\nfive_k(\muH) \nonumber\\
&\qquad\qquad\qquad + O^\nfive_k(\muH) O^\nfive_b(\muH) + O^\nfive_k(\muH) O^\nfive_{\bar b}(\muH) \Big] \nonumber\\
&+C_{b\bar b}(\mh,\muH) \[O^\nfive_b(\muH) O^\nfive_{\bar b}(\muH) + O^\nfive_{\bar b} (\muH) O^\nfive_b(\muH)\]  \nonumber\\
&+C_{bb}(\mh,\mum) \[O^\nfive_b(\muH) O^\nfive_b(\muH) + O^\nfive_{\bar b} (\muH) O^\nfive_{\bar b}(\muH)\]
\end{align}
where we have introduced a $bbH$ operator $O_{bbH}$, and $\mh$ is the Higgs mass.
Note that we have used the identities $C_{bk}=C_{\bar bk}=C_{k b}=C_{k\bar b}$,
$C_{b\bar b}=C_{\bar bb}$ and $C_{bb}=C_{\bar b\bar b}$.
These two results are the straightforward extensions of the results in \eqs{4Fmatching}{5Fmatching},
where, as discussed in \subsec{FOtransition}, we have made the choice to absorb power corrections 
into the coefficients for the light channels.

As in the DIS case, in our order counting we take the bottom PDF to be an object of order $\as$.
As discussed in \subsec{powercounting}, this is appropriate for a hard scale of the order 
of the Higgs mass, $\muH\sim \mh$ (more generically for $\muH \lesssim 1\TeV$).
Therefore, up to NLO, the fixed-order result our cross section matches into for $\mum\to \muH$ is given by (omitting the arguments for simplicity) 
\begin{align}
 \text{LO (FO, 4F)}  && \sigma  &= \quad a_H^2 \Cm^{\MSb \, (2)}_{ij} f_{i}^{\nfour} f_{j}^{\nfour} \nonumber\\
 \text{NLO (FO, 4F)} && & \quad + a_H^3 \Cm^{\MSb \, (3)}_{ij} f_{i}^{\nfour} f_{j}^{\nfour} \nonumber\\
  &&  & \quad + \ldots \label{eq:bbhxs_FO}
\end{align}
For $\mum< \muH$ the resummed and matched cross section is written as\footnote{For ease of notation, we do not distinguish
between bottom and anti-bottom PDFs, and also on whether they come from one or the other proton, and compensate 
for this with numerical factors.}
\begin{align}
 \text{LO+LL} && \sigma &= \quad a_H^2 \bar C^{(2)}_{ij} \f_{i} \f_{j}
    +a_H 4 C^{(1)}_{bg} \f_b \f_g
    + \phantom{a_H}2C^{(0)}_{b\bar b} \f_b \f_b
  \nonumber\\
 \text{NLO+NLL} &&  & \quad + a_H^3 \bar C^{(3)}_{ij} \f_{i} \f_{j}
    +a_H^2 4 C^{(2)}_{bk} \f_b \f_k
    +a_H 2C^{(1)}_{b\bar b} \f_b \f_b
  \nonumber\\
 && & \quad + \ldots, \label{eq:bbhxs_res}
\end{align}
where as in \eq{res-final-exp} we have left implicit the strict expansion of the products of effective PDFs and coefficient functions. 
We notice that in both cases, using the perturbative order counting introduced in \subsec{powercounting},
LO(+LL) is in fact order $\as^2$ and NLO(+NLL) includes the order $\as^3$ corrections.
In \subsec{results}, we discuss the implementation and results of a strict expansion of \eq{bbhxs_res}
as well as a more practical implementation keeping higher-order cross terms and using standard 5F PDFs.

\begin{figure}[t]
\centering
\includegraphics[width=0.9\textwidth,page=1]{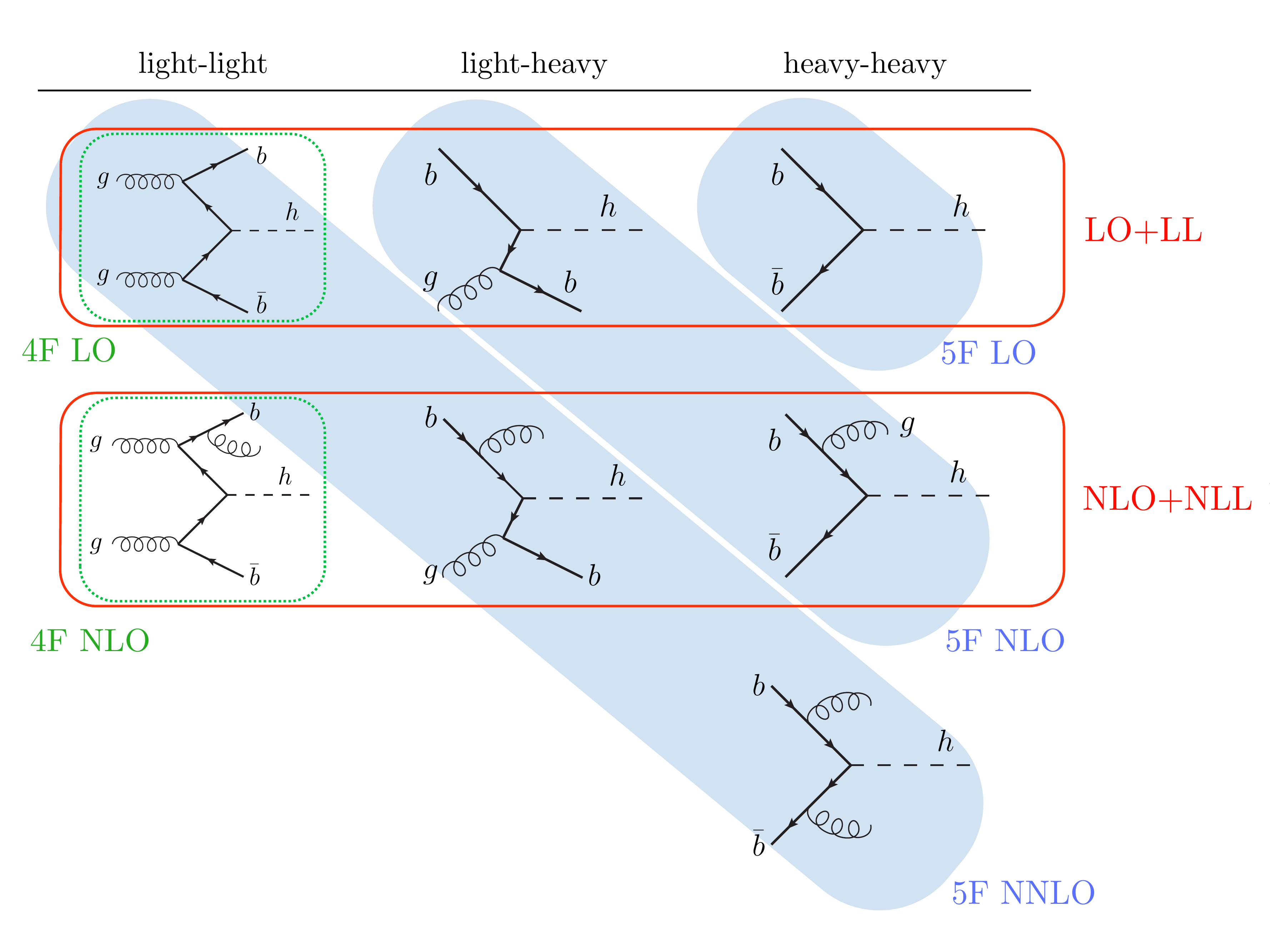}
\caption{Sample diagrams appearing in the computation of the Higgs production cross section in association with $b$ quarks.
  Diagrams are grouped according to the different countings adopted in our resummed result, \eq{bbhxs_res},
  and in the 5FS result, \eq{bbhxs_5F}.
  The 4FS counting coincides with the resummed counting in the fixed-order limit where only the diagrams in the first column are considered.}
\label{fig:bbh-diagrams}
\end{figure}

The 4FS result corresponds to the result of \eq{bbhxs_FO} used for all scale hierarchies and where the decoupling 
scheme is used for the $b$-quark renormalization of $\as$.
The 5FS result on the other hand corresponds to the massless limit of \eq{bbhxs_res}, replacing $\f_{i,b}$ with $f_{i,b}^\nfive$
and with the perturbative order counting performed only on the coefficient functions (i.e. assuming the bottom PDF of order $1$).
This has the expansion
\begin{align}
\text{LO (5F)}   && \sigma &= \quad 2C^{(0)}_{b\bar b} f_b^\nfive f_b^\nfive \nonumber\\
\text{NLO (5F)}  &&   & \quad +a_H 2C^{(1)}_{b\bar b} f_b^\nfive f_b^\nfive +a_H 4 C^{(1)}_{bg} f_b^\nfive f_g^\nfive \nonumber\\
\text{NNLO (5F)} &&   & \quad +a_H^2 \(2C^{(2)}_{b\bar b}+2C^{(2)}_{bb}\) f_b^\nfive f_b^\nfive +a_H^2 4 C^{(2)}_{bk} f_b^\nfive f_k^\nfive
    +a_H^2 C^{(2)}_{ij} f_{i}^\nfive f_{j}^\nfive \nonumber\\
 && & \quad + \ldots, \label{eq:bbhxs_5F}
\end{align}
where $C^{(2)}_{ij}$ are the massless coefficients (namely the massless limit of $\bar C^{(2)}_{ij}$).
In \fig{bbh-diagrams} we illustrate the different countings diagrammatically. This highlights that one can regard our results as a resummation-improved 4FS result.

The massless coefficients $C^{(0)}_{b\bar b}$, $C^{(1)}_{b\bar b}$, $C^{(1)}_{bg}$ and $C^{(2)}_{bk}$
required to reaching NLO+NLL accuracy in our result, \eq{bbhxs_res}, are the same as those of the massless
5FS computation and can be found explicitly in \mycite{Harlander:2003ai}.
Trivially extending \eq{defn-cbar} to the case of two initial-state legs, the matching coefficients 
$\bar C_{ij}^{(2)}$ and $\bar C_{ij}^{(3)}$ can be written as,
\begin{subequations}
\begin{align}
  \bar C_{gg}^{(2)} &= \Cm_{gg}^{\MSb,(2)} - 4\,\Mm_{bg}^{(1)}\, C_{bg}^{(1)} - 2\,\Mm_{bg}^{(1)}\, \Mm_{bg}^{(1)}\, C_{b\bar b}^{(0)},\\
  \bar C_{q\bar q}^{(2)} &= \Cm_{q\bar q}^{\MSb,(2)}, \\
  \bar C_{gg}^{(3)}
                       &= \Cm_{gg}^{\MSb, (3)} - 2\,\Mm_{gg}^{(1)}\, \bar C_{gg}^{(2)}
  \nonumber\\ & \quad
                - 4\,\Mm_{bg}^{(1)}\,C_{bg}^{(2)} 
                - 4\,\( \Mm_{bg}^{(2)} + \Mm_{bg}^{(1)}\,\Mm_{gg}^{(1)} \)\,C_{bg}^{(1)}
  \nonumber\\ & \quad
                - 2\,\Mm_{bg}^{(1)}\,\Mm_{bg}^{(1)}\,C_{b\bar{b}}^{(1)} 
                - 4 \Mm_{bg}^{(2)}\,\Mm_{bg}^{(1)} \, C_{b\bar{b}}^{(0)}, \\
  \bar C_{qg}^{(3)} &= \Cm_{qg}^{\MSb, (3)} - 2\,\Mm_{bq}^{(2)}\,C_{bg}^{(1)} - 2\,\Mm_{bg}^{(1)}\,C_{bq}^{(2)}  
                         - 2\,\Mm_{bq}^{(2)}\,\Mm_{bg}^{(1)}\,C_{b\bar{b}}^{(0)}, \\
  \bar C_{q\bar q}^{(3)} &= \Cm_{q\bar q}^{\MSb, (3)}.
\end{align}
\end{subequations}
The $\MSb$ massive coefficients can be obtained from the decoupling-scheme coefficients
as described in \app{ren}:
\beq
\Cm_{ij}^{\MSb, (2)} = \Cm_{ij}^{(2)},\qquad
\Cm_{ij}^{\MSb, (3)} = \Cm_{ij}^{(3)} -2\frac{4T_F}{3}\ln\frac{\muH^2}{m^2} \Cm_{ij}^{(2)}.
\eeq
We have implemented analytic expressions for the coefficients $\Cm_{ij}^{(2)}$ in an in-house code
and extract the numerical result for $\Cm_{ij}^{(3)}$ from \texttt{Madgraph5\_aMC@NLO}~\cite{Alwall:2014hca},
after generating the process $p p \to b \bar{b} H$ at NLO.
We have explicitly checked that our implementations, including pole scheme to $\MSb$ scheme changes for $Y_b$ in
$\Cm_{ij}$, exactly reproduce the inclusive results of the \texttt{bbh@nnlo} code \cite{Harlander:2003ai}
and of the recent $b\bar{b}H$ studies of \mycite{Wiesemann:2014ioa}.

\subsection{Setup}\label{subsec:setup}

Here we summarize the set of input parameters we use to produce the results of \subsecs{scaledep}{results}.
Unless indicated otherwise we always use the setup detailed below.

\begin{description}
\item[Collider energy] We provide predictions for the LHC at $\sqrt{s}=8\TeV$.
\item[PDFs] We have created PDF sets using a modified version of \texttt{APFEL} \cite{Bertone:2013vaa} for the evolution from a fixed low scale where the parametrization of a known PDF set (MSTW2008) has been used.
The main reasons for our modifications were the implementation of a general value for the threshold matching scale $\mum$
as well as the generation of the effective PDFs $\f ^{\{k\}}$ required in a strict expansion of \eq{bbhxs_res}.
Further details are given in \app{PDFs}.
\item[Higgs mass] We use $\mh = 125\GeV$ as default.
\item[Bottom mass]
For all results where the bottom mass is fixed to its physical value, we use a pole mass of $\mb = 4.75\GeV$
for the kinematic mass scale that enters in the 4F matrix elements and in the low-scale matching coefficients $\Mm_{ij}$.
For the Yukawa coupling we use the $\MSb$ mass $\overline{m}_b(\overline{m}_b) = 4.16\GeV$ as input, see also below.
The use of different bottom masses for the Yukawa coupling and in the matrix elements is not unsual -- this
has been the setup of the 4FS calculations of Ref.~\cite{Dittmaier:2003ej,Dawson:2003kb}.
What is different to previously used setups (and to the LHCHXSWG) is that the two values we use are not related
to each other via a one-loop conversion. This is not a problem, since the two perturbative series in which they enter are unrelated.%
\footnote{In the future, a better approach would be to replace the pole mass in the threshold corrections by a proper short-distance mass scheme with a well-defined conversion from the $\MSb$ scheme.}
What is relevant in our case is that we consistently use common values in both the resummation and fixed-order parts of the calculation.
The numerical values above are chosen to have reasonable physical values and to enable an as consistent as possible comparison with the default 4FS and 5FS results.

In our results where we vary $\mb$ to study the dependence on the bottom mass, $\mb$ and $\overline{m}_b(\overline{m}_b)$ are varied consistently, with the conversion between the two at one loop as required for our NLO calculation.

\item[Yukawa couplings] All the Yukawa couplings are set to zero except the bottom quark Yukawa, $Y_b$.
The bottom Yukawa is renormalized in the $\MSb$ scheme and its running is set to 4 loops. In our numerical
studies we always evaluate it at the hard scale $\muH$, which is the appropriate scale for the resummation of large logarithms $\ln(\muH/\mb)$ associated with the $b\bar b H$ vertex and hence leads to better perturbative convergence~\cite{Braaten:1980yq}.

\item[Bottom loops] Contributions to Higgs production where the Higgs couples to a closed $b$-quark loop
are usually included in the gluon-fusion cross section, since their most important effect is due to the
interference of the bottom loop with the top loop. As usual, we exclude these contributions from our $b\bar b H$
computation, such that the result has no double counting with the gluon-fusion cross section.
Our result still includes bottom loop contributions, but only in diagrams with two bottoms in the final state,
(not included in the gluon-fusion cross section) as part of the NLO correction to the $gg$ channel in our result.

\item[$Y_b\cdot Y_t$ interference]
In our results we neglect the interference contribution proportional to $Y_t\cdot Y_b$ by setting $Y_t=0$.
In the SM, this correction is known to be important and reduces the inclusive 4FS NLO
cross section by roughly $10\%$ at the LHC for $m_H=125\GeV$~\cite{Dittmaier:2003ej,Dawson:2003kb,Wiesemann:2014ioa},
while in BSM scenarios with large $\tan\beta$ its relative contribution can be much smaller.
This interference has been computed in the 4FS where it first enters at NLO via diagrams containing a top-quark loop, 
whilst in the 5FS up to NNLO this interference does not contribute~\cite{Harlander:2003ai}.
For comparisons between 4FS and 5FS predictions it is often preferred that the interference terms are dropped \cite{Harlander:2011aa,Dittmaier:2011ti}
since the latter are not present in the 5FS. To better compare with the results in the literature we also make this choice here.
However, we emphasize that the $Y_t\cdot Y_b$ terms can be straightforwardly and consistently included as an additional
nonsingular fixed-order piece in our result. To do so, we can simply allow for a nonzero top Yukawa in the
fixed-order coefficients $\Cm_{ij}$. No changes to the resummed part of our result are required at the order we are working.

\end{description}

\subsection{Scale dependence and theory uncertainties}\label{subsec:scaledep}

\begin{figure}[t]
\begin{center}
\includegraphics[width=0.9\textwidth]{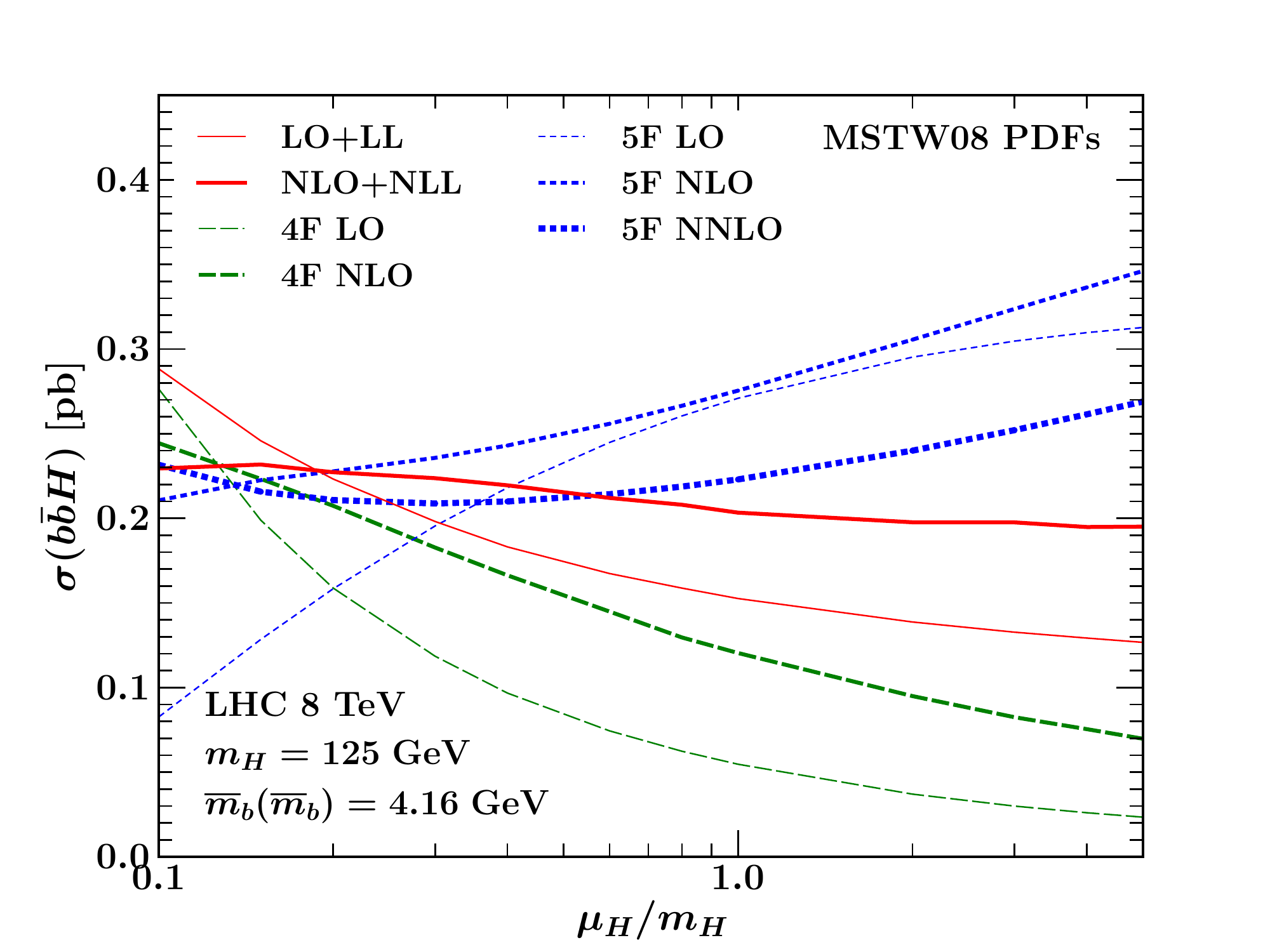}
\end{center}
\caption{$b\bar{b}H$ cross section under hard-scale variation.}
\label{fig:bbh-muf-variation}
\end{figure}

In this subsection, we discuss in detail the perturbative uncertainties in our results.
We begin by looking at the hard scale dependence.
We fix $\mb$ to its physical value, set $\mum=\mb$, and plot in \fig{bbh-muf-variation} the 
cross section obtained according to the 4FS (at LO and NLO, obtained using the code of
\cite{Wiesemann:2014ioa}), the 5FS (at LO, NLO, and NNLO, obtained using \texttt{bbh@nnlo} \cite{Harlander:2003ai}) 
and our result (at LO+LL and NLO+NLL).

As expected, a clear reduction of the scale dependence is observed in all results when moving to higher orders.
We also notice that the patterns of scale dependence of the 4FS (green dashed) and 5FS (blue dotted) results are opposite to 
each other with the former decreasing and the latter increasing with increasing $\muH$ (except at NNLO).
This is due to the fact that at LO the scale dependence is dominated by $\as$ for the 4FS result
(which clearly increases at small scales), while for the 5FS result it is driven only by
the bottom PDF, which vanishes at the bottom threshold and therefore drops rapidly as the scale decreases.
Therefore, over a wide range of hard scales the two results differ significantly.

In contrast, the framework we have presented in \sec{EFT} leads to cross sections (red solid)
that are less sensitive to the choice of the hard scale, even at LO.
The reason behind this is a large compensation between the contributions from the $b\bar b$, $bk$, and $ij$ channels. 
This is due to the fact that each $b$-initiated contribution compensates the collinear subtraction in a gluon 
(or light quark) initiated contribution and close to the heavy-quark threshold these terms are all of the same order. 
This leads to a scale dependence in the (N)LO+(N)LL results that has a similar pattern to that of the unresummed 4FS 
result, however the resummation of collinear logarithms significantly stabilizes the dependence on $\muH$.
As with the 4FS, the 5FS results also have a greater dependence on $\muH$ compared to our resummed results. The reason for this
is that the 5FS predictions adopt a standard perturbative counting and thus the compensation observed in the EFT results 
is not present. 

Additionally, \fig{bbh-muf-variation} illustrates that a smaller scale $\muH\sim \mh/4$ leads to a more stable
perturbative expansion for all the results, and also leads to better agreement between the different approaches.
The reason for this has been studied in \mycite{Maltoni:2012pa} by a careful investigation of the actual size of the
logarithms that arise in the 4FS prediction.

\begin{figure}[t]
\begin{center}
\begin{tabular}{cc}
\hspace{-1cm} \includegraphics[width=8.7cm]{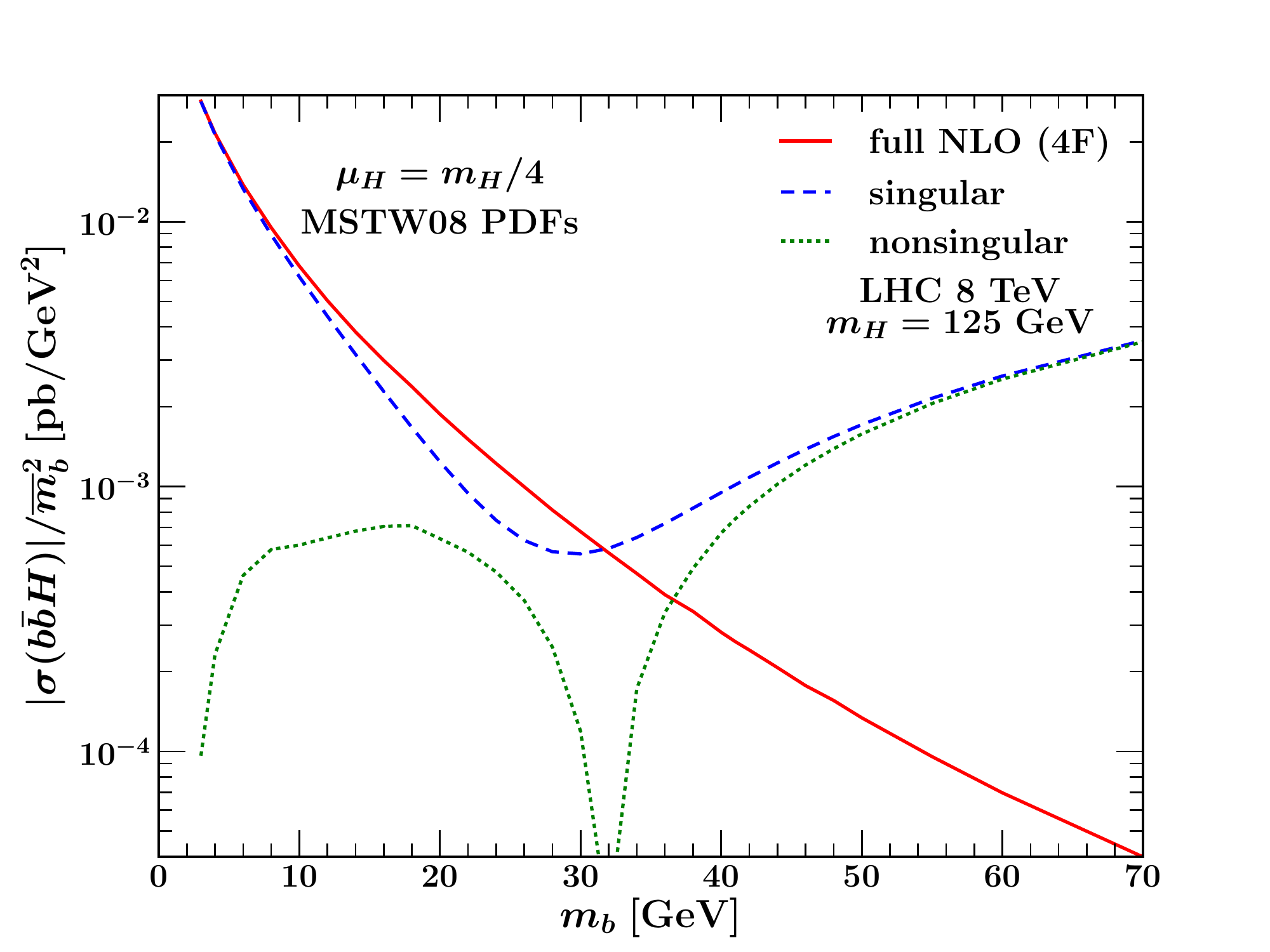} &
\hspace{-1.3cm} \includegraphics[width=8.7cm]{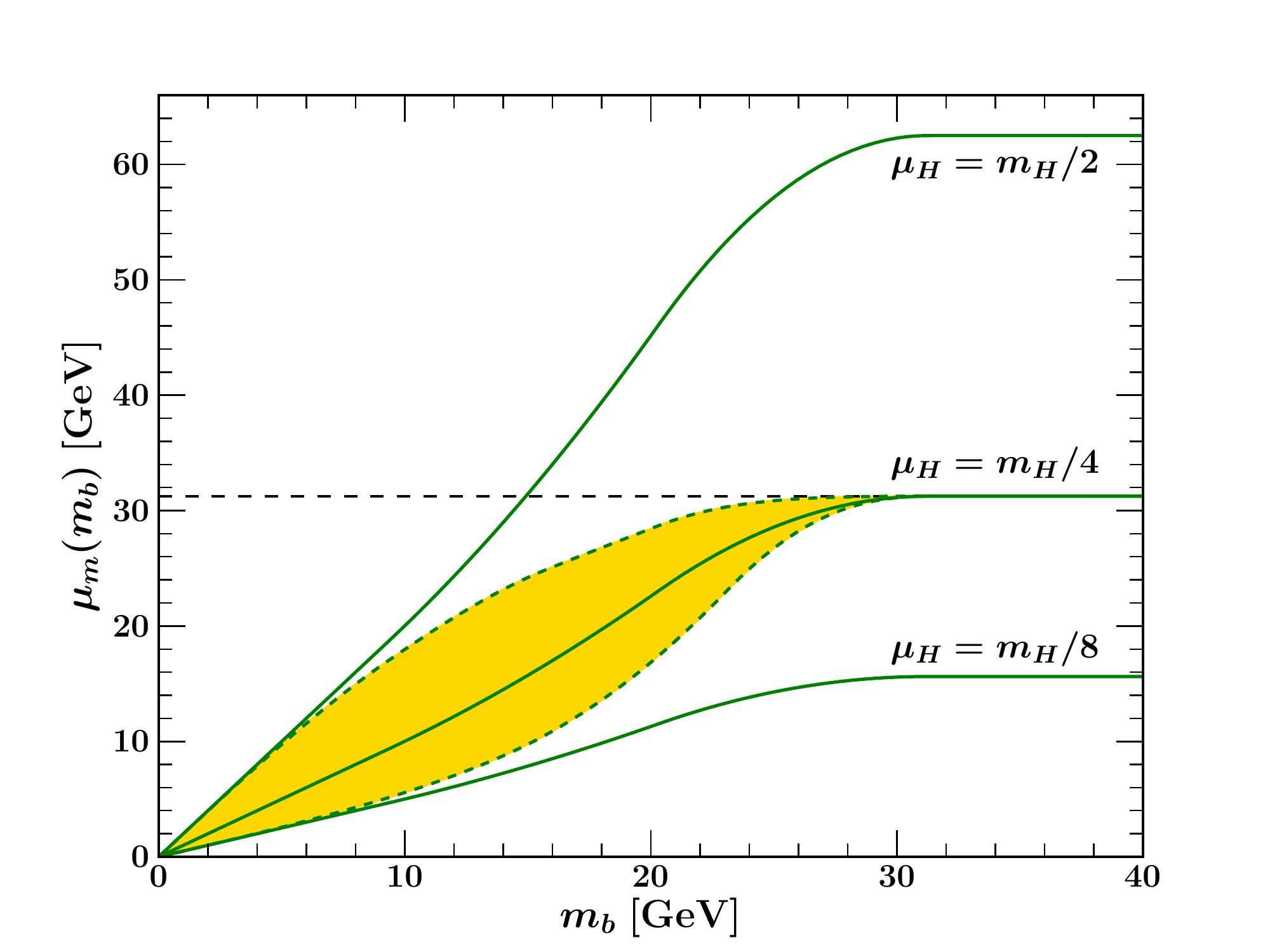}
\end{tabular}
\end{center}
\caption{Comparison of the magnitude of the singular, nonsingular, and full NLO cross sections when varying $\mb$ (left) and 
  profile scale variations (right). See text for further details.}
\label{fig:bbh-sing-nons}
\end{figure}

Next, we discuss the choice of $\mum$ and its associated perturbative uncertainties.
For this purpose, it is important to identify the kinematic region where the resummation is important and where it must be turned off.
To this end, in the left plot of \fig{bbh-sing-nons} we show the fixed NLO result and its decomposition
into singular \eq{resum-xs-mum} and nonsingular \eq{nonsing} contributions, for a fixed value of the Higgs mass $\mh=125$~GeV
and as a function of the bottom mass. In this plot we vary the bottom mass but have divided the cross sections by the bottom Yukawa coupling to better highlight the perturbative structure.

In the $\mb\to 0$ region, the singular terms clearly dominate, while the nonsingular corrections are suppressed by at least an order of magnitude and tend to zero. This is the resummation region, where the canonical choice $\mum = \mb$ is appropriate to resum the large logarithms $\ln(\mb/\mh)$ in the singular corrections.

With increasing $\mb$ the singular contribution starts deviating from the full result,
crosses it at around $\mb\sim 30$~GeV~$\sim\mh/4$, and becomes much larger than the full result in the large-$\mb$ region.
This large-$\mb$ region corresponds to the fixed-order region, which exhibits a delicate balance between singular and nonsingular contributions, with a large cancellation between the two yielding the full result. This means that the distinction into singular and nonsingular is meaningless here. To not spoil this cancellation it is imperative that the resummation is switched off completely, which is done by taking $\mum = \muH$. The fixed-order region starts at $\mb\gtrsim\mh/4$, where the magnitude of both singular and nonsingular is larger than the full result, so there is clearly an $\ord{1}$ cancellation between them.
We have verified that this pattern holds at both LO and NLO and upon variation of the hard scale
in the range $\mh/16<\muH<\mh$.
We can therefore safely take $\mb\sim\mh/4$ as the point where we should turn off the resummation for any configuration
we might consider.

A smooth transition between the canonical value $\mum = \mb$ in the resummation region and $\mum = \muH$ in the fixed-order region
is achieved by using profile scales~\cite{Ligeti:2008ac, Abbate:2010xh}, where the scale $\mum$ is promoted to a function of $\mb$,
which smoothly interpolates between these two limits. The use of profile scales is a common practice when performing resummation in EFTs based on RGEs.
Following \mycites{Berger:2010xi, Stewart:2011cf, Stewart:2013faa}, we choose different sets of profiles that allow us to separately estimate fixed-order and resummation uncertainties, which in the end are added in quadrature.

For our central scales we use
\begin{align} \label{eq:mum-profile}
 \muH = \mh/4
 \,, \qquad
 \mum(\mb,\mh,\muH) = \muH\, f_{\text{run}}\(\frac{\mb}{\mh/4}\),
\end{align}
with the profile function
\begin{align} \label{eq:profile-function}
f_{\text{run}}(x) = 
    \begin{cases}
      x,  &  \text{ if}\; 0 \leq x \leq x_1 \\
      x+\frac{(2-x_2-x_3)(x-x_1)^2}{2(x_2-x_1)(x_3-x_1)}, &  \text{ if}\; x_1 < x \leq x_2 \\
      1-\frac{(2-x_1-x_2)(x-x_3)^2}{2(x_3-x_1)(x_3-x_2)}, &  \text{ if}\; x_2 < x \leq x_3 \\
      1,  & \text{ if} \; x_3 < x
    \end{cases}
\end{align}
and we have chosen the appropriate values of $\{ x_1=0.3, x_2 = 0.65, x_3 = 1.0 \}$.
In this way, the resummation slowly turns off as $\mb$ increases, becoming
completely switched off for $\mb\geq\mh/4$, which corresponds to the point identified above.
The right-hand plot of \fig{bbh-sing-nons} illustrates these profile functions: the solid green curves
correspond to \eq{mum-profile} as a function of $\mb$ for the hard scale choices
$\muH = \{ \mh/8, \mh/4, \mh/2 \}$.
Note that at small $\mb$ the standard scale $\mum=\mb$ is recovered for
the central profile scale with $\muH=\mh/4$.
The hard scale variation by a factor of two leaves the ratio $\mum/\muH$ fixed and therefore does not
change the resummation. At the same time for large $\mb$ it recovers the usual fixed-order scale variation.
Hence, we use these variations to estimate the fixed-order uncertainty $\Delta_{\text{FO}}$.

The variation of the central profile, while keeping the hard scale $\muH$ fixed, is performed
by multiplying the central profile by a factor, 
\begin{align}
\mum^{\text{vary}}(\mb,\muH,\mh,\alpha) &= f^{\alpha}_{\text{vary}}\(\frac{\mb}{\mh/4}\) \, \mum(\mb,\muH,\mh) \nonumber\\
&= \muH\, f^{\alpha}_{\text{vary}}\(\frac{\mb}{\mh/4}\)\, f_{\text{run}}\(\frac{\mb}{\mh/4}\), 
\end{align} 
where $\alpha \in \[-1,1\] $ and 
\begin{align}
f_{\text{vary}}(x) &= 
    \begin{cases}
      2 \left(1-\frac{x^2}{x^2_3} \right),  &  \text{ if}\; 0 \leq x \leq \frac{x_3}{2} \\
      1 + 2\left(1-\frac{x}{x_3} \right)^2,  &  \text{ if}\; \frac{x_3}{2} < x \leq x_3 \\
      1,  &  \text{ if}\; x_3 < x. 
    \end{cases}
\end{align}
The multiplicative factor $f_{\text{vary}}(x)$ tends to 2 in the limit $x\to 0$ and tends to 1 in the limit $x\to x_3$
(as before, we use $x_3=1$).
The effect of this factor (when varying $\alpha \in \[-1,1\] $) is to vary the arguments of the resummed logarithms in the
small $\mb$ region by a factor of two, while keeping the hard scale fixed. Hence, we can use these variations to estimate
the resummation uncertainty $\Delta_{\text{resum}}$. In the limit  $x\to x_3$ (or $\mb \to \muH$) the effect of this variation tends to
zero, as it must, and thus the resummation uncertainty vanishes in the fixed-order result as it should. In the transition region between the resummation and fixed-order regions, this variation effectively captures the uncertainty in the transition.
In the right-hand plot of \fig{bbh-sing-nons} the yellow band enclosed by the dotted green curves
shows the effect of this variation on the central profile $\muH=\mh/4$.

A key advantage of the setup we have discussed above is that it provides a concrete way by which to estimate
the theoretical uncertainties. Our result for the cross section is obtained via a two-step matching procedure
and the variation of the scales at which this matching is performed is a natural way to arrive at a realistic
error estimate.
To obtain our estimate of the total theoretical uncertainty, we take $\Delta_{\text{FO}}$ and $\Delta_{\text{resum}}$ as the
maximum variation among each of their respective profile variations. We then obtain $\Delta_{\text{tot}}$ by adding the two in quadrature,
\begin{align} \label{eq:uncertainties}
\Delta_{\text{tot}}^2 = \Delta_{\text{FO}}^2 + \Delta_{\text{resum}}^2.
\end{align}
We emphasize that since all variations we perform amount to variations of scales (albeit more intricate
than standard scale variations), the resulting perturbative uncertainties $\Delta_{\text{FO}}$, $\Delta_{\text{resum}}$ and $\Delta_{\text{tot}}$ decrease when increasing the perturbative order of a calculation.

Finally, we note that profiling the scale $\mum$ is nontrivial for general values of $\mb$ and $\muH$.
Since $\mum$ corresponds to the scale at which PDFs are matched from a theory involving bottom quarks to 
a theory with no bottom quarks, each point of a profile function for $\mum$ corresponds to a different PDF
set (with $\mum$ as the bottom threshold). To produce the $\mb$-variation plots in \subsec{results} we have produced
20 PDF sets for each of the five profiles in \fig{bbh-sing-nons}. For the results at the physical $\mb$ value presented
in \subsec{santander}, we are always in the canonical region ($x < x_1$ in the profile function), which means we are only required
to generate PDF sets for the values $\mum \in \{0.5\mb, \mb, 2\mb\}$.

\subsection{Cross section and power corrections as a function of $\mb$} \label{subsec:results}

\begin{figure}[t]
\begin{center}
\begin{tabular}{cc}
\hspace{-1cm} \includegraphics[width=8.7cm]{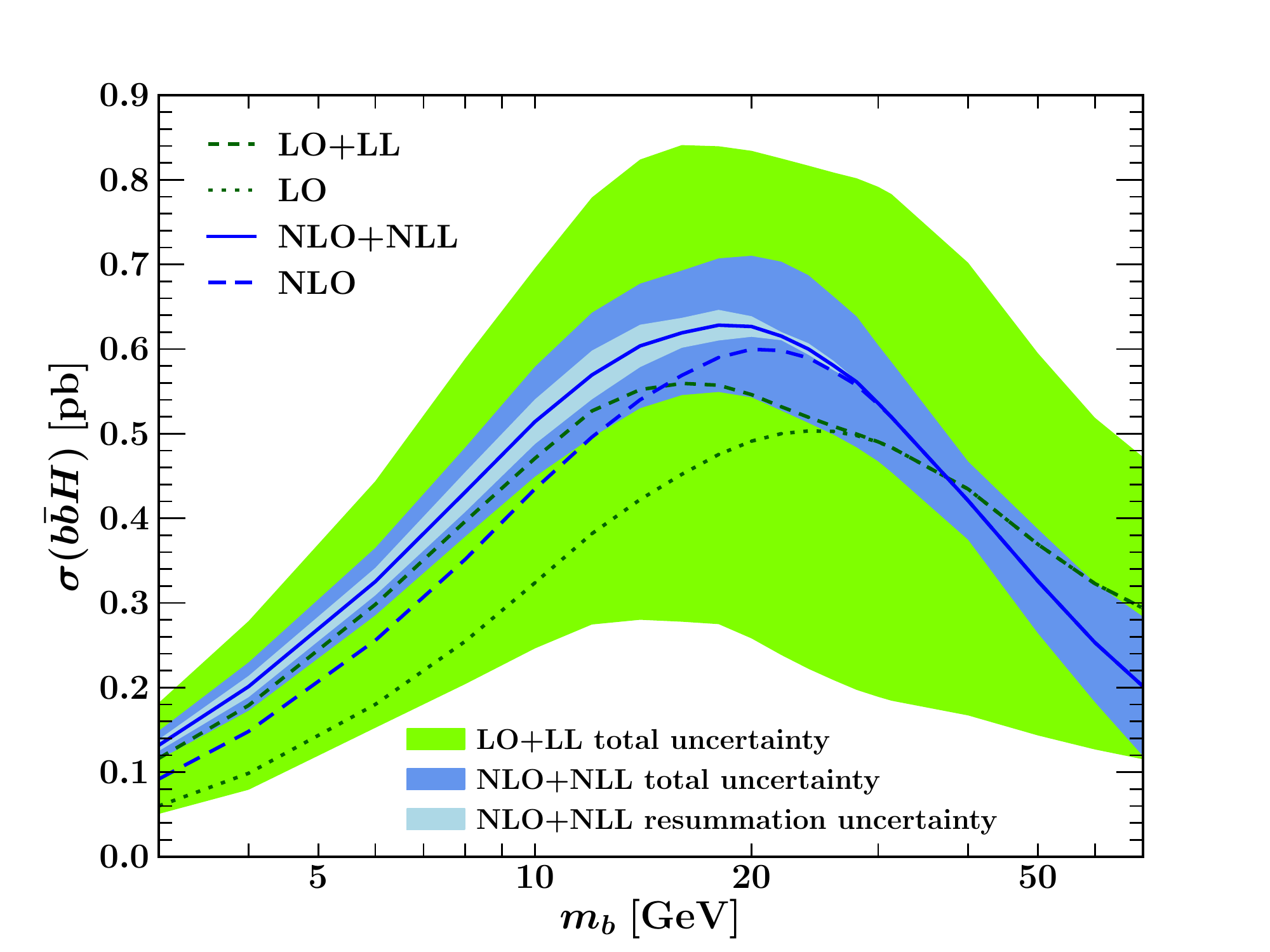} &
\hspace{-1.3cm} \includegraphics[width=8.7cm]{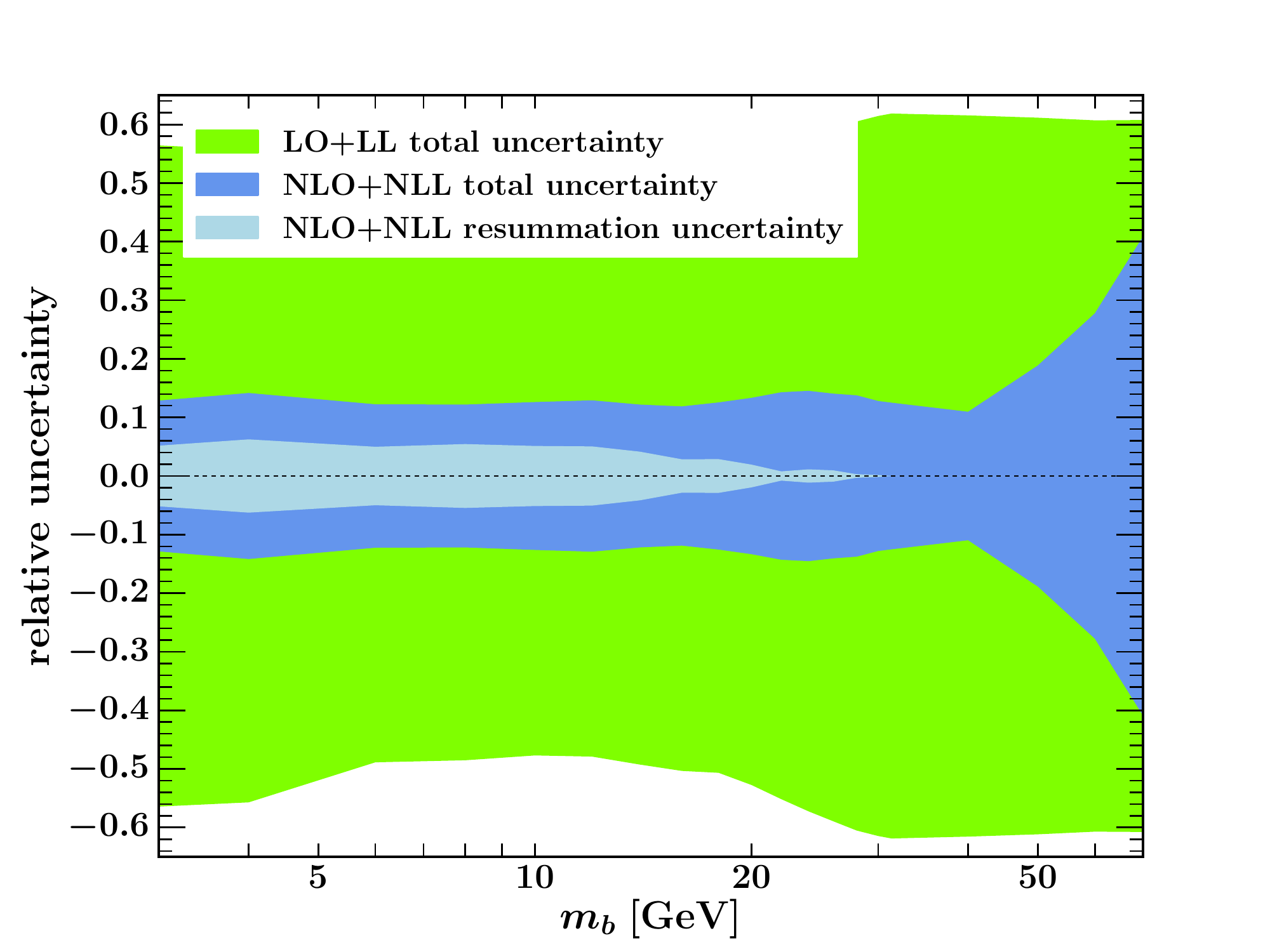}
\end{tabular}
\end{center}
\caption{(N)LO+(N)LL cross section for $b\bar{b}h$ as a function of $\mb$ (left) and relative uncertainties (right).
A strict expansion of the cross section \eq{bbhxs_res} has been performed.}
\label{fig:bbh-mb-scan}
\end{figure}

In this subsection we study the cross section as a function of $\mb$. The reason for this is to confirm
that the result obtained in the framework presented in this paper does indeed smoothly interpolate between
resummation and fixed-order regions. It also serves as an important validation of the method we 
employ to estimate uncertainties. 
In the left-hand plot of \fig{bbh-mb-scan} we show the LO+LL (dashed dark green) and NLO+NLL (solid navy) cross sections 
including error bands (green and blue bands respectively). 
We also plot central values for the associated fixed-order cross sections at LO (dotted dark green) and NLO (dashed navy). 
The right-hand plot of \fig{bbh-mb-scan} displays the relative size of the total LO+LL uncertainty (green band) and 
of the NLO+NLL resummation (light blue band) and total (navy band) uncertainties.
We emphasise that the results in \fig{bbh-mb-scan} are from an implementation of the strict expansion of the cross 
section in \eq{bbhxs_res}.

The first feature to point out is that at large $\mb$ both LO+LL and NLO+NLL results tend to their fixed-order 
counterparts (i.e., tend to the LO and NLO cross sections). This clearly shows that the framework we have introduced
indeed fulfills this desired property in the limit of large $\mb$. The fact that this transition occurs smoothly
is a natural result of the strict expansion used here (that is, there are no higher-order cross terms
in our result that might spoil the full cancellation between resummation pieces). The smooth transition between 
the low and high $\mb$ regions is also a direct consequence of the order counting we adopt.
If we were not to regard the $b$-quark PDF as an order $\as$ object,
then the strict expansion we have performed would not be possible and a discontinuity would 
be present in the region of $\mb \sim \muH$ arising from the higher-order terms. Finally, the smooth transition to 
the fixed-order results indicate that our method for including the power-suppressed $\ord{\mb^2/\mh^2}$ terms
to the strict EFT result works perfectly, and that as we have discussed in \subsec{FOtransition} it is indeed the case that 
(at least to the order we work to) it is possible to consistently include all power-corrections present in the 
fixed-order result.

Regarding the estimates of the perturbative uncertainty, \fig{bbh-mb-scan} reveals that the error bands we assign
are indeed reasonable and robust over the full range of $\mb$, with the NLO+NLL band fully contained within
that of the LO+LL.
The right-hand plot of \fig{bbh-mb-scan} indicates that the total uncertainty is dominated by the fixed-order 
scale uncertainty in the large-$\mb$ limit, and the resummation uncertainty vanishes, as it should,
in the limit $\mum \to \muH$. However, with decreasing $\mb$ we see that the resummation uncertainty 
becomes nonnegligible, forming an important component of the total error. The total uncertainty
is of the order of 12--14\% over most of the range of $\mb$ considered here, and grows as $\mb$ is increased
beyond the scale $\muH$.

\begin{figure}[t]
\begin{center}
\includegraphics[width=11.5cm]{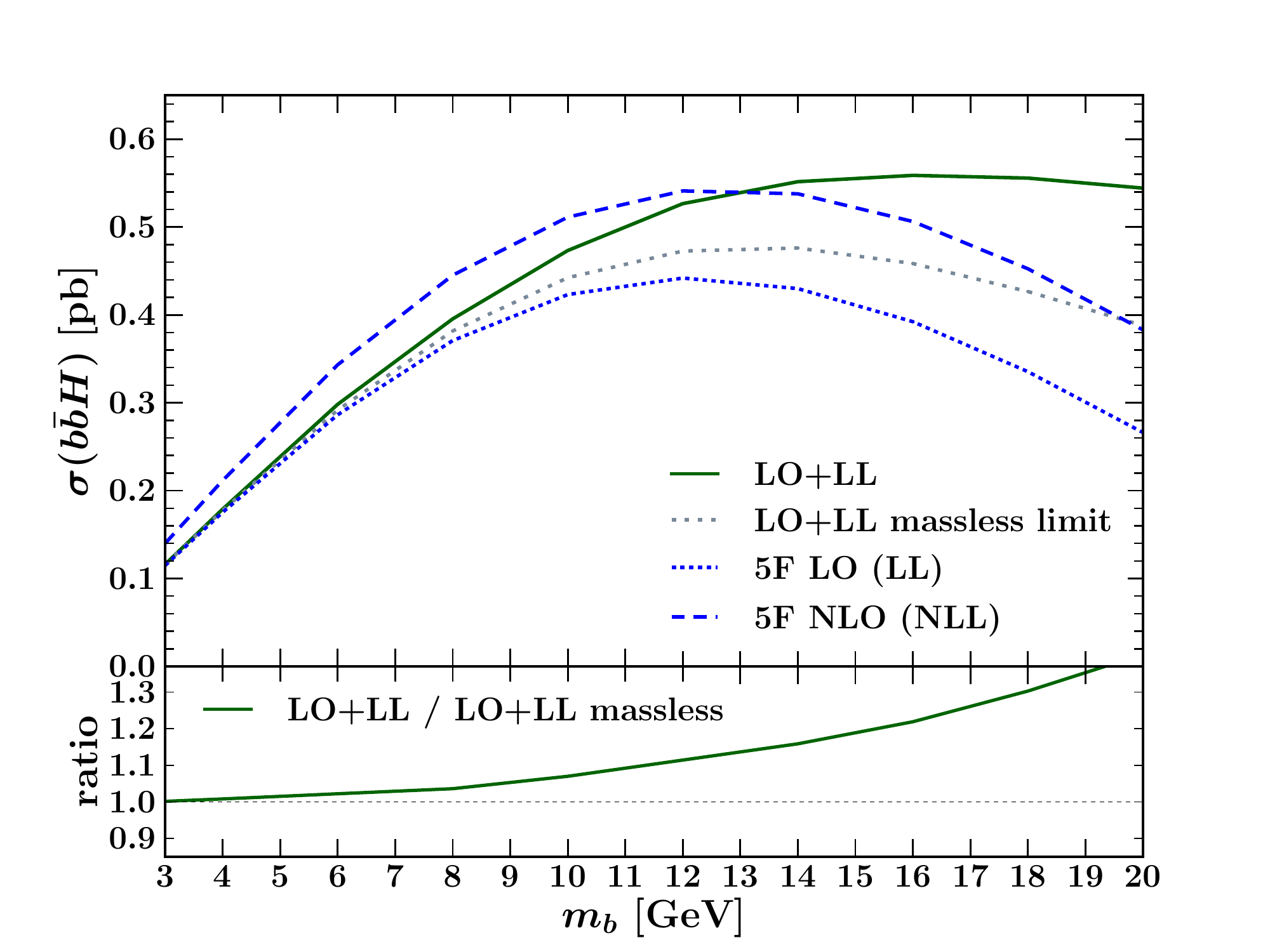}
\end{center}
\caption{A comparison of the LL+LO cross section where nonsingular corrections $\sim \mb^2/\mh^2$ are included (solid dark green)
and have been set to zero (dotted gray). The latter cross section is essentially a re-arranged 5F computation.
The lower panel indicates the size of such power corrections through the ratio of the two cross sections. For reference we
have included the LO (dotted blue) and NLO (dashed blue) 5F predictions in the upper panel.}
\label{fig:bbh-massless-comparison}
\end{figure}

Our result also allows us to consistently quantify the size of power corrections of $\ord{\mb^2/\mh^2}$.
This relies on the observation that in the small $\mb$ limit, due to the vanishing of the nonsingular
contributions, our result essentially becomes a re-arranged 5F computation. 
This means that all terms required to obtain a consistently matched prediction can in principle be extracted 
from a calculation that sets $\mb=0$ from the outset. The comparison between such a re-arranged 5F calculation 
and the result where the power corrections are included allows us to study the size of the latter.
To illustrate this, in \fig{bbh-massless-comparison} we compare the LO+LL prediction where
power corrections in $m_b$ have been included (solid blue) to the same prediction made with strictly massless coefficient functions (dotted gray).%
\footnote{Given that the term $C_{gg}^{(3)}$ is not required to make a prediction at NNLO for $b\bar{b}H$ in the 5F scheme, 
this ingredient in the massless limit is not known analytically and therefore we could not make the same comparison
for the NLO+NLL result. Nevertheless, by construction, exactly the same pattern that is observed for the LO+LL
result is expected to hold for the NLO+NLL result.}  
It is clear that in the small $\mb$ limit the nonsingular $\ord{\mb^2/\mh^2}$ terms are unimportant and that the matched result
can simply be constructed, with negligible errors due to missing power corrections, from massless coefficient functions.
This argument indicates that should S-ACOT or FONLL with standard perturbative counting be applied to the case of $b\bar{b}H$, 
then the resulting cross section will likely be almost the same as that of the 5F prediction.

It is also apparent that these power-corrections do increase in importance, their size exceeding 10\%, for $\mb \gtrsim 10$~GeV.
Therefore, in such parameter regions including them is vital for a faithful description of the cross section.
In \fig{bbh-massless-comparison} we have also plotted the 5F LO and NLO results, which deviate visibly from the LO+LL result as
$\mb$ grows, indicating that in such regions a massless 5F prediction becomes an inadequate description of the process. 

\begin{figure}[t]
\begin{center}
\includegraphics[width=11.5cm]{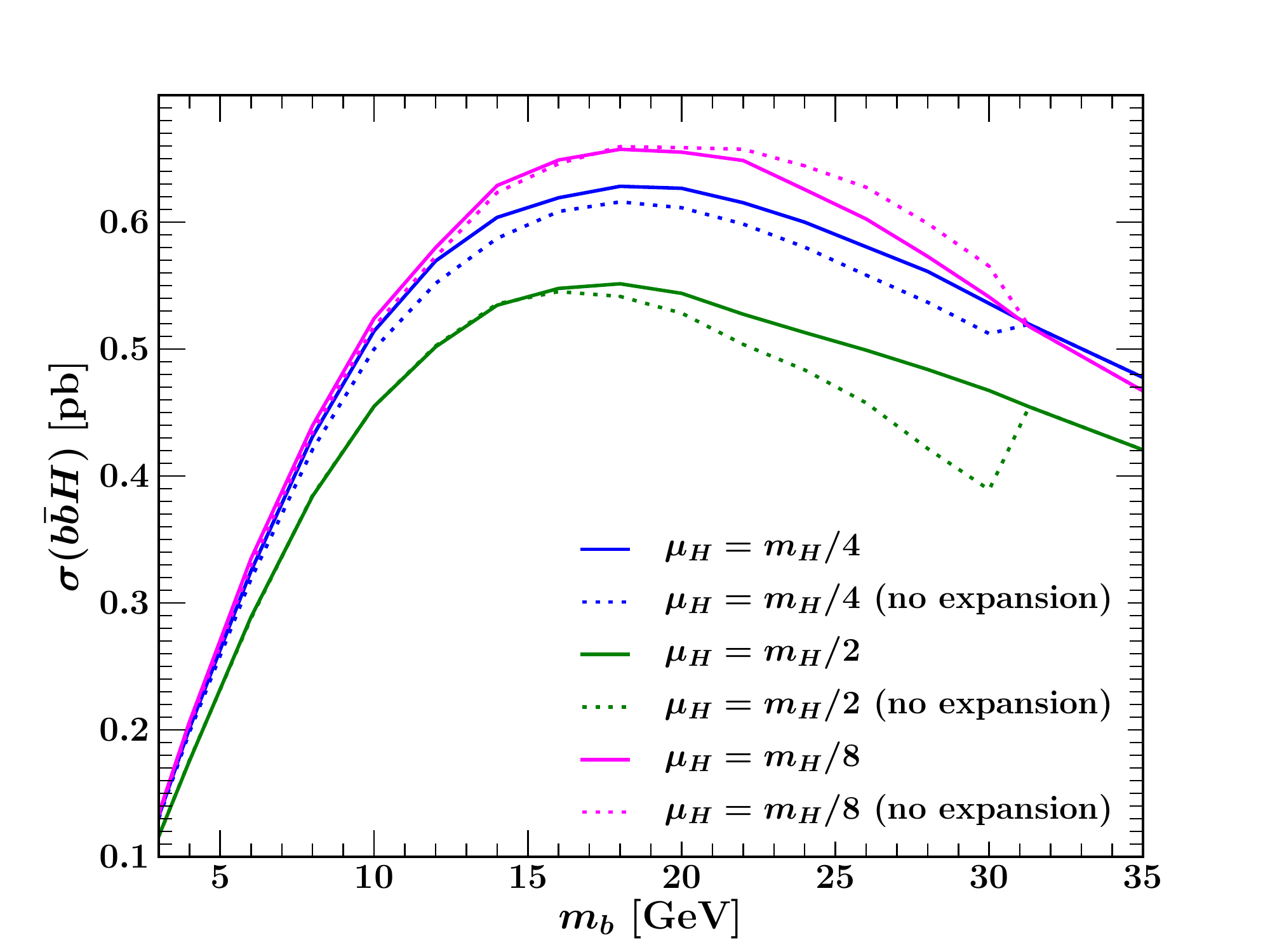}
\end{center}
\caption{NLO+NLL cross section for $b\bar{b}H$ using the nonexpanded implementation of \eq{bbhxs_res}. 
The solid and dotted curves correspond to the expanded and nonexpanded predictions respectively and 
magenta, blue and green colours correspond to the hard scale choices $\muH = m_H/8, \, m_H/4, \, m_H/2$.}
\label{fig:bbh-exp-vs-noexp}
\end{figure}

Finally, we make some brief comments regarding the difference between the cross section obtained under a strict perturbative 
expansion of \eq{bbhxs_res} (or equivalently \eq{res-final}) compared to that obtained by not expanding the two sets 
of matching coefficients. As mentioned earlier, not performing a strict expansion will generically 
lead to a discontinuity in the limit $\mum\to\muH$ due to the non-cancellation of spurious higher-order interference terms.
This is illustrated in \fig{bbh-exp-vs-noexp} where we
have plotted the NLO+NLL cross section predictions for three hard scale choices under a strict perturbative expansion (solid)
and with no strict expansion (dotted), namely using standard PDFs $f^\nfive_{i,b}$.  
At large $\mb$ the solid and dotted curves display significant differences and in particular the latter showing 
discontinuities for $\mum \to \muH$. 
In the small $\mb$ limit however, it is clear that the differences between expanded and nonexpanded approaches become 
much smaller. In particular, this means that the total uncertainty band in the region of physical $b$-quark masses
is basically the same in the two approaches. 
This property can be used to greatly simplify the practical implementation in this region and we have 
exploited this to produce the results in \subsec{santander}. 
However, it is also important to point out that in general it is only an expanded result, akin to that of \eq{resum_exp_compact}, 
that guarantees a consistent and smooth matching between resummation and fixed-order regions. 
These observations may well be important when considering heavy-quark initiated processes where the value of $m/Q$ is not 
as small as in the setup we study here.

\subsection{LHC phenomenology}
\label{subsec:santander}

Here we return to the phenomenologically relevant case of a physical bottom mass and consider the 
cross section as a function of the Higgs mass. The previous subsection confirmed that the framework
we use leads to a cross section that consistently describes both resummation and fixed-order regions.  
It is of course of interest to compare in a meaningful manner our (N)LO+(N)LL predictions to other
predictions available, namely predictions in the 4F and 5F schemes, as well as to the
Santander matching prescription used to combine these two. The latter is a practical formula that combines 
the 4FS and 5FS predictions for the total inclusive cross section through a weighted average of the two \cite{Harlander:2011aa},
\begin{align} \label{eq:santander-matching}
\sigma^{\text{matched}} = \frac{\sigma^{\text{4FS}} + \omega \sigma^{\text{5FS}}}{1+\omega}, \;\;\;\text{where}\;\;\;
\omega = \mylog{\frac{\mh}{\mb}} -2
\,.\end{align}  
This construction is such that the combined result tends to that of the 4FS when the collinear logarithms, $\ln(\mh/\mb)$ 
are small, and to that of the 5FS when the logarithms are large (i.e., in the limit $\mb/\mh \to 0$). The choice of the 
weight $\omega$ is motivated by the fact that this leads to roughly equal weights being assigned for the 4FS and 5FS numbers
around $\mh \sim 100$ GeV, which is the region of `best' agreement between the 4FS and 5FS predictions (see for example \mycite{Campbell:2004pu}). 
Beyond this motivation the choice of $\omega$ is arbitrary and there is no strong theoretical argument preventing the 
choice of alternative weights or different ways of averaging the two cross section predictions. Moreover, the practical
formula combines two predictions made using different PDF and $\overline{m}_b(\overline{m}_b)$ inputs, 
which is somewhat inconsistent.  
The estimate of the uncertainty on a Santander matched prediction is given by the error band obtained by applying the 
formula \eq{santander-matching} to the upper and lower uncertainty curves of the 4F and 5F predictions. 

Regarding the set of inputs we use, we have chosen to stick as closely as possible to those used in the
LHCHXSWG~\cite{Dittmaier:2011ti,Dittmaier:2012vm,lhchxswg:bbh} and also those used in recent studies of
$b\bar{b}H$ production~\cite{Wiesemann:2014ioa}. 
Explicitly, we use the default MSTW2008 PDF sets, at 
the appropriate order for the LO, NLO and NNLO 5FS predictions, whilst we use the fixed-flavour $n_f=4$ set for the 4FS
predictions. For the 4FS and Santander matched results we explore the effect of using $\overline{m}_b(\overline{m}_b) \neq 4.16$~GeV (i.e., a different
$\overline{m}_b(\overline{m}_b)$ than that used in 5FS predictions), as done by the LHCHXSWG.%
\footnote{The reason the choice $\overline{m}_b(\overline{m}_b) = 4.34 \text{ GeV } \neq 4.16$~GeV is made in some 4FS predictions is that this $\MSb$ mass corresponds to a 1-loop conversion of the used pole mass. However, by using a fixed pole mass as input one reintroduces the pole mass renormalon ambiguity into the cross section through $Y_b$.}
The central choice of hard scale is $\muH=(\mh+2\mb)/4$, with $\mb=4.75$~GeV,
and we vary this hard scale by a factor of two to obtain the 
fixed-order uncertainty. The errorbars for the 4FS and 5FS predictions are obtained by setting $\muf = \mur = \muH$, 
that is we do not consider $\muf \neq \mur$ variations here.
The bands for the Santander matched cross sections are obtained with the Santander prescription. 

\begin{figure}[t]
\begin{center}
\includegraphics[width=14.5cm]{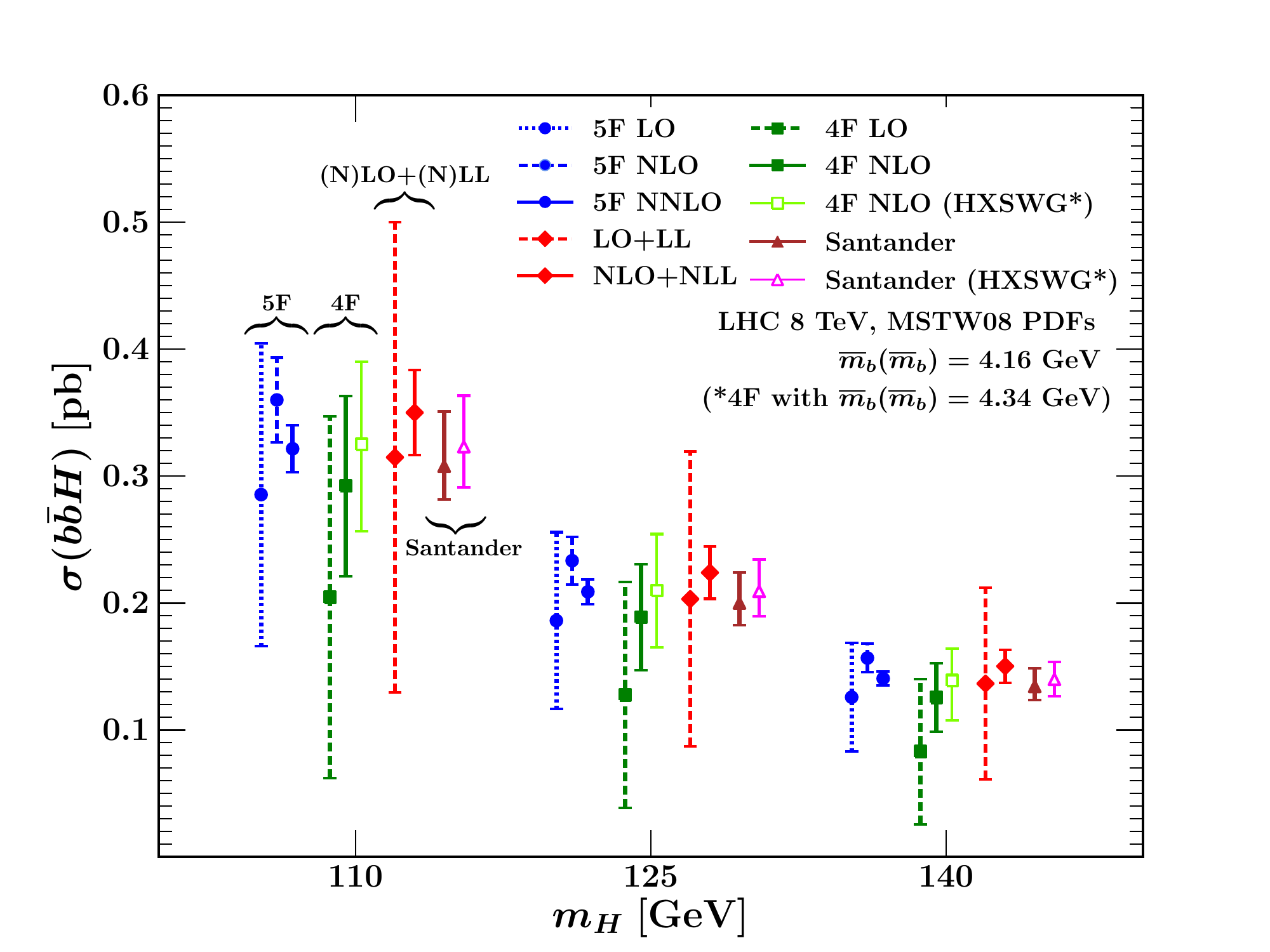}
\end{center}
\caption{A comparison between all available cross section results for $b\bar{b}H$. }
\label{fig:bbh-mH-scan-comparison}
\end{figure}

In \fig{bbh-mH-scan-comparison} we plot the 5FS (blue points) and 4FS (green points) cross sections,
the (N)LO+(N)LL matched predictions (red points) as well as the Santander matched cross sections (brown and magenta points) 
for $\mh \in \{ 110, 125, 140 \}$ GeV. 
The error bands for the (N)LO+(N)LL predictions have been obtained as 
discussed in \eq{uncertainties}. We note here that since the ratio $\mb/(\mh/4) < 0.3$, the profile region we are in is actually
always linear, namely $\mum(\mb,\mh, \muH) \propto \mb$.
Compared with the green 4FS NLO point, the light green 4FS NLO point has been obtained by setting $\overline{m}_b(\overline{m}_b) = 4.34$~GeV.
This (somewhat artificially) shifts the cross section upwards by $\sim 10\%$ by increasing $Y_b(\muH)$. 
The magenta Santander point has been computed with the 4FS component using this increased value of $\overline{m}_b(\overline{m}_b)$
and is therefore seen to be higher than the point that uses a consistent $\MSb$ mass in both 4FS and 5FS components.
We have checked that the results of \fig{bbh-mH-scan-comparison} are fully consistent with those in the literature.

As seen in the previous subsection the error band of the NLO+NLL predictions is contained within that of the LO+LL.
The LO+LL uncertainty band is quite wide (and larger than the 4FS or 5FS LO bands) --- something that is to be expected 
from a LO prediction and a maximal variation of the two matching scales, \eq{uncertainties}.
Rather than being of concern, this observation gives us confidence that we do not underestimate the inherent uncertainties
present and furthermore allows us to trust the size of the NLO+NLL band.
We additionally observe that most of the LO+LL scale uncertainty is driven by the $\mum$ variation,
while at NLO+NLL the uncertainty band is dominated by the $\muH$ variation.
On the other hand, the 4FS displays a less significant reduction in its uncertainty band at NLO,
and with its large correction shows relatively poor perturbative convergence due to the presence of unresummed logarithms.
The 5FS band shrinks more visibly with increasing order, however the NNLO central value lies outside
the NLO band.%
\footnote{We note that the 5FS bands (and to a much lesser extend the 4FS bands) would be larger if we had 
considered $\mur \neq \muf$, potentially improving the convergence pattern described above. 
We have not done this in order to directly compare to our method of estimating uncertainties.}

The (N)LO+(N)LL results lie significantly higher than their respective 4FS (N)LO counterparts. This is due to the resummation 
contained in the former results. We also notice that the NLO+NLL results lie slightly above the NNLO 5FS predictions. The
reason for this is that the former results contain the (positive) effects of light channels ($gg$, $qg$ and $q\bar q$) at $\ord{\as^3}$
whilst the latter results contain (negative) two-loop corrections to the $b\bar{b}$ channel (see \fig{bbh-diagrams}).

By construction the Santander matched result lies between the 4FS NLO and 5FS NNLO predictions (irrespective of 
the precise inputs for $\overline{m}_b(\overline{m}_b)$ used in the 4FS calculation). 
Our NLO+NLL result is therefore higher than the Santander matched results, and specifically
we find an increase of 6\% with respect to the magenta LHCHXSWG Santander point, and of 12\%
with respect to the brown Santander point (which is directly comparable to the NLO+NLL result).
We also notice that the error bands of the NLO+NLL prediction are roughly of the same size as those obtained through
Santander matching indicating that the size of the latter is likely to be realistic. 
There is a sizeable overlap in the uncertainty bands of the Santander and the NLO+NLL predictions,
however the central Santander points lie towards the bottom edge (magenta) and outside (brown) of the NLO+NLL error band.
We also point out that while our NLO+NLL result is stable upon hard scale variation, the Santander matched result
would be significantly smaller for larger hard scales, as is clear by inspection of \fig{bbh-muf-variation}.

\begin{figure}[t]
\begin{center}
\includegraphics[width=14.5cm]{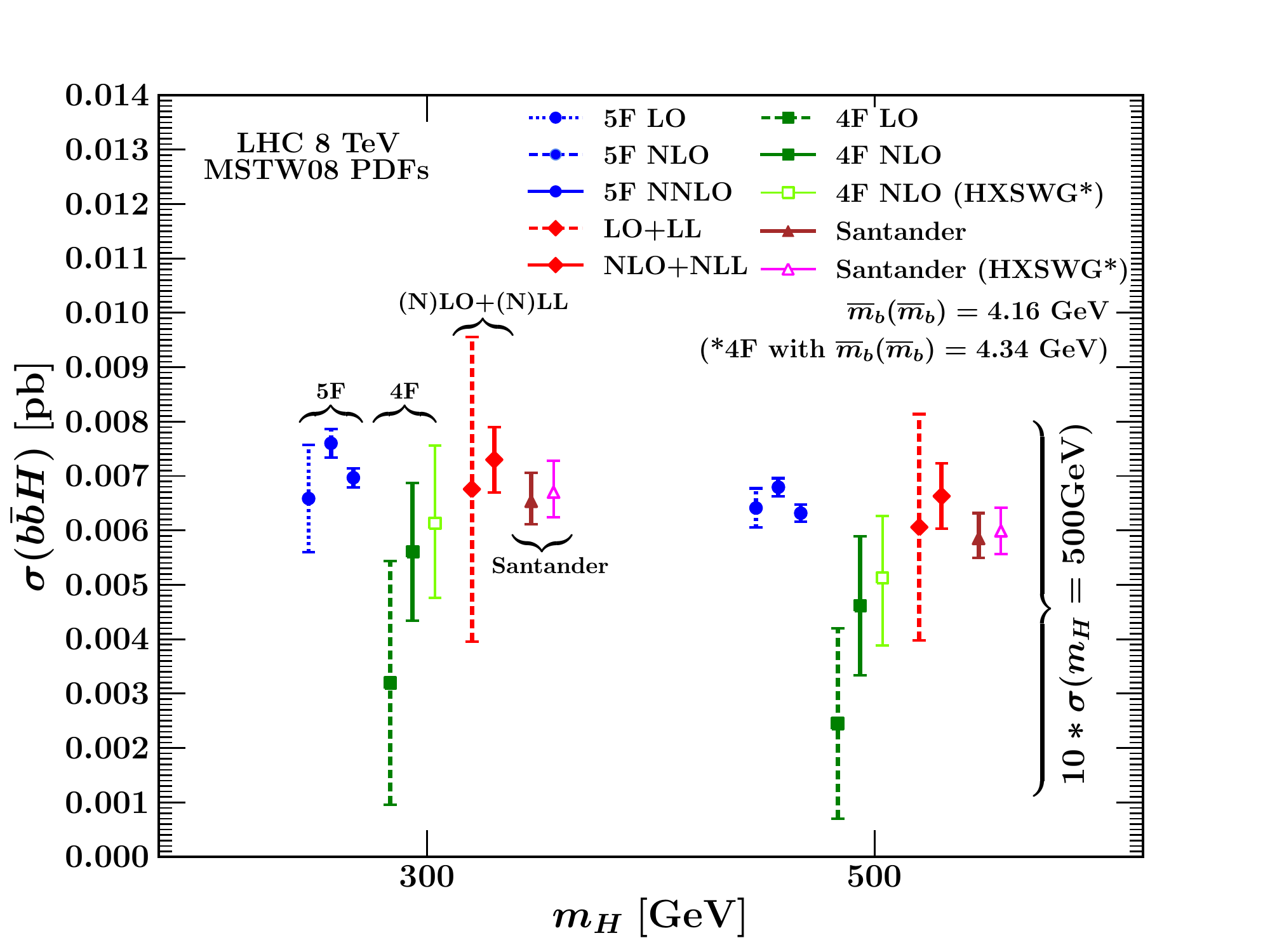}
\end{center}
\caption{Same as Fig.~\ref{fig:bbh-mH-scan-comparison} but for large Higgs masses. }
\label{fig:bbh-mH-large-scan-comparison}
\end{figure}

In \fig{bbh-mH-large-scan-comparison} we present the same comparison as in \fig{bbh-mH-scan-comparison}
for larger Higgs masses, $\mh=300$~GeV and $\mh=500$~GeV.
The overall pattern observed for a light Higgs remains qualitatively unchanged.
The NLO corrections in the 4FS have grown noticeably, such that the 4FS NLO results are now outside the 4FS LO bands,
which is likely due to the unresummed logarithms, which get larger at higher Higgs masses.
While at lower $\mh$ the central values of the 4FS NLO and 5FS NNLO are relatively close to each other and
have overlapping uncertainty bands, at higher $\mh$ they are further apart and have only barely
overlapping uncertainty bands. Consequently, the Santander average becomes even less reliable.
Encouragingly, the LO$+$LL and NLO$+$NLL results at large $\mh$ continue to display the good perturbative behaviour and convergence
pattern present at lower $\mh$ values. They also remain systematically higher than the Santander matched predictions.

Finally, we briefly comment on the effect of including the known $C^{(2)}_{b\bar b}$ term\footnote
{Additionally, we include the lowest order $bb$-channel term $C^{(2)}_{bb}$, appearing at the same order, which however gives a negligible effect.}
(i.e., the two-loop corrections to the $b\bar b$-channel),
which is formally of higher order in our approach. This is the only known contribution that we have not included in the NLO+NLL result.
The NLO+NLL result for $\mh=125$~GeV shown in \fig{bbh-mH-scan-comparison} is $\sigma_{\text{NLO+NLL}} = 0.224 \pm 0.021$~pb. 
With the addition of the higher-order $C^{(2)}_{b\bar b}$-term this cross section becomes
$\sigma_{\text{NLO+NLL+}C_{b\bar b}^{(2)}} = 0.211 \pm 0.010$~pb,
to be compared with the NNLO 5FS cross section of $\sigma_{\text{5F,NNLO}} = 0.209 \pm 0.010$~pb.
Clearly the addition of this higher-order term reduces the NLO+NLL cross section and additionally reduces its uncertainty band,
bringing both central value as well as the size of the error bands closer to those of the 5FS NNLO result.
This term is likely to be important at very high scales $\muH\gtrsim1$~TeV, where the 5FS perturbative counting
is expected to be more appropriate. However, we feel that including this term for a SM Higgs is rather ad-hoc given that there 
are a number of additional terms that would also contribute at the same order, but which are not known and have not been included here.
Furthermore, the sizeable reduction of the error bars is slightly discomforting given that we have no control on the effects 
of these missing terms. We note that the error bars of the original NLO+NLL cross section nicely cover the effect of 
adding the $C^{(2)}_{b\bar b}$ term, which provides further support that the uncertainty bands presented are indeed reasonable.

\section{Conclusions}
\label{sec:conclusions}

We have presented a systematic EFT setup to derive heavy-quark initiated cross sections at hadron colliders.
Our framework includes the resummation of potentially large logarithms $\sim\ln(\mb/Q)$.
Furthermore, it consistently includes power corrections $\sim \mb^2/Q^2$, reproducing the full fixed-order (4FS) result.
As such our final result gives predictions that are accurate in both of the limits $\mb\ll Q$ and $\mb\sim Q$
as well as in the transition region in between.

Our result is obtained via a two-step matching procedure, and variation of the scales at which
these two matchings are performed allows us to obtain a robust estimate of the perturbative uncertainties.
The construction of the coefficient functions of our result bears several similarities with existing 
VFNSs for DIS.
A key difference in our approach is the different perturbative order counting.
In particular, we argue that it is more appropriate to count the effective $b$-quark PDF, which is generated
perturbatively at the scale $\mum$, as a perturbative object of $\ord{\as}$. This organization of the perturbative
series leads to perturbatively stable results and allows for a smooth transition
between fixed-order and resummation regions.
The simplicity of the EFT approach for DIS makes its generalization to hadron-hadron collisions 
straightforward.

We have applied our framework to the case of the $b\bar{b}H$ cross section. We first studied
the cross section as a function of $\mb$, which served to demonstrate that our resummed and matched result
satisfies all required properties.
We then presented numerical results of phenomenological interest for the LHC.
We have compared our results to the 4FS and 5FS result, as well as
the Santander prescription, which combines the two results by taking a weighted average.
Since our predictions are derived from a field-theory setup, consistently combining
the 4FS and 5FS limits, it can be regarded as a definite improvement over this prescription.
At NLO+NLL, we find a slightly reduced perturbative uncertainty compared to the
Santander average. The Santander central value is lower than the NLO$+$NLL result
by about 12\% for $m_H=125$~GeV and lies outside our uncertainty band
when consistent $\MSb$ masses are used in both the 5FS and 4FS ingredients of the Santander result.
The difference is reduced to 6\% when using a larger $\MSb$ mass in the 4FS result,
as used by the LHCHXSWG, with the central value lying at the lower edge of our uncertainty band.
We observe that the NLO+NLL result is stable upon variation of the hard scale, unlike the 
Santander matched results which would vary significantly under such variation.

The framework presented in this paper can be straightforwardly applied to
other processes involving initial-state $b$-quarks, for example single-top or $V+b$-jet production.
Additionally, extending the framework to study more differential observables of interest, such as the transverse
momentum of the Higgs in $b\bar{b}H$-production or jet-vetoed cross sections is possible. 
In making the calculation less inclusive, more scales appear in the perturbative expansion, which can be efficiently dealt with in an EFT setup.
Finally, it would be very interesting to adopt our perturbative counting in DIS in the context of
PDF fits, where the used variable flavor number scheme typically plays a central role in determining the accuracy
and the goodness of the fit itself.

\vspace{0.5cm}
\noindent
\paragraph{Note added:} While this paper was being finalized \mycite{Forte:2015hba} appeared, which obtains
the $b\bar{b}H$ cross section in the FONLL approach.
As discussed in \sec{comparison}, the FONLL approach adopts a different perturbative counting to 
the one we have presented here. In particular, the result of \mycite{Forte:2015hba} is computed 
at FONLL-A accuracy, that is, it combines the 4FS LO result with the 5FS NNLO result.
In comparison to our NLO+NLL result, it does not include the 4FS NLO contributions from $\bar C^{(3)}_{gg}$, 
$\bar C^{(3)}_{q\bar{q}}$, and $\bar C^{(3)}_{qg}$. However, it does include the $C^{(2)}_{b\bar{b}}$ 
and $C^{(2)}_{bb}$ 5FS NNLO terms, which are higher-order terms in our approach (and whose impact on our result is 
discussed at the end of \subsec{santander}).
As a result, the FONLL-A result of \mycite{Forte:2015hba} is very close to the 5FS NNLO, as one might expect, since the difference
to the latter is the inclusion of the numerically small $\ord{\as^2\,m_b^2/m_H^2}$ power corrections from the 4FS LO result.
Finally, \mycite{Forte:2015hba} does not include an estimate of the resummation and matching uncertainty
(analogous to our $\mum$ variation).

\begin{acknowledgments}
We thank Stefano Forte, Andr\'e Hoang, Fabio Maltoni, Piotr Pietrulewicz, Luca Rottoli  and Marius Wiesemann for useful discussions.
The work of MB is supported by an European Research Council Starting Grant ``PDF4BSM: Parton Distributions in the Higgs Boson Era''.
The work of FT was supported by the DFG Emmy-Noether Grant No.\ TA 867/1-1.
The work of AP was in part supported by the Collaborative Research Center SFB676 of the DFG, ``Particles, Strings, and the Early Universe"
\end{acknowledgments}

\appendix

\section{Renormalization}
\label{app:ren}

In this appendix we briefly discuss aspects of renormalization important to our work.
In the $\MSb$ scheme ultraviolet (UV) divergences associated with
quark lines are subtracted in the same way for all quarks, independently of their mass
(mass doesn't play a role in the UV).
However, a direct consequence of the $\MSb$ scheme is that $\as$ runs with
$n_f=6$ flavors, irrespectively of the energy scale.
A more physical renormalization scheme for heavy quarks is the so called CWZ or \emph{decoupling scheme}~\cite{Collins:1978wz},
where all ``light'' quarks are renormalised with $\MSb$ counter terms,
while UV divergences associated with the heavy-quark loops are subtracted at zero momentum.
This ensures decoupling of the heavy quarks at energies much smaller than their mass.
Since the concept of light and heavy depends on the actual scale, in practice a variable flavor number renormalization scheme
is used, where the number of active (light) quarks renormalised in $\MSb$ depends on the hard scale
and changes at the crossing of the heavy quark thresholds. We have used this in Sect.~\ref{sec:EFT} 
when treating the fixed-order and resummation regions differently regarding the renormalization of $b$-quark
loops.

At the threshold scale $\mu=\mum$, matching conditions relate the value of $\as$ above and below that scale.
Denoting with a superscript the number of flavors used in the evolution of $\as$, we have
\begin{align}
\frac{\as^\nfive(\mum^2)}{\as^\nfour(\mum^2)} &=
1+ \frac{\as^\nfour(\mum^2)}\pi \frac{T_F}{3}\ln\frac{\mum^2}{m^2}
+\ldots
\label{eq:alphasmatching}
\end{align}
where $m$ is the heavy quark pole mass.
These conditions are currently known through 4 loops~\cite{Chetyrkin:1997sg},
but for our applications we just need the 2 loop expression of \eq{alphasmatching}.

A generic observable $F(\as)$ can be written as a perturbative expansion equivalently in both schemes,
\beq
F(\as) = \sum_k {\as^\nfour}^k(\mu) F^{\nfour(k)}(\mu) = \sum_k {\as^\nfive}^k(\mu) F^{\nfive(k)}(\mu),
\eeq
and to all orders they are identical.
The relation between the coefficients $F^{\nfour(k)}$, whose heavy quark UV divergences are renormalised in the decoupling scheme,
and $F^{\nfive(k)}$, renormalised in $\MSb$, can be simply obtained by using \eq{alphasmatching}
to write $\as^\nfour$ in terms of $\as^\nfive$ in the first sum (or viceversa), re-expanding and matching order by order.

\section{Construction of PDFs with variable threshold}
\label{app:PDFs}

In this work we have considered PDFs in which the heavy quark thresholds are not fixed to
the heavy quark mass, but can vary. Moreover, the effective PDFs defined in \eq{5FPDF_exp_resum}
are required with mixed evolution and matching accuracies.
If we consider the ``universal'' PDFs as those at a small scale $\muL$,
then the dependence on the threshold and the details of the order at which each ingredient is retained
are all in the perturbative evolution.
Typically PDFs are used through the \texttt{LHAPDF} library, where evolved PDFs
at any (available) scale are built from interpolation grids, previously created assuming
a particular evolution with specific heavy quark thresholds.

For our purposes, performing the evolution each time picking the desired value of the heavy quark thresholds
seems advisable. However, this approach faces speed problems, since
performing the evolution is much more time consuming than interpolating a grid.
Therefore, for the work in this paper we have created \texttt{LHAPDF} grids
with different choices for the $b$-quark threshold, $\mum$,  and with the required
combinations of the perturbative ingredients at different orders.
To do so, we have used the public code \texttt{APFEL}, which has the ability of creating grids
after performing its own evolution. To accommodate the possibility of choosing a threshold
different from the heavy quark mass, we have modified the code adding the $\mum$-dependent matching
conditions through NNLO from Ref.~\cite{Buza:1996wv}.
Furthermore, we modified the code to produce PDF grids with the required combinations of 
orders of the matching coefficients, $\Mm_{ij}$. 
Practically, we proceeded as follows:
\begin{itemize}
\item We start with a central member of a public PDF set, namely MSTW2008 with $\as(m_Z)=0.1171$, both at NLO and at NNLO.
  The value of the bottom pole mass used in this work has been taken to be $m_b=4.75$~GeV for consistency with the chosen PDF set.
\item We compute all the required PDFs as well as $\as$ from this set at an initial scale $\mu_0=m_c=1.4$~GeV.
\item We use \texttt{APFEL} to perform the forward evolution and create the corresponding grids for the different
  setups we are interested in.
\end{itemize}
Note that we choose to perform the evolution at (N)NLL, starting from a (N)NLO set, for
producing the PDFs that we used in our (N)LO+(N)LL results.
While this is somewhat in constrast with the discussion in \subsec{powercounting},
using the evolution at one higher order has two advantages.
The first is that the order of the evolution and the highest order in the matching functions
are consistent, so that when we do not use the strict expansion we can use standard PDFs, as explained
below \eq{rescount}. The second is that our final NLO+NLL result in \sec{bbh} is more directly comparable to
the 5FS NNLO result.

\begin{figure}[t]
\begin{center}
\includegraphics[width=0.495\textwidth,page=10]{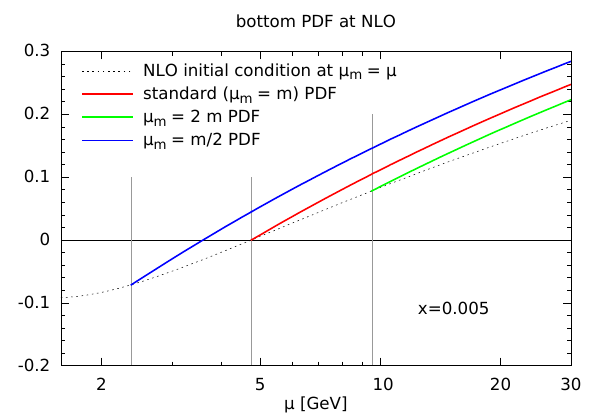}
\includegraphics[width=0.495\textwidth,page=11]{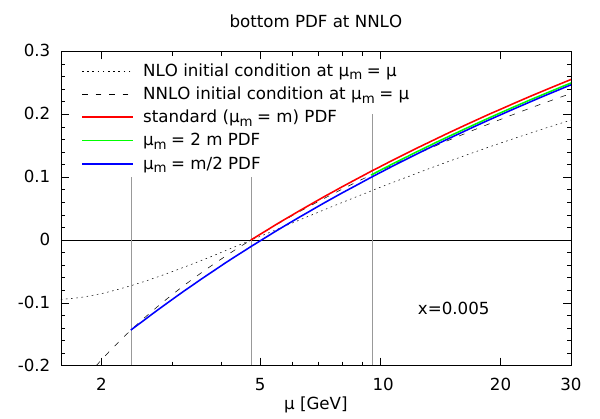}
\end{center}
\caption{Bottom PDF as a function of the scale at fixed $x=0.005$ for different values of the bottom threshold $\mum$
  at NLO (left) and NNLO (right).}
\label{fig:bottomPDFthresholds}
\end{figure}
As an illustration of what the modified code is capable of, in \fig{bottomPDFthresholds} we show the standard 
5F bottom PDF $f_b^\nfive$ for a fixed value of $x=0.005$. We plot this as a function of the scale $\mu$ close to 
the bottom threshold for three different values of the threshold itself, $\mum=m_b/2,m_b,2m_b$, as used in the 
phenomenology section \ref{subsec:santander}.
Changing the threshold of course shifts the value of the PDF at $\mum$, and in particular makes this value nonzero
at NLO (at NNLO it is already nonzero even for $\mum=m_b$).
The initial condition, shown as the gray dotted or dashed lines, is given by the product of 4F PDFs with the matching conditions.
We observe that at NNLO the PDFs obtained with different values of the threshold are almost identical at
a large scale, indicating that a NNLO cross section is only likely to have a mild dependence on $\mum$.
We also note that the initial condition itself behaves in a nice perturbative way at large scales,
where we expect higher order corrections to be small, while it deviates strongly at smaller scales,
where $\as$ is larger and higher-order corrections are not negligible.

\section{Massive kinematics in hadron-hadron collisions}
\label{app:collider_kin}

In the case of an incoming massless parton its four-momentum is considered to be
a fraction of the four-momentum of the proton. In case of massive incoming parton this 
formulation would lead to a mass that scales with the momentum fraction, which is clearly inconsistent. 
To overcome this, the proper approach~\cite{Collins:1998rz} is to use light-cone coordinates, where 
momenta can be written as
\beq
p = \(p^+, p^-, \vec p_t\), \qquad p^\pm = (p^0\pm p^3)/\sqrt{2},
\eeq
where $p^i$ are the usual Minkowski components. We choose to orient the beam axis along the third spatial direction.
A collinear particle with mass $m$ has $\vec p_t = \vec0$, and the mass can be expressed as $m^2=2p^+p^-$.
We can write the momenta of massless protons\footnote{The proton mass can always be neglected compared to its momentum at the LHC.}
in the hadronic center-of-mass frame as
\beq
P_1 = \(P_1^+, 0, \vec 0\),\qquad 
P_2 = \(0, P_2^-, \vec 0\)
\eeq
with $P_1^+=P_2^-=\sqrt{S/2}$, and $S=(P_1+P_2)^2$ is the total invariant mass squared of the colliding protons.
The collinear component of a parton's momentum scales with the largest light-cone 
component of the proton, so two incoming partons with momentum fractions 
$x_1$ and $x_2$ have momenta
\beq
p_1 = \(x_1 P_1^+, \frac{m_1^2}{2x_1P_1^+}, \vec 0\),\qquad
p_2 = \(\frac{m_2^2}{2x_2P_2^-}, x_2 P_2^-, \vec 0\).
\eeq
The accessible values of $x_{1,2}$ are determined by kinematic constraints.
Imposing the condition that the parton energy cannot be greater than the proton energy,
we find the constraints
\beq\label{eq:condition_x}
x_i \geq \frac{m_i^2}{S} + \Ord{\frac{m_i^4}{S^2}}, \qquad
x_i \leq 1-\frac{m_i^2}{S} + \Ord{\frac{m_i^4}{S^2}}, \qquad
i=1,2.
\eeq
Since the masses $m_{1,2}^2$ are negligible with respect to $S$,
these conditions reproduce the massless limit constraint $0\leq x_{1,2}\leq 1$.
The partonic invariant mass squared is given by
\beq\label{eq:s_y}
s = (p_1+p_2)^2 = x_1x_2 S + m_1^2 + m_2^2 + \frac{m_1^2m_2^2}{x_1x_2 S}.
\eeq
In order to produce a final state of invariant mass $M$ the inequality
\beq\label{eq:inequality}
s\geq M^2
\eeq
must be satisfied.
If we consider at least one of the two partons to be massless (say $m_2=0$),
this inequality has a single solution $x_1x_2\geq(M^2-m_1^2)/S$.
This is already problematic, since for small invariant masses $M\to m_1$
very small values of $x_{1,2}$ are accessible.
The situation becomes worse if both partons are massive,
as in the case of the subprocess $b\bar b\to H$. In this case there are two solutions of \eq{inequality},
namely (setting $m_1=m_2=m$ for simplicity)
\beq
x_1x_2 \geq \frac{M^2}{4S}\[1+\sqrt{1-\frac{4m^2}{M^2}}\]^2
\qquad\text{and}\qquad
x_1x_2 \leq \frac{M^2}{4S}\[1-\sqrt{1-\frac{4m^2}{M^2}}\]^2,
\eeq
where the second solution represents a new region of very small $x_i$ that is inaccessible in the massless case.
The physical interpretation is as follows. As the momentum fraction of a parton is reduced,
its energy reaches a minimum at $x_i=m_i/\sqrt{S}$ (where it is not moving in the center-of-mass frame), 
and then starts increasing upon further reduction of $x_i$. 
Therefore, at very small $x_i$, a parton is very energetic again thus making it possible to produce
a high invariant mass final state. In this configuration both heavy quarks have become
``anticollinear'' with respect to their respective protons.

It is clear from this simple kinematical argument that a massive extension of the standard factorisation theorem
cannot just work in its usual form in presence of two incoming hadrons (otherwise, there would be a huge contribution
from unconstrained small-$x$ PDFs).
When there is just one proton, as in DIS, the problematic configuration described above never takes place,
the effect of the parton's mass being a further restriction to the accessible values of $x$ with respect to the massless case,
and standard collinear factorisation works even in presence of massive partons~\cite{Collins:1998rz}.
In the hadron-hadron collider case, only a systematic expansion in the heavy quark mass
such as the one presented in this work allows the description of heavy quarks in the initial state.

\phantomsection
\addcontentsline{toc}{section}{References}

\bibliographystyle{jhep}
\bibliography{QCD}

\end{document}